\documentclass[preprint]{aastex}
\usepackage{emulateapj5,apjfonts}

\begin{document}

\def\plotthree#1#2#3{\centering \leavevmode
\epsfxsize=0.30\columnwidth \epsfbox{#1} \hfil
\epsfxsize=0.30\columnwidth \epsfbox{#2} \hfil
\epsfxsize=0.30\columnwidth \epsfbox{#3}}

\title{Extraplanar Emission-Line Gas in Edge-On Spiral Galaxies. \\
I. Deep Emission-Line Imaging}

\author{Scott T. Miller\altaffilmark{1,2} and Sylvain
Veilleux\altaffilmark{2,3,4}}

\affil{Department of Astronomy, University of Maryland, College Park,
MD 20742; \\ stm, veilleux@astro.umd.edu}

\altaffiltext{1}{Current address: Department of Astronomy,
Pennsylvania State University, 525 Davey Lab., University Park, PA
16802; stm@astro.psu.edu}

\altaffiltext{2}{Visiting Astronomer, Kitt Peak National Observatory
and Cerro Tololo Inter-American Observatory, National Optical
Astronomy Observatory, which is operated by the Association of
Universities for Research in Astronomy, Inc. (AURA) under cooperative
agreement with the National Science Foundation; Visiting Astronomer,
Anglo-Australian Observatory, P.O.  Box 296, Epping, NSW 1710,
Australia; Visiting Astronomer, Isaac Newton Group of
Telescopes, Santa Cruz de La Palma, Tenerife, Spain}

\altaffiltext{3}{Current address: 320-47 Downs Lab., Caltech, Pasadena, 
CA 91125 and Observatories of the Carnegie Institution of Washington, 
813 Santa Barbara Street, Pasadena, CA 91101; veilleux@ulirg.caltech.edu}

\altaffiltext{4}{Cottrell Scholar of the Research Corporation}

\begin{abstract}
The extraplanar diffuse ionized gas (eDIG) in 17 nearby, edge-on disk
galaxies is studied using deep Taurus Tunable Filter H$\alpha$ and
[N~II] $\lambda$6583 images and conventional interference filter
H$\alpha$ + [N~II] $\lambda$6548, 6583 images which reach flux levels
generally below $\sim$ 1 $\times$ 10$^{-17}$ erg s$^{-1}$ cm$^{-2}$
arcsec$^{-2}$. [N~II] $\lambda$6583/H$\alpha$ excitation maps are
available for 10 of these objects.  All but one galaxy in the sample
exhibit eDIG.  The contribution of the eDIG to the total H$\alpha$
luminosity is relatively constant, of order 12 $\pm$ 4\%.  The
H$\alpha$ scale height of the eDIG derived from a two-exponential fit
to the vertical emission profile ranges from 0.4~kpc to 17.9~kpc, with
an average of 4.3 kpc. This average value is noticeably larger than
the eDIG scale height measured in our Galaxy and other galaxies.  This
difference in scale height is probably due in part to the lower flux
limits of our observations. The ionized mass of the extraplanar
component inferred by assuming a constant filling factor of 0.2 and a
constant pathlength through the disk of 5 kpc ranges from 1.4 $\times$
10$^7$ M$_{\odot}$ to 2.4 $\times$ 10$^8$ M$_{\odot}$, with an average
value of 1.2 $\times$ 10$^8$ M$_{\odot}$. Under these same
assumptions, the recombination rate required to keep the eDIG ionized
ranges from 0.44 to 13 $\times$ 10$^6$ s$^{-1}$ per cm$^{-2}$ of the
disk, or about 10 to 325\% of the Galactic value.  A quantitative
analysis of the topology of the eDIG confirms that several galaxies in
the sample have a highly structured eDIG morphology.  The distribution
of the eDIG emission is often correlated with the locations of the
H~II regions in the disk, supporting the hypothesis that the
predominant source of ionization of the eDIG is photoionization from
OB stars located in the H~II regions. A strong correlation is found
between the IR (or FIR) luminosities per unit disk area (basically a
measure of the star formation rate per unit disk area) and the
extraplanar ionized mass, further providing support for a strong
connection between the disk and eDIG components in these galaxies. The
excitation maps confirm that the [N~II]/H$\alpha$ ratios are
systematically higher in the eDIG than in the disk.  Although
photoionization by disk OB stars is generally able to explain these
elevated [N~II]/H$\alpha$ ratios, a secondary source of ionization
appears to be needed when one also takes into account other line
ratios; more detail is given in a companion paper (our Paper II). 
\end{abstract}


\section{Introduction}

The deposition of a large amount of mechanical energy at the centers
of starburst and active galaxies may severely disrupt the gas phase of
these systems and result in large-scale winds that may extend beyond
galactic scale and influence the surrounding intergalactic medium
(e.g., Chevalier \& Clegg 1985; Schiano 1985; Tenorio-Tagle \&
Bodenheimer 1988; Tomisaka \& Ikeuchi 1988; Suchkov et al. 1994;
Heckman, Armus, \& Miley 1990; Veilleux et al. 1994; Cecil et
al. 2001, 2002).  In more quiescent systems, however, the local energy
inputs from individual star-forming regions are insufficient to
produce large-scale winds that can blow out through the halo of the
host galaxy. In this case, a ``galactic fountain'' may result: hot,
enriched gas flows up into the halo where it cools down and eventually
falls back onto the disk, presumably concentrated in clouds formed
through thermal instabilities (e.g., Shapiro \& Field 1976; Kahn 1981,
1991; Habe \& Ikeuchi 1980; Salpeter 1985; Houck \& Bregman 1990;
Shapiro \& Benjamin 1991).  The kinematics of at least some
high-velocity HI clouds in our Galaxy may be explained in this way
(e.g., Bregman 1980; Rosen \& Bregman 1995).  Most likely, quiescent
fountaining only represents the low-energy end of a broad range of
processes taking place in star-forming disks.  The existence of
``supershells'' with 100 -- 1000 pc diameters in the Milky Way and a
few nearby galaxies (e.g., Heiles 1979, 1984; Brinks \& Bajaja 1986;
Norman \& Ikeuchi 1989; Deul \& den Hartog 1990; Normandeau et
al. 1996) implies violent events that require the mechanical energy of
a few hundred supernovae ($\sim$ 10$^{53}$ ergs).  The injection of
this large amount of energy in the disk ISM may lead to the formation
of vertical ``chimneys'' which funnel the hot gas into the halo,
entraining with it the neutral and ionized ISM from the disk (e.g.,
Norman \& Ikeuchi 1989).  Galactic fountains, chimneys, and
large-scale winds may all combine to create a widespread circulation
of matter and energy between the disks and halos of galaxies which may
be responsible for the metallicity-radius relation within galaxies and
the mass-metallicity relation between galaxies (e.g., Larson \&
Dinerstein 1975; Vader 1986; Franx \& Illingworth 1990).

Evidence for a strong interstellar disk -- halo connection in our
Galaxy is seen at nearly all wavelengths. Pulsar dispersion measures
(e.g., Gu\'elin 1974; Taylor \& Manchester 1977), free-free absorption
of low-frequency galactic radio emission (Ellis 1982), ultraviolet
absorption lines (e.g., York 1983) and faint, optical emission lines
(Reynolds, Scherb, \& Roesler 1973; Reynolds, Roesler, \& Scherb 1974;
Reynolds 1980) point to the existence of a widespread, ionized
component of interstellar gas extending to large heights above the
plane of the Galaxy. Often called the Reynolds layer, this component
consists of a low-density (0.2 -- 0.4 cm$^{-3}$), 2 -- 3 kpc thick
layer of warm ($\sim$ 10$^4$ K), ionized material that has an
emission-line spectrum which differs from that of typical H II regions
(e.g., Reynolds, Roesler, \& Scherb 1977; Reynolds 1985a, 1985b, 1988,
1990; Kulkarni \& Heiles 1988; Reynolds \& Tufte 1995; Haffner,
Reynolds, \& Tufte 1999). The Reynolds layer appears to act as a
interface between the dense star-forming ISM in the Galaxy midplane
and the hot, diffuse Galactic corona (e.g., Rosen \& Bregman 1995; de
Avillez \& Berry 2001).

Support for dynamically active disks in external galaxies comes from a
wide variety of observations, including the detection of supershells
in the disks of nearby gas-rich systems (e.g., LMC; Kim et al. 1999),
thick radio disks in edge-on spirals (e.g., Ekers \& Sancisi 1977;
Allen et al. 1978; Hummel \& van der Hulst 1989; Hummel \& Dettmar
1990; Hummel, Beck, \& Dahlem 1991a; Hummel, Beck, \& Dettmar 1991b;
Hummel et al. 1991c; Dahlem, Dettmar, \& Hummel 1994; Golla \& Hummel
1994; Dahlem, Lisenfeld, \& Golla 1995; Irwin, English, \& Sorathia
1999; T\"ullman et al. 2001), and extraplanar dust and line-emitting
filaments in edge-on disk systems (e.g., Dettmar 1992; Dahlem 1997 and
references therein). The most prominent example of extraplanar diffuse
ionized gas (eDIG) outside of our Galaxy is seen in the nearby edge-on
Sb galaxy NGC~891, considered by many to be a near twin of our own
Galaxy.  Optical line-emitting filaments that extend out to distances
of $\sim$4.5~kpc above the disk of the galaxy are seen superposed on
diffuse emission extending over an area of $\sim$5~kpc in diameter and
with scale height of 2 -- 3~kpc (e.g., Dettmar 1990; Rand, Kulkarni, \&
Hester 1990; Sokolowski \& Bland-Hawthorn 1991; Dettmar \& Schultz
1992; Pildis, Bregman, \& Schombert 1994a; Rand 1997, 1998; Hoopes,
Walterbos, \& Rand 1999; Howk \& Savage 2000; Collins \& Rand
2001). The eDIG in NGC~891 shares a resemblance with the Reynolds
layer in our own Galaxy and, for this reason, NGC~891 is usually the
Rosetta stone to which all other galaxies are compared when discussing
eDIG.

Results from emission-line imaging surveys suggest that few edge-on
spirals show evidence of a widespread diffuse ionized medium like that
of our Galaxy and NGC~891, but these observations often reveal
kpc-scale filaments and plumes which may represent the bright features
of a more extensive but fainter complex of high-$\vert$z$\vert$
material (see, e.g., Dettmar 1992; Pildis, Bregman, \& Schombert
1994b; Veilleux, Cecil, \& Bland-Hawthorn 1995; Rand 1996; Hoopes et
al. 1999).  Few constraints exist on the full extent of the eDIG in
these galaxies.  This question is directly relevant to Ly$\alpha$
absorption systems seen in quasar spectra since the eDIG may
contribute significantly to the gaseous cross-section of star-forming
disk galaxies.  Giant ($\sim$ 15 kpc) H$\alpha$-emitting halos have
been purported to exist in the edge-on galaxies NGC~4631 and NGC~4656
(Donahue et al. 1995), but unpublished Fabry-Perot observations do not
seem to confirm this finding (S. N. Vogel 2003, private
communication). Dahlem et al. (1996) have detected a soft X-ray halo
in NGC~3628 which extends to 25~kpc, but this galaxy is apparently
undergoing a starburst-driven wind event which is considerably more
energetic than the phenomena taking place in quiescent star-forming
disks (see also Bregman \& Pildis 1994; Wang et al. 1995; Bregman \&
Houck 1997; Dahlem, Weaver, \& Heckman 1998).

In an attempt to refine our knowledge on the eDIG of nearby galaxies,
we have carried out an emission-line imaging survey of a moderately
large sample of edge-on disk galaxies down to flux levels of several
$\times$ 10$^{-18}$ erg s$^{-1}$ cm$^{-2}$ arcsecond$^{-2}$,
often about an order of magnitude fainter than that of most published
surveys.  The criteria used in selecting the sample of galaxies are
described in \S 2. The methods used to acquire and reduce the images
are discussed in \S 3 and \S 4.  The empirical results from the image
analysis are given in \S 5 while a quantitative analysis of the eDIG
morphology is carried out in \S 6.  In \S 7, the morphological
properties of the eDIG are compared with the properties of the host
galaxies to shed new light on the origin of the eDIG in these
objects. A summary of the results is given in \S 8. A companion paper
(Miller \& Veilleux 2003; hereafter Paper II) describes the results
from long-slit spectroscopy of the eDIG.

\section{Sample}

The galaxies in the sample are listed in Table 1. These 17 objects
were selected based on inclination, proximity, angular size, and lack
of companions. The sample was limited to edge-on galaxies with an
inclination larger than $\sim$ 80$^\circ$ with the majority of them
having $i$ $>$ 85$^\circ$.  This was done to facilitate separation of
the disk emission from that of the halo.  Even with this sample,
inclination biases may affect some of the eDIG parameters. At
85$^\circ$, a galaxy with a radius of 10~kpc will have a projected
radius of $\sim$~0.9~kpc in the plane of the sky, and at 80$^\circ$,
the radius will be projected $\sim$~1.7~kpc in the plane of the sky.
Inclination effects are discussed in more detail in \S 5 and \S 7,
when relevant.

Nearby ($z$ $<$ 0.1) galaxies were selected to provide good spatial
resolution. Most of the sample galaxies have a spatial resolution
better than 300~pc, enough to resolve an average H~II region (see,
e.g., Hodge 1987, although recent HST observations suggest that H~II
regions can be, on average, as small as 50~pc; e.g., Scoville et
al. 2001).  One should note, however, that the galaxies in our sample
cover a relatively broad range in distance (1.6 -- 63 Mpc).  The
impact of possible distance-related biases on the results of our study
is discussed in \S 5 and \S 7. NGC~55 and NGC~973 are by far the
nearest and furthest galaxy in our sample and are sometimes excluded
from the analysis to reduce the range in distance covered by the
sample.

The sample galaxies were also chosen so that they would easily fit
within the field of view of the imager (D$_{25}$ $<$ 7\arcmin\ for all
galaxies, with the exception of NGC~55) and allow us to use the shift
and stare method to build ``super'' sky flats (see \S 4.1).  Finally,
the sample was generally limited to galaxies with no known
companion(s) within 5$\arcmin$ in order to avoid contamination from
tidal features such as warps, bridges, or tails which may extend far
from the galactic plane (NGC~2820 and NGC~3432 are the only two clear
exceptions in the sample; IC~1815 is located 4$\farcm$6 from NGC~973,
but there are no obvious signs of interaction between these two
objects).

\section{Data Acquisition}

\subsection{Conventional Narrow-band Imaging}

Ten of the sample galaxies were observed with traditional techniques
at either the Kitt Peak National Observatory (KPNO) or the Cerro
Tololo Inter-American Observatory (CTIO) facilities during three
observing runs.  Complete details of the observations can be found in
Table~2, including the telescopes and filters used for these
observations as well as the total exposure times, number of exposures,
field of view, and spatial scale (after binning, when relevant; see
\S 4).  Narrow-band filters were chosen such that the redshifted
H$\alpha$ emission of each galaxy fell near the center of the filter
bandpass.

An important goal of this survey is to image the galaxies down to a
surface brightness at H$\alpha$ of $\sim$ 1 $\times$~10$^{-17}$ erg
s$^{-1}$ cm$^{-2}$ arcsec$^{-2}$ or below.  In order to reach this
faint flux level, great care was taken in both the acquisition and
reduction of the data.  One of the greatest problems in accurate data
reduction is the inconsistency in the sky background of each image.
Since the total exposure time for each galaxy is on the order of
hours, it is difficult to maintain a constant sky background over this
length of time.  Both the change in the zenith angle of the object
(and thus the airmass) as well as the sky photometry can lead to
inconsistencies in the background which need to be carefully
corrected.

To reduce long-term photometric variations, short-exposure H$\alpha$
and Gunn $r$-band continuum observations were obtained alternatively.
The exposures were long enough, though, to insure that each frame was
dominated by sky noise.  The CCD used for the KPNO 0.9-m observations
(but not for the CTIO Curtis Schmidt observations) was spatially
binned 2 $\times$ 2 to increase the signal rate per pixel from the
diffuse emission.  The Poissonian sky background noise dominates over
the read-out noise, and so binning the data 2~$\times$~2 does not
affect the signal-to-noise ratio of the data for a given exposure time
but allowed us to take shorter exposures, thus enabling us to take
more images in one night. This is important for the shift and stare
method (see \S 4.1) and helps average out temporal variations and
reduce the number of cosmic rays per frame.  The resulting spatial
resolution of the KPNO data is still better than that of the CTIO
Curtis Schmidt observations (Table 2).

\subsection{Taurus Tunable Filter Imaging}

Ten galaxies were also observed with the The Taurus Tunable Filter
(TTF) at the Anglo-Australian Telescope (TTF) and William Herschel
Telescope (WHT). This instrument is able to achieve the same desired
flux levels (i.e. $\la$~1 $\times$ 10$^{-17}$ erg s$^{-1}$
cm$^{-2}$ arcsec$^{-2}$) in much less time.  Developed at the
Anglo-Australian Observatory, the TTF design is based on the concept
of a conventional Fabry-Perot (FP) interferometer.  Using two parallel
plates of high reflectivity, light travels through the interferometer
such that at each radius from the optical axis a specific wavelength
is imaged, resulting in the wavelength varying quadratically with
distance from the optical axis.  The TTF is able to maintain plate
parallelism down to $\sim$ 2 $\mu$m and scan over a range of $\sim$
10$\mu$m.  This allows the instrument to work at low orders of
interference and thus ensures that the wavelength varies little ($\la$
18 \AA) over the full field of view.  In other words, it is now
possible to obtain quasi-monochromatic images with a single exposure.
In addition, the TTF has the ability of adjusting the bandwidth of the
filter from $\sim$ 6 to 60 \AA, thus making it possible to obtain very
narrow-band images.  This allowed us to separate the H$\alpha$ and
[N~II] $\lambda$6583 emission lines, a difficult task with
conventional narrow-band filters.

The TTF at the AAT also has the ability to switch between frequencies
at a very rapid rate, up to $\sim$ 100 Hz if needed.  By shuffling the
charge of the CCD in sync with the frequency switching, it is possible
to observe an object at two distinct frequencies (and two distinct
bandwidths) {\it at almost the same time}.  This technique allowed us
to average out any sky variations over both frames.  For our
observations, redshifted H$\alpha$ and [N~II]$\lambda$6583 were
observed alternatively every 2 minutes for several cycles, thus
yielding very low flux levels and more precise, two-dimensional
[N~II]/H$\alpha$ maps.  Unfortunately, the WHT did not have charge
shuffling and frequency switching capabilities at the time of the
observations, but the use of the TTF in obtaining very narrow-band
quasi-monochromatic images still allowed us to reach the faint flux
levels desired in less time than it would have taken using
conventional methods.  A more complete description of the TTF
cabilities can be found in Bland-Hawthorn \& Jones (1998; see also the
TTF web page http://www.aao.gov.au/ttf).

Three of the galaxies observed with the TTF are in common with the
KPNO/CTIO sample (NGC~55, NGC~2424, and NGC~3044).  This overlap
allows us to assess the impact of lower spatial resolution on the KPNO
and CTIO data. Full details about the TTF runs are listed in Table~3.
The etalon was tuned to a bandpass of $\sim$~13~--~16~\AA, allowing us
to obtain separate H$\alpha$ and [N~II] images for these galaxies.  An
intermediate blocking filter [$\Delta\lambda$(AAT)~=~210~\AA;
$\Delta\lambda$(WHT)~=~280~\AA] was used to block out all but one
order of interference through the TTF.  The charge shuffling and
frequency shifting capabilities at the AAT was used to obtain
simultaneous images of H$\alpha$ and [N~II], while continuum images
were taken separately.  At the WHT, each image was taken individually.
Because the galaxies filled a significant fraction of the frame, we
could not use the shift and stare method, but the capabilities of the
TTF in reaching faint flux levels more than compensated for the
relatively modest field of view of $\ga$ 8$\arcmin$.  The position of
the galaxies in each frame was shifted slightly between observations
to allow us to reject bad pixels when the frames were combined
together. 

A neon calibration cube was obtained before and after each galaxy
observation to determine the wavelength calibration in relation to
spatial position and etalon gap spacing.  The free spectral range
({\em i.e.} the wavelength range between orders) was 230~\AA\ for the
AAT and 280~\AA\ for the WHT; both were larger than, or equal to, the
blocking filter width and thus ensured that only one order was imaged.

\section{Data Reduction}

\subsection{KPNO and CTIO data}

The data sets were reduced using standard IRAF packages.  A mean bias
frame was calculated and subtracted from each image making sure that
no additional noise was added to the images.  ``Super'' sky flats were
created using the shift and stare method. As opposed to dome flats and
normal sky flats, which are not taken at the same time as the objects
(and thus cannot truly correct for the conditions present during the
actual observation), ``super'' sky flats were constructed using the
object images themselves.  For each H$\alpha$ -- continuum exposure
pair, the frame was shifted such that the position of the galaxy was
in a different section of the image in each exposure.  Using this
method, about 20 to 30 exposures were acquired for each galaxy.

Bright objects were masked out using the IRAF FOCAS add-on package,
resulting in a more accurate representation of the flatfield response
{\em at the time of the observations}.  The images were then combined
together, rejecting anamolously high and low pixels that were not
masked out by the FOCAS routines.  Because each observation of the
galaxy was positioned in different regions of the frame, each pixel in
the combined frame is well sampled (i.e., not masked out due to a
bright object).  When possible, separate sky flats were created for
each object and each night so that variations in the observations were
minimized.  These accurate sky flats enabled us to achieve lower flux
levels than possible with conventional flats. Note, however, that
contamination by Galactic [N~II] $\lambda$6583 emission from the
Reynolds layer (which is known to vary on sub-arcminute angular scale)
and scattering in the telescope and filter limit the accuracy of this
technique.

The images were flatfielded using the sky flats and then aligned.
Finally, the images were corrected for airmass and the sky background
was subtracted.  The sky background was determined interactively using
regions uncontaminated by the galaxies or other bright objects in
order to ensure that it was not affected by low surface brightness
emission.  The images were then flux calibrated using observations of
standard flux calibrators (Stone 1977, Stone \& Baldwin 1983, Landolt
1992a, 1992b).  The IRAF COMBINE package was used to not only add
the individual frames together but also remove cosmic rays.

In order to remove the presence of continuum emission in the
narrow-band images, the galaxies were also observed with an
intermediate-band ($\sim$~800~--~900~\AA\ FWHM) Gunn $r$ filter.
These images were obtained and reduced following the same procedures
used for the narrow-band images, and then subtracted from the
narrow-band images in order to obtain the H$\alpha$ +
[N~II]$\lambda$6583 images. The delicate procedure of continuum
subtraction is described in more detail in \S 4.3 below.

This method of data reduction has allowed us to observe down to an
average flux level of $\sim$ 1 $\times$~10$^{-17}$ erg s$^{-1}$
cm$^{-2}$ arcsec$^{-2}$ (the 2$\sigma$ flux limit for each galaxy is
listed in Table~2).

\subsection{Taurus Tunable Filter Data}

The TTF data were reduced using the Zodiac data reduction package
(which was first developed at the University of Hawaii/IfA by George
Miyashiro) as well as standard IRAF packages, following the procedures
described in Jones, Shopbell, \& Bland-Hawthorn (2002). The bias
structure was relatively constant over the images, therefore a
constant bias was subtracted from each.  The data were flatfielded
using traditional sky~flats/dome~flats as required.  Bright objects
were masked out and a mean sky value was calculated in annular bins
around the optical axis.  An azimuthally symmetric sky frame was
produced from this sky radial profile for each image and subtracted
off.  In all cases, the field of view of the TTF was large enough to
derive accurate sky levels using regions free of galaxy emission.  The
data were flux calibrated using observations of standard flux
calibrators (Stone \& Baldwin 1983; Massey et al. 1988).  Cosmic rays
were detected and removed, and the images for each galaxy were then
aligned and summed together.

In order to remove continuum emission from each galaxy, Cousins $R$-band
continuum images were obtained as well.  These images were reduced in
a similar fashion to the TTF images and subtracted from them in order
to obtain individual H$\alpha$ and [N~II]$\lambda$6583 images for each
galaxy. The continuum subtraction method is described next. 

\subsection{Continuum Subtraction}

Obtaining continuum subtracted emission-line images is not
straightforward.  The obvious approach is to scale the continuum image
such that the foreground stars are perfectly subtracted;
unfortunately, there are problems with this simple approach.
First, the stellar population of the galaxies differs from that of the
field stars, and so the scaling factor for the field stars is
different from that of the galaxies. Second, and more importantly, the
stellar population in the bulge of the galaxy differs from that in the
disk of the galaxy, so ideally the continuum needs to be scaled
differently in these two regions.  In this study, we decided to use an
iterative process in which the scaling factor of the continuum image
is determined from the overall morphology of the residual
emission-line image.  Radial cuts of the continuum-subtracted image
are examined to determine if oversubtraction has occurred.
Oversubtraction of the continuum image is noticeable as a dip 
in the vertical profile below the background of the emission-line map.

\section{Empirical Results}

The continuum-subtracted emission line images of each galaxy in the
sample are presented in Figure 1 -- 17 and discussed in the
Appendix. [N~II] $\lambda$6583/H$\alpha$ excitation maps are also
presented for 10 of the galaxies in the sample. The present section
discusses the overall trends found in the sample. For a more detailed
discussion of each object, the reader should refer to the Appendix.

\subsection{Integrated H$\alpha$ Luminosities}

The integrated H$\alpha$ luminosity for each galaxy (measured within
the field of view) is listed in Table~4.  In an effort to compare
values between galaxies of different sizes, we also derive the
H$\alpha$ luminosity/D$^2_{25}$, or the total H$\alpha$ luminosity
divided by the square of the apparent isophotal diameter (at a surface
brightness level $\mu_B$ = 25.0 B mag arcsec$^{-2}$).  This quantity
gives only a basic indication of the H$\alpha$ luminosity per unit
disk area for each galaxy, since only a fraction of the total
H$\alpha$ luminosity is detected in these highly inclined
galaxies. Due to uncertainties in the continuum subtraction and flux
calibration, typical uncertainties for the H$\alpha$ luminosity are
$\sim$ 10 -- 25\%.  The KPNO and CTIO data were corrected for
contamination by the nearby [N~II] lines.  While studies indicate that
the [N~II]/H$\alpha$ ratio increases with increasing $\vert$z$\vert$
(e.g., see \S 5.6 and Paper II and references therein), a constant
ratio of [N~II]$\lambda$6583/H$\alpha$ = 0.5 and
[N~II]$\lambda$6548/H$\alpha$ = 0.17 was assumed for the entire DIG in
order to simplify the calculations.

The H$\alpha$ luminosities in the galaxies of our sample range from
3.4 $\times$ 10$^{39}$ erg s$^{-1}$ to 83 $\times$ 10$^{39}$ erg
s$^{-1}$, with an average value of 32 $\times$ 10$^{39}$ erg s$^{-1}$,
and L(H$\alpha$)/D$^2_{25}$ ranges from 0.8 $\times$ 10$^{37}$ erg
s$^{-1}$ kpc$^{-2}$ to 40 $\times$ 10$^{37}$ erg s$^{-1}$ kpc$^{-2}$,
with an average value of 6.2 $\times$ 10$^{37}$ erg s$^{-1}$
kpc$^{-2}$ (see Table 4 and Fig. 18). These results can be compared
with published measurements in three cases: NGC ~55 (Hoopes,
Walterbos, \& Greenwalt 1996; Ferguson, Wyse, \& Gallagher 1996; Otte
\& Dettmar 1999), NGC~2188 (Domg\"orgen \& Dettmar 1997), and NGC~3044
(Rossa \& Dettmar 2000).  For NGC~55, our value of 8.9 $\times$
10$^{39}$ erg s$^{-1}$ is only about half of what Hoopes et al. and
Ferguson et al. measure for the H$\alpha$ + [N~II] luminosity (2.6 and
2.0 $\times$ 10$^{40}$ erg s$^{-1}$, respectively), but taking the
[N~II] contamination into account brings their values closer to our
measurements.  Rather than list the luminosity, Otte \& Dettmar list
an H$\alpha$ flux of 2.8 $\times$ 10$^{-11}$ erg s$^{-1}$ cm$^{-2}$.
Converting their flux to a luminosity (using a distance to NGC~55 of
1.6 Mpc), one obtains a luminosity of 8.58 $\times$ 10$^{39}$ erg
s$^{-1}$, very similar to ours.  For NGC~2188, the difference in
H$\alpha$ luminosity between our value of 9.91 $\times$ 10$^{39}$ erg
s$^{-1}$ and Domg\"orgen \& Dettmar's value of 1.1 $\times$ 10$^{40}$
erg s$^{-1}$ lies within the uncertainties of the measurements.
Finally, our H$\alpha$ luminosity for NGC~3044 of 8.94 $\times$
10$^{40}$ erg s$^{-1}$ differs slightly but not significantly from
Rossa \& Dettmar's value of 1.27 $\times$ 10$^{41}$ erg s$^{-1}$.

\subsection{Detection Frequency, H$\alpha$ Luminosity, and Scale Height of
the eDIG}

In order to study the eDIG, one first needs to define this term. This
is non-trivial since the boundary between the disk H~II regions and
the eDIG is not clearly defined; the line emission smoothly changes
from one component to the other with no obvious break in the emission
profile.  The typical size of an H~II region which is resolvable from
the ground is on the order of 300~pc (Hodge 1987), so a vertical
cutoff at this height might be appropriate. However, since the sizes
of H~II regions undoubtedly vary with location within the galaxy as
well as with galaxy type, this definition is not adequate.

The scale height of the continuum emission is a better guideline.
This scale height was derived at each radius in the galaxy using an
exponential fit to the continuum image, excluding the inner 300~pc so
that dust obscuration would not affect the results.  This scale height
was then used as a first iteration for the lower boundary of the
extraplanar gas.  A mask was created such that all continuum emission
within a vertical scale height of the midplane of the disk was
excluded, and this mask was multiplied by the H$\alpha$ image so that
only the extraplanar gas outside of one vertical (continuum) scale
height remained.  For comparison, an H~II region mask was also created
using the H$\alpha$ equivalent width criterion described in
Bland-Hawthorn, Sokolowski, \& Cecil (1991). It was found that the
scale-height mask did not completely exclude all of the H~II regions
defined by the equivalent-width mask.  The extent of the scale-height
mask was varied until it did exclude the H~II regions; it was found
that a mask based on a vertical extent of about 1.25 $\times$ the
continuum scale height worked well in that it excluded the H~II
regions but did not extend much further past them vertically.  This
method was adopted to define the eDIG component in all of the galaxies
of the sample. The resulting vertical height of the masked region is
roughly constant with radius within each galaxy, but varies among the
different galaxies from $\sim$ 500~pc to 3~kpc.  This broad range in
continuum scale heights is mainly due to variations in the
inclinations of the galaxies.

Extraplanar DIG is detected in nearly all of the galaxies in the
sample.  Only one galaxy, ESO~209-9, does not provide enough evidence
to warrant further investigation; the lack of obvious eDIG in this
object may be due to an unfortunate combination of observational factors
(e.g., relatively high flux limits and poor spatial
resolution). Although most of the galaxies in the sample do not show
widespread diffuse ionized emission as pervasive as in NGC~891 (one
possible exception is NGC~2188, where extraplanar emission is detected
over a radius of 4~kpc; this emission may even extend further since
the detection was limited by the field of view of the AAT setup; see
also Domg\"orgen \& Dettmar 1997), many of them show localized regions
of eDIG emission often characterized by distinct filaments or plumes
extending up a few kpc from the plane of host galaxy.  The topology of
the eDIG is analysed more quantitatively in \S 6.

The eDIG emission has an average luminosity of 4.4 $\times$ 10$^{39}$
erg s$^{-1}$ and ranges from 0.12 $\times$ 10$^{39}$ erg s$^{-1}$ to
13 $\times$ 10$^{39}$ erg s$^{-1}$, and comprises 12 $\pm$ 4\% of the
total luminosity of the galaxies (Table 4 and Fig. 18; the eDIG
fraction is 13 $\pm$ 5\% if we exclude NGC~55 and the 6 galaxies in
the sample with $i < 85^\circ$).  All but four galaxies (NGC~55,
ESO~362-11, NGC~3432, NGC~4013), have values that lie within 10 --
16\%, suggesting that the luminosity of the ionized H$\alpha$ halo is
a fairly constant fraction of the total H$\alpha$ luminosity of the
galaxy.  All four galaxies with low eDIG/total luminosity ratios
clearly show the presence of extraplanar ionized gas, so it is not
clear why the ratios for these galaxies are lower than average.

Exponential curves were fit to the vertical profile of each H$\alpha$
image to determine the scale height of the line-emitting gas. In most
galaxies, the inner 300~pc are excluded in order to mask out
contamination by H~II regions or dust lanes (a few galaxies require
excluding more than the inner 300~pc; e.g., NGC~3628).  Both one- and
two-exponential fits were used to model the diffuse emission.  The
fits are shown in Figure~19.  Wang et al. (1997) discuss Rand's (1997)
use of a two-exponential fit and suggest that the two components
represent a quiescent component and a disturbed component of the DIG,
which they identify kinematically.  They use this distinction between
the two to define the fainter, more disturbed component as the
``extraplanar'' component of the DIG; we adopt the same definition
from here on out. The eDIG scale heights for the galaxies in the
present sample are listed in Table~5 and range from 0.4~kpc to
17.9~kpc, with an average of 4.3 kpc (Fig. 18; the average eDIG
scale height is 5.0 kpc if we exclude NGC~55 and the 6 galaxies in
the sample with $i < 85^\circ$). This average value is
noticeably larger than the eDIG scale height in our Galaxy (0.75 kpc;
Reynolds 1989, 1990) or in NGC~891 (0.5 kpc; Dettmar 1990).  In a
study of nearby edge-on galaxies, Hoopes et al. (1999) find scale
heights ranging from 450~pc to 1.7~kpc, again smaller on average than
the values found in our galaxies. This difference in scale height is
due to our definition of the eDIG and to the fainter flux levels
reached by our observations.  The better sensitivity of our
observations allows us to study a previously undetected eDIG component
above galaxy disks.

The emission measure, $\rm{EM} = \int n^2_e dl$, is related to the
H$\alpha$ intensity by
\begin{equation}
\rm{EM} = 2.75 \times T^{0.9}_4~I_{H\alpha}~~~cm^{-6} pc
\end{equation}
with I$_{H\alpha}$ given in Rayleighs (1 R = 5.66 $\times$ 10$^{-18}$
erg s$^{-1}$ cm$^{-2}$ arcsec$^{-2}$ at H$\alpha$), T$_4$ is the gas
temperature in units of 10$^4$ K, and $dl$ is given in units of
parsecs (e.g., Reynolds 1990).  The emission measure in the midplane
of the galaxy is derived using the H$\alpha$ intensity at
$\vert$z$\vert$ = 0, extrapolated from the two-exponential fit to the
vertical profiles (see Table 5). This quantity is an emission measure
determined in the midplane of the disk from the observed, edge-on
orientation and is an average over all radii since it is derived from
the average vertical profile of each galaxy.

The average value for the emission measure in the midplane associated
with the extraplanar gas (based on the second component of the
two-exponential fits) is 14 pc cm$^{-6}$, but ranges in value from
0.39 pc cm$^{-6}$ to 69 pc cm$^{-6}$ (Fig. 18).  For comparison, Hoopes
et al. (1999) find midplane emission measures of $\sim$ 2 -- 10
pc~cm$^{-6}$ (except for NGC~891, which has a midplane emission
measure of $\sim$ 25 -- 45 pc~cm$^{-6}$).  A careful examination of
our data shows that the galaxies with the largest scale heights have
some of the lowest midplane emission measures (e.g., ESO~240-11 and
NGC~5965), suggesting again that the present low-flux-level
observations are detecting a very faint, extraplanar region of diffuse
ionized gas.  Galaxies with large midplane emission measures (e.g.,
NGC~2188, NGC~55, \& NGC~1507) have small scale heights ($\sim$ 400 --
800~pc).  Only one galaxy, NGC~3044, has sizeable midplane emission
measures [16 (North) and 62 (South) pc~cm$^{-6}$] and scale heights
[2.8 (North) and 1.3 (South) kpc].  The midplane emission measures and
scale heights of the galaxies in our sample have a Pearson's
probability P[null] $\approx$ 0.024 for the null hypothesis of zero
correlation, indicating a weak anti-correlation between these two
quantities. This anti-correlation is consistent with the predictions
of hydrodynamical simulations of star-forming disk galaxies.  For
instance, Rosen \& Bregman (1995) find that as the energy flow rate
from the disk to the halo is increased, the midplane electron density
($\propto$ EM$^{0.5}$) decreases while the gas scale height increases.

\subsection{Midplane Density \& Mass of the eDIG}

Given the emission measure at the midplane of a galaxy, the electron
density at this location can be estimated using EM = $n_e^2fl$, where
$l$ is the length of the line of sight through the volume of gas and
$f$ is the filling factor (assumed to be constant with height).  For
simplicity, we assume $f$ = 0.2 at all $z$, as determined in the
midplane of our Galaxy (Reynolds 1990), and use a line-of-sight length
$l$ = 5 kpc (Veilleux et al. 1995).  This gives an electron density of
\begin{equation}
n_e(z) = 3.16 \times 10^{-2}{\rm EM}(z)^{0.5} (\frac{0.2}{f})^{0.5} (\frac{5\;{\rm kpc}}{l})^{0.5} {\rm cm}^{-3}.
\end{equation}
The calculated electron density can then be used to determine the mass
of the ionized gas.  Using M = $\int \rho d$V and assuming
$n_e$ $\sim$ $n_p$, we get
\begin{equation}
{\rm M}~\approx~\int n_e(z) {\rm m}_p d{\rm V},
\end{equation}
where m$_p$ is the mass of a proton.  Since $n_e$ $\propto$
EM$^{0.5}$, the scale height of the electron density is twice that of
the emission measure.  If we assume that this scale height represents
the extraplanar extent of the gas and calculate the eDIG mass for each
side of a galaxy, equation (3) then becomes
\begin{equation}
{\rm M}~\approx~2~\pi~n_e(z=0) {\rm m}_p H_z {\rm R}^2,
\end{equation}
where $H_z$ is the scale height of the extraplanar gas emission and
$R$ is the optical radius of the galaxy.  The electron density at $z =
0$ and ionized mass are listed in Table~5 and their distributions are
shown in Figure 18.  Values in the Table are calculated for both the
one-exponential fit and both components of the two-exponential fits,
but Figure 18 only shows the distribution of the one-sided eDIG masses
derived from the second component of the two-exponential fits.

The midplane electron densities of the eDIG derived from the
two-exponential fits range from 0.020~cm$^{-3}$ to 0.26~cm$^{-3}$,
with an average value of 0.11~cm$^{-3}$, and the total (two-sided)
extraplanar ionized masses range from 1.4 $\times$ 10$^7$ M$_{\odot}$
to 2.4 $\times$ 10$^8$ M$_{\odot}$, with an average value of 1.2
$\times$ 10$^8$ M$_{\odot}$ (the average value for the total EDIG mass
becomes 1.5 $\times$ 10$^8$ M$_{\odot}$ if we exclude NGC~55 and the 6
galaxies in the sample with $i < 85^\circ$). For comparison, Rand
(1996) derives values for $n_e(z=0)$ in the range of 0.16 --
0.3~cm$^{-3}$, on average slightly larger than the values derived in
our study.  Rossa \& Dettmar (2000) measure eDIG masses ranging from
1.8 to 13 $\times$ 10$^{6}$ M$_{\odot}$.  These values are
considerably smaller than the values derived here.  This discrepancy
with our data arises primarily from the fact that Rossa \& Dettmar use
a different method to calculate the mass of diffuse ionized gas.
Another source of discrepancy is the better sensitivity of our
observations which allows us to detect fainter eDIG emission.

\subsection{Recombination Rate Requirements}

The H$\alpha$ intensity, scale height, and electron density of the
eDIG can be used to estimate the recombination rate in a column
perpendicular to the galactic plane using 
\begin{equation}
\eta_{DIG} = \int \alpha(T) n^2_e(z) dz = \alpha(T)f<n^2_e(z = 0)> H_z,
\end{equation}
where the value of the Case B recombination coefficient, $\alpha_B$(T)
is equal to 2.58 $\times$ 10$^{-13}$ cm$^3$ s$^{-1}$ at T = 10$^4$ K
(Martin 1988), $f$ is the filling factor, which we again assume to be
constant at 0.2, $n_e$ is the electron density of the eDIG at $z$ = 0,
and $H_z$ is the scale height of the eDIG emission.  Substituting the
eDIG electron densities and scale heights listed in Table 5, we find
that $\eta_{DIG}$ varies from 0.44 to 13 $\times$ 10$^6$ s$^{-1}$ per
cm$^{-2}$ of the disk, with an average of 3.6 $\times$ 10$^6$ s$^{-1}$
per cm$^{-2}$ (the average value for the recombination rate becomes
4.6 $\times$ 10$^6$ s$^{-1}$ if we exclude NGC~55 and the 6 galaxies
in the sample with $i < 85^\circ$). For comparison, the diffuse
interstellar medium in our Galaxy requires a recombination rate of 4
$\times$ 10$^6$ s$^{-1}$ per cm$^{-2}$ of galactic disk (Reynolds
1984).

\subsection{[N~II] $\lambda$6583/H$\alpha$ Excitation Maps}

[N~II] $\lambda$6583/H$\alpha$ ratio maps have been derived for all
ten objects in the sample for which TTF data are available. They are
presented in the Appendix. Artifacts of the data acquisition and
reduction are affecting these maps. In a few cases, a slight mismatch
in the central wavelengths of the H$\alpha$ and [N~II] images is
apparent, causing artificial [N~II]/H$\alpha$ gradients along the
galaxy disks where the velocity gradients are steepest (NGC~2188,
NGC~3044, ESO~362--11). Reflective ghosts are also a problem in two
objects, causing the arc-like gradient in the excitation map of NGC~55
and substructures in the map of NGC~891. One therefore has to be
careful when interpreting these maps.  Nevertheless, general
statements about the excitation properties of the eDIG can be made
from these data.

It is clear that the [N~II]/H$\alpha$ ratios in the eDIG are generally
larger than in the disk. Arguably the cleanest case for this effect is
NGC~4013, where the [N~II]/H$\alpha$ ratios are H~II region-like in
the disk but reach values close to unity in the eDIG. A vertical
[N~II]/H$\alpha$ gradient is also easily detected in NGC~891, and is
visible but more shallow in NGC~1507, NGC~2188, NGC~3044 and possibly
ESO~362-11. These results confirm the conclusions of previous studies in
these and other galaxies (e.g., Keppel et al. 1991; Dettmar \& Schultz
1992; Veilleux et al. 1995; Golla, Dettmar, \& Domg\"orgen 1996;
Domg\"orgen \& Dettmar 1997; Rand 1998, 2000; T\"ullman \& Dettmar
2000; see also the Appendix and Paper II).  We also confirm the
systemic trend of increasing [N~II]/H$\alpha$ ratio with decreasing
H$\alpha$ surface brightness found in NGC~55 by Ferguson, Wyse, \&
Gallagher (1996) and Otte \& Dettmar (1999). A similar effect is seen
in NGC~7817.

Photoionization models (e.g., Mathis 1986; Domg\"orgen \& Mathis 1994;
Sokolowski 1994; Bland-Hawthorn et al. 1997; Mathis 2000) have had
some success explaining the increase in the [N~II]/H$\alpha$ line
ratio observed in the eDIG.  This increase is attributed to a decrease
with height of the ionization parameter ($U$), a measure of the ratio
of the ionizing photon number density ($\Phi$) to the electron density
($n_e$).  Under the assumption of ionization equilibrium, $\Phi$
$\propto$ $n^2_e$ at all heights.  Therefore, $U$ $\propto$ $\Phi/n_e$
$\propto$ $n_e$, so the ionization parameter should fall off
exponentially with height.  As $U$ decreases, lower ionization species
are favored, leading to an increase in the [N~II]/H$\alpha$
ratio. More quantitatively, however, photoionization models have
difficulties reproducing [N~II] $\lambda$6583/H$\alpha$ in excess of
unity unless one takes into account the multi-phase nature of the ISM,
the possible depletion of certain gas-phase abundance of metals onto
dust grains, and the absorption and hardening of the stellar radiation
field as it propagates through the dusty disk (Sokolowski 1994;
Bland-Hawthorn et al. 1997). The line ratios measured from the present
data are generally less than $\sim$ 1.5 and can therefore be explained
from OB photoionization. More line ratios are needed to better
constrain the source of ionization of the eDIG.  This issue is
re-examined in Paper II, where we present the results from deep
long-slit spectroscopy of several edge-on spirals.  In that paper, we
find that photoionization by massive OB stars alone generally has
difficulties explaining all of the line ratios in the extraplanar
gas. Hybrid models that combine photoionization by OB stars and
another source of ionization such as photoionization by turbulent
mixing layers or shocks provide a better fit to the line ratios in the
eDIG.

\section{Morphological Analysis of the eDIG}

\subsection{Spatial Correlation between Disk and Extraplanar Emission}

The extraplanar emission is summed vertically to derive the brightness
profile of the eDIG as a function of galactocentric radius.  The same
is done with the emission coming from the H~II regions in the
disk. Next, a cross-correlation analysis is carried out between 
the radial profiles of the disk emission and eDIG emission in the
hopes of quantifying possible spatial correlations between these two
components. Note, however, that dust in the disk of the galaxies is
almost certainly obscuring some of the disk H$\alpha$ emission and
thus skewing the results. The results of this correlation analysis
should therefore be used with caution and in combination with other
morphological indicators.

Figure 20 displays the radial disk and eDIG emission profiles for each
galaxy. Also shown in this figure are the results of the
cross-correlation analysis for each side of the extraplanar emission.
The correlation coefficient is plotted as a function of the radial
shift between the disk and eDIG profiles. A correlation coefficient of
1 (--1) signifies a perfect (anti-)correlation. All galaxies
demonstrate some degree of correlation between the disk and halo
emission.  The correlation coefficients generally peak at or near zero
radial shift, but the peak value varies from one galaxy to the next.
The values of the coefficients at zero radial shift are used in the
following discussion to assess the degree of correlation.  These
values are listed in Table~6.

The profile of the correlation coefficient also provides a rough
measure of the topology of the extraplanar gas.  Galaxies with
extraplanar filaments ionized by nearby disk H~II regions are expected
to produce more strongly peaked correlation profiles than galaxies in
which the extraplanar gas is ionized by a more diffuse source.
Galaxies with widespread extraplanar emission may show significant
correlation with the disk, but their correlation profiles should be
broad since the diffuse emission is not necessarily tied to specific
locations in the disk.

The average amplitude of the correlation peaks at zero radial shift is
found to be 0.60 $\pm$ 0.24. The majority of the galaxies in the
sample therefore show significant correlation between the disk and
halo components. Roughly 50\% of the galaxies in the sample show
sharply peaked correlation profiles with FWHM/2 $\lesssim$ 3.0~kpc
indicative of a highly structured eDIG morphology.  These results
suggest that the eDIG is often tied to the H~II regions within the
disk, providing support to the hypothesis that the dominant source of
ionization of the eDIG originates in the disk (e.g., Domg\"orgen \&
Mathis 1994; Mathis 2000).  This issue is discussed in more detail in
\S 7 and Paper II.

\subsection{Deviation from Average Vertical Profile}

To further quantify the topology of the eDIG, the 1-$\sigma$ deviation
of the eDIG emission from the average vertical profile was calculated.
First, the average vertical profile of the H$\alpha$ emission was
determined by averaging over all (galactocentric) radii.  This average
vertical profile was then subtracted from the emission profile at each
radius to look for deviations from the average.  For this procedure,
the vertical emission profile at a given radius was renormalized such
that the integrated flux measured at each radius is equal to the
integrated flux associated with the average emission profile.  All of
the data points were also Poisson-weighted in order to allow
comparisons between galaxies of different brightnesses, and to reduce
the impact of the bright eDIG emission near the disk relative to the
emission at high $\vert z\vert$.

The 1-$\sigma$ deviation derived from this analysis is a rough
indicator of the level of fluctuations in the extraplanar emission.
High deviations are expected in galaxies with filamentary
morphologies, since bright knots or plumes of emission as well as
regions void of eDIG differ greatly from the average extraplanar
emission.  A galaxy with widespread diffuse eDIG emission would yield
a low 1-$\sigma$ deviation since the extraplanar emission is more
uniform across the halo.

The values of the 1-$\sigma$ deviations are listed in Table~7.  For
most galaxies, the results here are consistent with our findings in \S
6.1.  Galaxies which show large 1-$\sigma$ deviations also demonstrate
a high level of correlation between disk and halo emission and sharp
peaks in their correlation plots, reinforcing the conclusion that they
exhibit structured morphology.  Prime examples include the southeast
side of NGC~2820 and the northern side of NGC~3044 (especially in the
CTIO image). Galaxies with structured morphologies (based on the
1-$\sigma$ deviations listed in Table 7) also appear to have greater
eDIG masses (P[null] value $\sim$ 0.01).  Other interesting
correlations are discussed in the next section.

\section{The Influence of the Host Galaxy on the eDIG}

Results from \S 5 and \S 6 indicate that the galaxies in our sample
show a broad range in eDIG properties (H$\alpha$ luminosity,
midplane emission measure and electron density, scale height, mass,
and 1-$\sigma$ deviation from exponential fit). To gain insight into
the origin and possible sources of energy and ionization of the eDIG,
it is important to find out if these quantities depend on the
properties of the host galaxies. To address this issue, the properties
of the eDIG listed in Tables~4 and 6 are compared with the host galaxy
properties listed in Table~1.  The results of these comparisons are
listed in Table~8.  A least-squares linear fit analysis is conducted
for each data set, and the Pearson correlation coefficient $r$ for the
fit is computed as well as the Pearson's probability (P[null] value)
of zero correlation. A correlation coefficient $r$ = 1.0 denotes a
perfect correlation, and a P[null] value less than $\sim$ 0.01
indicates a significant correlation. To examine the impact of possible
biases associated with distance and inclination, the least-squares
linear fit analysis was conducted twice: once for the entire sample
and another time for a subset of 10 objects that excludes NGC~55 and
the 6 galaxies in the sample with $i < 85^\circ$ (Table 8).

The lack of obvious correlation between inclination and any of the
eDIG parameters (including the scale height) suggests that effects
associated with low inclination are fairly modest and do not bias the
results of our study. Distance biases are, however, present in our
sample.  The eDIG scale height shows a significant tendency to
increase with increasing distance (P[null] $\approx$ 0.001 for the
entire sample and $\sim$ 0.005, when considering the subset of objects
that excludes NGC~55 and the 6 low-inclination galaxies). This trend
is attributed to the limited spatial resolution of the observations.
The further away a galaxy lies, the harder it is to detect the
extraplanar emission and to accurately measure the scale height. Our
method may be overestimating this quantity in the more distant
objects.  Deep observations with higher spatial resolution will be
needed to reduce this distance-related effect.

The strongest, physically meaningful correlation in the data set is
found when comparing the infrared (or far-infrared) luminosities per
unit disk area with the (one-sided) eDIG masses.  These luminosities
are an indicator of the amount of dust that is being heated by nearby
stars in these galaxies, and therefore a good indicator of the star
formation rate (e.g., Kennicutt 1998).  Note the lack of a significant
correlation between the H$\alpha$ and F/IR luminosities per unit
area. This confirms that the H$\alpha$ luminosities are strongly
affected by dust extinction in the disks and are not a good measure of
the star formation rates in these highly inclined galaxies.

The strong correlation between F/IR luminosities per unit disk area
and the eDIG masses indicates that the star formation rate per
unit disk area influences the amount of eDIG.  It may not be
surprising {\em a posteriori} to find a strong correlation between
these two quantities since a large star formation rate per unit disk
area implies the presence of a large concentration of young, massive
OB stars, and thus a large flux of UV photons which may be reaching
and ionizing the extraplanar gas. The larger star formation rates in
the infrared-bright disk galaxies also imply larger supernova rates
and larger energy injection rates in the disk ISM, thus creating a
greater number of superbubbles which expand and ultimately clear
pathways for the gas and ionizing photons to escape into the halo
(e.g., ``porosity'' parameter of Heiles 1990).  This is not the first
time that a correlation has been suggested to exist between the eDIG
properties and the star formation rate (e.g., Dettmar 1992; Rand,
Kulkarni, \& Hester 1992; Veilleux et al. 1995), but the size of the
current galaxy sample allows us to put this correlation on a more
solid statistical footing.

Interestingly, there is no evidence for a significant correlation (or
anti-correlation) between the F/IR luminosities per unit disk area and
the midplane emission (P[null] = 0.024 and 0.030 when considering the
entire sample) or scale height of the eDIG (P[null] = 0.14 and 0.19).
This is perhaps surprising since the F/IR luminosity per unit disk
area is a direct measure of the rate at which energy is injected into
the disk (the greater the star formation rate, the greater the number
of supernovae events per year, and the greater the flow of energy into
the disk and halo), and numerical simulations predict that the eDIG
layers reach larger scale heights and become more tenuous as the
injected energy rate is increased (e.g., Rosen \& Bregman 1995; de
Avillez \& Berry 2001).  This comparison with the models assumes that
the conditions in the disk ISM are the same for all of the galaxies in
the sample, but this is almost certainly not be the case.  The
predicted correlations may be masked by large galaxy-to-galaxy
variations in the properties of the disk ISM (e.g., mass, density,
clumpiness, etc). This may also explain the lack of any obvious
correlation between the eDIG topology (as measured by the peak
correlation coefficients of Table 6 or the 1-$\sigma$ deviations
listed in Table 7) and the F/IR luminosities per unit disk area.

\section{Summary}

The present paper reports the results from a deep H$\alpha$ imaging
survey of 17 nearby, edge-on disk galaxies down to flux limits of
several $\times$ 10$^{-18}$ erg s$^{-1}$ cm$^{-2}$ arcsec$^{-2}$. These
data are used to study the properties of the extraplanar diffuse
ionized gas (eDIG) in these objects. The results from this analysis
are then compared with the global properties of these galaxies to
constrain the nature and origin of the eDIG.  The main conclusions of
this study are as follows:

\begin{itemize}

\item[$\bullet$] All but one galaxy in the sample exhibit eDIG.  The
contribution of the eDIG to the total H$\alpha$ luminosity is fairly
constant with an average value of 12 $\pm$ 4\%.

\item[$\bullet$] The scale height of the eDIG derived from a
two-exponential fit ranges from 0.4~kpc to 17.9~kpc, with an average
of 4.3 kpc. This average value is noticeably larger than the eDIG
scale height in our Galaxy and other galaxies.  This difference in
scale height is probably due to our two-exponential fitting method and
to the fainter flux levels reached by our observations.

\item[$\bullet$] The midplane electron densities of the eDIG based on
the H$\alpha$ intensities at $\vert$z$\vert$ = 0 derived from the
two-exponential fits of the entire sample range from 0.020~cm$^{-3}$
to 0.26~cm$^{-3}$, with an average value of 0.11~cm$^{-3}$. These
numbers assume a constant filling factor with heights of 0.2, as
determined in the midplane of our Galaxy (Reynolds 1990), and use a
constant pathlength through the disk of $l = 5$ kpc (Veilleux et
al. 1995). Using these electron densities, we find that the total
(two-sided) extraplanar ionized masses range from 1.4 $\times$ 10$^7$
M$_{\odot}$ to 2.4 $\times$ 10$^8$ M$_{\odot}$, with an average value
of 1.2 $\times$ 10$^8$ M$_{\odot}$. Under these same assumptions, the
recombination rate required to keep the eDIG ionized ranges from 0.44
to 13 $\times$ 10$^7$ s$^{-1}$ per cm$^{-2}$ of the disk, equivalent
to 10\% to 325\% of the Galactic value. 

\item[$\bullet$] The lack of obvious correlation between inclination
and any of the eDIG parameters indicates that inclination biases do
not significantly affect the results of our study. The limited spatial
resolution of our observations and broad range in distance of our
sample galaxies appears to introduce a distance-related bias on the
eDIG scale height. Our method may be overestimating this quantity in
the more distant objects of our sample.

\item[$\bullet$] A quantitative analysis of the topology of the eDIG
indicates that about half of the galaxies in the sample have a highly
structured eDIG morphology.  There is often a correlation between the
intensity of the eDIG emission and the locations of the H~II regions
in the disk, providing support to the hypothesis that the predominant
source of ionization of the eDIG is photoionization from OB stars
located in the H~II regions.

\item[$\bullet$] Comparisons between the various properties of the
eDIG and the global properties of the host galaxies indicate that the
strongest, physically meaningful correlation is seen between the F/IR
luminosities per unit disk area (basically a measure of the star
formation rate per unit disk area) and the extraplanar ionized mass,
further supporting the existence of a strong connection between the
eDIG and the disk of the galaxy. Contrary to model predictions, no
significant correlation is seen between the midplane electron density
or scale height of the eDIG and the infrared luminosity per unit disk
area of the host galaxy. This apparent discrepancy may be due to large
galaxy-to-galaxy variations in the properties of the disk ISM.

\item[$\bullet$] [N~II] $\lambda$6583/H$\alpha$ ratio maps in 10
galaxies show (or, in some cases, confirm) the existence of vertical
excitation gradients in six of them, and a general trend to detect
larger [N~II]/H$\alpha$ ratios at lower H$\alpha$ surface brightnesses
in two additional objects. The large ratios in the eDIG can generally
be explained through photoionization by OB stars in the disk as long
as one takes into account the multi-phase nature of the ISM, the
possible depletion of certain gas-phase abundance of metals onto dust
grains, and the absorption and hardening of the stellar radiation
field as it propagates through the dusty disk. This issue is examined
in more detail in Paper II, where long-slit spectra are used to deduce
that hybrid models which combine photoionization by OB stars and
another source of ionization such as photoionization by turbulent
mixing layers or shocks provide a better fit to the line ratios in the
eDIG.
\end{itemize}

\clearpage

\acknowledgments

The authors are indebted to the Time Allocation Committees of KPNO,
CTIO, AAT, and WHT for their generous allocation of telescope time for
this project. The authors wish to thank J. Bland-Hawthorn for
countless discussions on the issues discussed in this paper, and for
assistance in using the Taurus Tunable Filter for both the AAT and WHT
observing runs.  The authors thank the referee, Ren\'e Walterbos, for
several suggestions which significantly improved this paper, as well
as Andrew Wilson, Stacy McGaugh, and Kim Weaver who read and commented
on an early version of this paper. SV is indebted to the California
Institute of Technology and the Observatories of the Carnegie
Institution of Washington for their hospitality, and is grateful for
partial support of this research by a Cottrell Scholarship awarded by
the Research Corporation, NASA/LTSA grant NAG 56547, and NSF/CAREER
grant AST-9874973.  STM was also supported in part by NSF/CAREER grant
AST-9874973. This work has made use of NASA's Astrophysics Data System
Abstract Service and the NASA/IPAC Extragalactic Database (NED), which
is operated by the Jet Propulsion Laboratory, California Institute of
Technology, under contract with the National Aeoronautics and Space
Administration.

\clearpage

\centerline{Appendix: Notes on Individual Objects}
\vskip 0.2in

\vskip 0.2in
\centerline{NGC~7817} 
\vskip 0.2in

The H$\alpha$ and [N~II] $\lambda$6583 images and corresponding
[N~II]/H$\alpha$ ratio map of NGC~7817 obtained with the WHT/TTF are
presented in Figure~1.  An SAbc galaxy with an inclination of
79$^\circ$ (the lowest in our survey), NGC~7817 presents a spiral arm
pattern which is clearly visible in the images.  A foreground star
projected just south (and slightly west) of the nucleus is
oversubtracted (resulting in the white ``spot'' noticeable on one of
the spiral arms).  Diffuse emission is visible between the bright H~II
regions in the disk of the galaxy, but little extraplanar emission is
detected.  While there may appear to be some faint eDIG on the
northwest side of the galaxy (EM $\sim$ 5 pc cm$^{-6}$), the modest
inclination of this galaxy makes it difficult to determine whether
this material is indeed high $\vert$z$\vert$ gas rather than emission
at moderately large galactocentric radius. The [N~II]/H$\alpha$ ratios
of the diffuse gas are noticeably larger than in the bright H~II
regions.

An H$\alpha$ luminosity/D$^2_{25}$ of 5.0 $\times$ 10$^{37}$ erg
s$^{-1}$ kpc$^{-2}$ is derived from the data, $\sim$ 11\% of which is
attributed to eDIG.  The extraplanar gas has a scale height of $\sim$
1.7~kpc on the northwest side and 1.9~kpc on the southeast side, and a
total eDIG mass of $\sim$ 1.1 $\times$ 10$^8$ M$_{\odot}$. The
extraplanar and disk emission components are strongly correlated, but
the difficulty in separating the disk from the halo makes this result
uncertain.  The rather large 1-$\sigma$ deviation of the northwest
halo gas may be due in part to the low inclination of this galaxy
rather than filaments or plumes extending up into the halo.

\vskip 0.2in
\centerline{NGC~55}
\vskip 0.2in

NGC~55 is nearest galaxy in our sample. It was observed at both CTIO
and AAT, but the AAT observation covers a smaller portion of the
galaxy.  The H$\alpha$ + [N~II] image obtained at CTIO and the
separate H$\alpha$ and [N~II] images obtained with the TTF on the AAT
are shown in Figure~2, along with a [N~II]/H$\alpha$ ratio map.  A
number of arcs and plumes are detected in the AAT image. These data
are also affected by reflective ghosts which complicate the
interpretation of the ratio map (the arc-like feature of large
[N~II]/H$\alpha$ across the map is an artifact of these ghosts). While
the resolution of the CTIO image is not as high, this image does show
the presence of a diffuse gas layer (EM = 2 -- 10 pc cm$^{-6}$) which
extends out about 500~pc from the center of the galaxy, over a radial
distance of $\sim$ 2~kpc (in roughly the same region mapped by the AAT
observations).

The CTIO image was corrected for [N~II] contamination as detailed in
\S 5.1. An H$\alpha$ luminosity/D$^2_{25}$ of 3.9 $\times$ 10$^{37}$
erg s$^{-1}$ kpc$^{-2}$ is derived from this image, $\sim$ 5\% of
which is attributed to eDIG.  The scale height of the extraplanar gas
has a value of 0.37~kpc on the NE side and a value of 0.45~kpc on the
SW side.  The total extraplanar ionized mass is 4.9 $\times$ 10$^7$
M$_{\odot}$.  In general, these are some of the lowest values measured
for all of our galaxies.  The disk and extraplanar emission components
are not well correlated and the 1-$\sigma$ deviation are fairly
modest.  This suggests that the widespread emission apparent in the
H$\alpha$ image is not well correlated with the H~II regions found in
the disk of the galaxy. The proximity of this galaxy makes this lack
of correlation particularly obvious.

NGC~55 has been imaged by Ferguson et al. (1996), Hoopes et
al. (1996), and Otte \& Dettmar (1999). Ferguson et al. detected a
number of loops and chimneys in the halo of NGC~55, up to distances of
$\sim$ 1.5~kpc (many of which are apparent in our images), as well as
a faint shell of emission at a height of 2.6~kpc (not detected in our
CTIO image, and outside the field of view of the AAT image).  They
measured a total H$\alpha$ + [N~II] luminosity of 2.0 $\times$
10$^{40}$ erg s$^{-1}$ and a diffuse gas fraction of 17 $\pm$ 3\% for
the inner 5~kpc of the galaxy.  Hoopes et al. measure an H$\alpha$ +
[N~II] luminosity of 2.6 $\times$ 10$^{40}$ erg s$^{-1}$ and a diffuse
gas fraction of $\sim$ 35\%.  Our measure of the diffuse gas fraction
agrees with that of Ferguson et al., but our measurement of the
H$\alpha$ luminosity is lower than both sets of observations; the
additional [N~II] emission in the Hoopes et al. and Ferguson et
al. measurements accounts for some of the discrepancy. The
measurements of Otte \& Dettmar (8.58 $\times$ 10$^{39}$ erg s$^{-1}$,
based on a distance to NGC~55 of 1.6 Mpc) are in closer agreement with
ours. The excitation map shown in Fig. 2 confirms the trend of
finding large [N~II]/H$\alpha$ ratios in features with low H$\alpha$
surface brightnesses seen in the long-slit spectra of Ferguson et
al. and Otte \& Dettmar.

\vskip 0.2in
\centerline{NGC~891}
\vskip 0.2in

The H$\alpha$ and [N~II] images of NGC~891 obtained with the TTF on
the WHT are presented in Figure~3.  As mentioned in Table 4, our
image of NGC~891 is contaminated by a few poorly subtracted foreground
stars and reflective ghost images.  Nevertheless, the data clearly
show a dust lane which intersects the galaxy, and diffuse emission
extending over a broad range of radii and up to heights on the order
of 2~kpc.

The H$\alpha$ luminosity/D$^2_{25}$ ratio for this galaxy is $\sim$
1.1 $\times$ 10$^{37}$ erg s$^{-1}$ kpc$^{-2}$.  A scale height of
$\sim$ 1.0~kpc was measured for both halos, slightly smaller than the
value derived by Rand, Kulkarni, \& Hester (1990; 1.2 kpc) using a
disk + bulge + halo decomposition, but larger than the value obtained
by Dettmar (1990; 0.5 kpc) using a one-exponential fit. The total
ionized mass of extraplanar material is estimated at 1.4 $\times$
10$^8$ M$_{\odot}$.  Based on the portion of NGC~891 which we were
able to analyze, we find little filamentary emission, but rather
widespread, diffuse emission above the disk.

The line ratio map shown in Figure 3 is affected by reflective
ghosts but clearly shows the presence of a vertical [N~II]/H$\alpha$
gradient in this object, in agreement with results of long-slit
studies (e.g., Keppel et al. 1991; Dettmar \& Schultz 1992; Rand 1998,
2000).

\vskip 0.2in
\centerline{NGC~973}
\vskip 0.2in

The H$\alpha$ + [N~II] image of NGC~973 is presented in Figure~4.
At a distance of 63~Mpc, this galaxy is the most distant in our
sample, and so has the worst linear resolution.  An extremely bright,
foreground star (not shown) lies close to the NE side of the galaxy,
making accurate sky subtraction in this region difficult.  The
residuals from another star located on the NE side of the galaxy are
also visible. As a consequence, estimates for the ``global''
properties of NGC~973 exclude the northeasternmost portion of the
galaxy.  Extraplanar emission with an average EM = 7 pc cm$^{-6}$
appears to extend from the nucleus to the SE halo of NGC~973.

The H$\alpha$ luminosity/D$^2_{25}$ ratio for NGC~973 is $\sim$ 0.8
$\times$ 10$^{37}$ erg s$^{-1}$ kpc$^{-2}$, $\sim$ 16\% of which is
attributed to extraplanar gas.  A two-exponential fit was not possible
for the SE halo of the galaxy, but a scale height of 5.0~kpc was
measured for the NW halo; this number is uncertain because it relies
on the detection of very faint emission.  The total ionized mass of
extraplanar material is only 3.9 $\times$ 10$^7$ M$_{\odot}$ and also
quite uncertain.  The disk and halo gas emission are strongly
correlated and there is no suggestion of widespread, diffuse emission
above the disk.

Pildis et al. (1994b) observed NGC~973, but were unable to detect any
extraplanar emission down to a limit of 4.6 $\times$ 10$^{-17}$ erg
s$^{-1}$ cm$^{-2}$ arcsec$^{-2}$ (or $\sim$ 5 x our detection limit). 

\vskip 0.2in
\centerline{NGC~1507}
\vskip 0.2in

The WHT/TTF H$\alpha$ and [N~II] images of NGC~1507 are presented in
Figure~5.  NGC~1507 is an SB(s)m galaxy with an inclination of
80$^\circ$.  Only a few individual H~II regions are visible in this
galaxy.  Gas with an average EM $\sim$ 20 pc cm$^{-6}$ appears to
extend up to 1~kpc in height (beyond the H~II regions) over a range of
radii $R$ = 2 -- 4~kpc on the southern side of the galaxy.

A H$\alpha$ luminosity/D$^2_{25}$ ratio of 39.8 $\times$ 10$^{37}$ erg
s$^{-1}$ kpc$^{-2}$ is derived from this image, $\sim$ 11\% of which
is attributed to extraplanar gas.  The extraplanar gas has a scale
height of $\sim$ 0.56~kpc on the west side and 0.73~kpc on the east
side, and a total extraplanar ionized mass of $\sim$ 7.3 $\times$
10$^7$ M$_{\odot}$.  A rather strong correlation is found between the
extraplanar and disk emission on the west side of the galaxy, and
Figure 20 indicates that the western extraplanar emission appears to
extend over $R$ = 1 -- 3~kpc in the southern portion of the galaxy.

Although a slight mismatch in the central wavelengths of the H$\alpha$
and [N~II] images is causing an artificial radial gradient in the
excitation map (Fig. 5), there is clear evidence for the
[N~II]/H$\alpha$ ratios in this galaxy to be systematically larger
outside of the bright H~II regions and at large $\vert z \vert$.

\vskip 0.2in
\centerline{ESO~362-11}
\vskip 0.2in

H$\alpha$ and [N~II] images of ESO~362-11 are shown in Figure~6 (the
bright spot in the southern halo is the residuals from a
bright foreground star).  This Sbc galaxy was observed at the AAT
with the TTF.  A dust lane clearly bisects the nucleus of ESO~362-11,
confirming its inclination angle of 90$^\circ$.  

A H$\alpha$ luminosity/D$^2_{25}$ ratio of 0.8 $\times$ 10$^{37}$ erg
s$^{-1}$ kpc$^{-2}$ is derived from this image, only $\sim$ 3\% of
which is attributed to extraplanar gas.  The extraplanar gas has a
scale height of $\sim$ 6.8~kpc on the south side and 3.3~kpc on the
north side, and a total extraplanar ionized mass of $\sim$ 1.2
$\times$ 10$^8$ M$_{\odot}$.  A rather strong correlation is found
between the disk and extraplanar emission on the south side of the
galaxy, but there appears to be a significant radial shift of $\sim$ 2
kpc. No obvious correlation is seen with the northern halo

The ratio map, also shown in Figure 6, suggests the presence of a
vertical [N~II]/H$\alpha$ gradient in this galaxy, although a slight
mismatch in the central wavelengths of the H$\alpha$ and [N~II] images
is clearly causing an artificial radial gradient in the data.

\vskip 0.2in
\centerline{NGC~2188}
\vskip 0.2in

Figure~7 presents H$\alpha$, [N~II], and [N~II]/H$\alpha$ maps of
NGC~2188 obtained with the TTF on the AAT. The field of view prevents
us from fully imaging this large galaxy; our results are based only on
the portion of the galaxy within the field of view. A number of
H$\alpha$ filaments and arcs of diffuse gas are apparent on the west
side of this galaxy.  Two prominent arcs of gas (with average EM
$\approx$ 60 pc cm$^{-6}$) are clearly present in the western halo,
roughly 4~kpc south of the nucleus.  The emission measure of the
filaments in the western halo is $\sim$ 50 pc cm$^{-6}$ at a height of
800~pc.

Based on the portion of NGC~2188 that lies within the field of view of the 
TTF, the H$\alpha$ luminosity/D$^2_{25}$ ratio is 9.8 $\times$ 10$^{37}$
erg s$^{-1}$ kpc$^{-2}$, 13\% of which is attributed to extraplanar
gas.  The scale height of the extraplanar gas is $\sim$ 0.8~kpc for
both the western and eastern halos, and the total extraplanar ionized
mass (within the field of view) is 9.5 $\times$ 10$^7$ M$_{\odot}$.  A
strong correlation is found between the disk and extraplanar emission
in NGC~2188, but the 1-$\sigma$ deviations from the average vertical
profile are small, suggesting that while there are a number of filaments
and plumes in the halo of the galaxy, there is also a pervasive
diffuse eDIG layer.  This diffuse layer can be seen in Figure~7
extending into the halo out to about 1~kpc above the disk.

Domg\"orgen \& Dettmar (1997) also imaged NGC~2188 in H$\alpha$ +
[N~II].  Our results closely match theirs.  They measure an H$\alpha$
luminosity of 1.1 $\times$ 10$^{40}$ erg s$^{-1}$, a value which
matches ours within the errors of the measurements.  They also detect
a number of filamentary structures whose positions are highly
correlated with H~II regions located in the southern portion of the
galaxy.

The [N~II]/H$\alpha$ ratio map is dominated by an artificial radial
gradient due to a wavelength mismatch between the H$\alpha$ and [N~II]
images. Nevertheless a shallow vertical gradient appears to be present
in this object, in general agreement with the spectroscopic results of
Domg\"orgen \& Dettmar.

\vskip 0.2in
\centerline{NGC~2424}
\vskip 0.2in

NGC~2424 is an SB(r)b galaxy with an inclination of 82$^\circ$.  This
object was observed at both the KPNO and WHT, and both sets of images are
shown in Figure~8.  The better resolution of the WHT/TTF image
allows us to trace the spiral arm pattern in the disk of the galaxy.
The relatively large distance of NGC~2424 prevents us from resolving
some of the H~II regions in the disk; these unresolved H~II regions
may be mistaken for diffuse gas.  Although the relatively small
inclination and large distance of this galaxy make the analysis
difficult, the WHT/TTF H$\alpha$ image seems to show an extended
extraplanar layer with average EM = 7 pc cm$^{-6}$.

The H$\alpha$ luminosity/D$^2_{25}$ ratio of this galaxy is $\sim$ 1.3
$\times$ 10$^{37}$ erg s$^{-1}$ kpc$^{-2}$, $\sim$ 13\% of which is
attributed to extraplanar gas.  The extraplanar gas has a scale height
of $\sim$ 10.8~kpc on the north side and 5.4~kpc on the south side
(based on the WHT observations), and a total ionized mass of $\sim$
2.2 $\times$ 10$^8$ M$_{\odot}$.  These numbers are subject to large
uncertainties. The relatively small 1-$\sigma$ deviations and lack of
significant correlation between extraplanar and disk emission suggest
that a rather uniform layer of extraplanar gas is associated with
NGC~2424.

The [N~II]/H$\alpha$ of NGC~2424 (Fig. 8) is unusual in that it does
not show any obvious vertical gradient. The central region harbors
[N~II]/H$\alpha$ ratios larger than unity.

\vskip 0.2in
\centerline{ESO~209-9}
\vskip 0.2in

The H$\alpha$ + [N~II] image of ESO~209-9 is shown in Figure 9.
Little H$\alpha$ emission is detected anywhere in the galaxy, let
alone extended emission into the halo. This is the only galaxy in the
sample with no obvious signs of extraplanar line emission. The modest
flux limit ($\sim$ 2 x 10$^{-17}$ erg s$^{-1}$ cm$^{-2}$
arcsec$^{-2}$) and spatial resolution (in pc) may explain the apparent
lack of eDIG in this object.

\vskip 0.2in
\centerline{NGC~2820}
\vskip 0.2in

NGC~2820 is part of an interacting group of galaxies which consists of
NGC~2805 -- NGC~2814 -- NGC~2820 and Mrk~108.  This is one of only two
galaxies in our sample with obvious interacting companions (the other
is NGC~3432). NGC~2820 was included in our study because radio
continuum data by van der Hulst \& Hummel (1985) show a thick disk of
radio emission associated with this object, a possible indication that
eDIG emission may also be present.  While no emission has been
detected connecting NGC~2805 to the other galaxies, van der Hulst \&
Hummel detected a radio continuum bridge between NGC~2820/Mrk~108 and
NGC~2814.  NGC~2820 is an SB(s)c galaxy with an inclination of
90$^\circ$.  The H$\alpha$ + [N~II] image is presented in Figure~10.
Two regions of extraplanar emission (about 2~kpc in height) are
observed in the southeast halo.  Fainter emission extending $\sim$
1~kpc up above the disk are also detected, one in the northwest halo at
a distance of $\sim$ 4.5~kpc southwest of the galactic center, the
other in the southeast halo at a distance of $\sim$ 5~kpc northeast of
the center (both of which have average EM $\approx$ 8 pc cm$^{-6}$).

An H$\alpha$ luminosity/D$^2_{25}$ of 2.0 $\times$ 10$^{37}$ erg
s$^{-1}$ kpc$^{-2}$ is derived, 16\% of which is attributed to
extraplanar gas.  A two-exponential fit does not reproduce adequately
the extraplanar gas profile for the southeast halo, but the northwest
halo gas has a scale height of 1.08 kpc and a total extraplanar
ionized mass of $\sim$ 3.3 $\times$ 10$^7$ M$_{\odot}$ associated with
it.  The very strong correlation found between the southeast halo and
disk gas (also evident in Fig.~20) indicates that the extraplanar
emission is strongly tied to the H~II regions in the disk of the
galaxy.  The large 1-$\sigma$ deviations of the SE extraplanar gas
indicates that the emission is not uniform across the galaxy, and
confirms that the filamentary structure correlates strongly with the
distribution of the H~II regions in the disk.

\vskip 0.2in
\centerline{NGC~3044}
\vskip 0.2in

Like NGC~2820, NGC~3044 is an SB(s)c galaxy seen essentially edge-on
(90$^\circ$).  NGC~3044 was observed with the WHT/TTF and at CTIO; the
resulting H$\alpha$, [N~II], and H$\alpha$ + [N~II] images are presented in
Figure~11.  Most remarkable in the WHT image is the arc of emission
(with an EM $\sim$ 20 pc cm$^{-6}$) seen rising above the disk into
the southern halo.  Located about 1~kpc west of the nucleus, the arc is
over 1~kpc in diameter and extends $\sim$ 1.5~kpc into the halo.  This
is most likely an expanding superbubble seen in projection.  Another
arc of H$\alpha$ emission is detected $\sim$ 7~kpc east of the nucleus
with an EM of 20 -- 35 pc cm$^{-6}$.  This feature extends $\sim$
1~kpc into the halo.  A more widespread layer of diffuse gas is also
detected on both sides of the galaxy, extending upward to a height of
$\sim$ 3~kpc.

The total H$\alpha$ luminosity/D$^2_{25}$ is $\sim$ 10 $\times$
10$^{37}$ erg s$^{-1}$ kpc$^{-2}$, 16\% of which is attributed to
extraplanar gas.  The scale height of the gas on the north side
(4.7~kpc) is greater than that on the south side (2.6~kpc; based on
the CTIO observations).  The ionized mass associated with this
extraplanar gas is 2.0 $\times$ 10$^8$ M$_{\odot}$, higher than most
galaxies.  The strong correlation between disk and north halo gas, but
the small 1-$\sigma$ deviation of the extraplanar emission from the
average vertical profile (based on the WHT observations) suggest that
the extensive extraplanar emission observed in the northern halo
originates from H~II regions in the disk of the galaxy, but is not
concentrated over them (although the H$\alpha$ intensity is somewhat
higher in these regions).  The locations of the apparent bubbles
extending into the southern halo do not appear to correlate with the
H~II regions in the disk of the galaxy, perhaps an indication that the
H~II regions in this region have had time to disappear since the
supernovae events took place in the disk.

Rossa \& Dettmar (2000) detect an extraplanar layer of DIG emission,
but while we detect eDIG up to a height of 3~kpc, they only detect it
up to a height of $\sim$ 1~kpc.  They also detect several plumes of
emission.  Rossa \& Dettmar also detect the extended arc structure
south of the disk of the galaxy.  They measure an H$\alpha$ luminosity
of 1.3 $\times$ 10$^{41}$ erg s$^{-1}$, slightly higher, but not
significantly so, than ours.  Collins et al. (2000) also observed
NGC~3044 and detected similar features including an extensive
extraplanar layer of DIG emission.

Mismatch in the central wavelengths of the H$\alpha$ and [N~II] images
are causing a radial gradient in the ratio map. Nevertheless, a slight
vertical gradient also appears to be present in [N~II]/H$\alpha$, in
agreement with the spectroscopic results of Otte \& Dettmar (1999).

\vskip 0.2in
\centerline{NGC~3432}
\vskip 0.2in

The H$\alpha$ + [N~II] image of NGC~3432, an SB(s)m galaxy with an
inclination of 84$^\circ$, is presented in Figure~12.  Our analysis
of the halo gas was complicated by the presence of three, very bright
foreground stars located near NGC~3432 which did not subtract out very
well.  The regions contaminated by these stars were excluded from the
calculations of the eDIG properties. Some eDIG is apparent over the
inner 2~kpc of the galaxy, extending out about 1.5~kpc in height and
ranging in EM from 2 to 10 pc cm$^{-6}$.  A faint arc of gas (possibly
a superbubble) with an average EM $\approx$ 10 pc cm$^{-6}$ is located
$\sim$ 3~kpc from the center of the galaxy, extending out into the NW
halo.  It is only $\sim$ 500~pc in diameter but reaches a height of
1~kpc.

The derived H$\alpha$ luminosity/D$^2_{25}$ is 8.6 $\times$ 10$^{37}$
erg s$^{-1}$ kpc$^{-2}$, only $\sim$ 7\% of which is attributed to
extraplanar gas.  The scale height of the gas is $\sim$ 1.0~kpc on
either side of the galaxy, and the ionized mass associated with this
extraplanar gas is 5.5 $\times$ 10$^7$ M$_{\odot}$.  The disk and halo
components of NGC~3432 are not strongly correlated with each
other. The relatively small 1-$\sigma$ deviations indicate relatively
smooth eDIG morphology.

English \& Irwin (1997) observed NGC~3432 in H$\alpha$ and 20~cm radio
continuum.  They detect one extraplanar filament extending away from
the galaxy (what we called in the first paragraph a ``faint arc of
gas''), but conclude that extraplanar emission detected in NGC~3432 is
due to tidal interactions with UGC~5983 (at the time of observation,
it was not apparent to us that NGC~3432 was an interacting galaxy, and
so was included in our survey), and not indicative of any disk-halo
interaction.  Other than this one filament, they do not detect any
other eDIG, but point out that their H$\alpha$ observations were not
done under photometric conditions and therefore were not very
sensitive to faint diffuse emission.

\vskip 0.2in
\centerline{NGC~4013}
\vskip 0.2in

NGC~4013 is an Sb galaxy with an inclination angle of 90$^\circ$,
observed at the WHT with the TTF.  The H$\alpha$ and [N~II] images are
shown in Figure~13.  A dust lane is clearly present on the southwest
side of the disk, and a poorly subtracted foreground star coincides
with the position of the nucleus of the galaxy.  Extraplanar emission
is clearly detected at heights of $\sim$ 2~kpc into the southeast halo
above the nucleus (EM = 45 pc cm$^{-6}$ at 1~kpc height) as well as
into the northwest halo at $\sim$ 2~kpc on the southwest side of the
galaxy (EM = 25 pc cm$^{-6}$ at 1~kpc height).  A bright knot of
H$\alpha$ emission is also detected in the northwest halo, at a
galactocentric radius of 1~kpc on the northeast side.

The total H$\alpha$ luminosity/D$^2_{25}$ for NGC~4013 is 3.2 $\times$
10$^{37}$ erg s$^{-1}$ kpc$^{-2}$, only 6\% of which is attributed to
extraplanar gas.  The scale height of the gas is $\sim$ 1.5~kpc on the
southeast side of the galaxy (the northwest side was poorly fit by a
two-exponential function), and the total ionized mass associated with
this extraplanar gas is 6.5 $\times$ 10$^7$ M$_{\odot}$.  There is
little or no correlation between the disk and halo gas, and no
substantial deviation from the average extraplanar vertical profile
with radius.  The presence of the foreground star near the nucleus of
the galaxy makes further analysis difficult.

NGC~4013 was also observed by Rand (1996).  He detects four filaments
on either side of the central region, forming an ``H'' structure which
extends about 2.5~kpc into the halo.  Rand also detects extraplanar
emission on the northeast side, extending up about 2.5~kpc, as well as
on the southwest side (but not as prominent).  Interestingly, we also
detect this less prominent emission on the southwest side, but not the
brighter emission on the northeast side.  Part of the apparent
discrepancy comes from the fact that Rand's data include a
contribution from [N~II] emission. The [N~II] image and
[N~II]/H$\alpha$ map indicate that the [N~II] emission is enhanced in
this region (we also observed this galaxy spectroscopically, and
confirm that [N~II]/H$\alpha$ $\approx$r 1.5 at $\vert$z$\vert$ $\sim$
1~kpc; see Paper II). The ratio map also shows a strong vertical
gradient. The [N~II] $\lambda$6583/H$\alpha$ ratios are H~II
region-like in the disk but reach values close to unity at
$\vert z\vert$ $\approx$ 1 kpc.

\vskip 0.2in
\centerline{NGC~4197}
\vskip 0.2in

NGC~4197 is an Sc galaxy at an inclination of 79$^\circ$; the
H$\alpha$ + [N~II] image is shown in Figure~14.  The relatively small
inclination of this object makes it possible to trace the general
spiral arm pattern in the disk of the galaxy.  A few bright foreground
stars are poorly subtracted, but only the one near the northeast edge
of the galaxy is close enough in apparent position to affect our
results.  Two faint plumes (EM $\sim$ 5 pc cm$^{-6}$) are visible,
both extending upward about 1 kpc in the southeast halo of the galaxy
(one near the center of the galaxy; the other out at $\sim$ 1~kpc).

We derive an H$\alpha$ luminosity/D$^2_{25}$ of 4.8 $\times$ 10$^{37}$
erg s$^{-1}$ kpc$^{-2}$, $\sim$ 13\% of which is attributed to
extraplanar gas.  The extraplanar gas has a scale height of $\sim$
4.9~kpc and a total extraplanar ionized mass of $\sim$ 1.0 $\times$
10$^8$ M$_{\odot}$. This scale height is uncertain as it is based on
the detection of very faint emission.  A strong correlation between
the SE halo emission and the disk suggests that the observed plumes
originate from H~II regions in the disk of the galaxy.

\vskip 0.2in
\centerline{NGC~5529}
\vskip 0.2in

A H$\alpha$ + [N~II] image of NGC~5529 is shown in Figure~15.  This
Sc type galaxy is reported to be perfectly edge-on.  It is also one of
the farthest galaxies in our sample, so resolving H~II regions is a
challenge.  The two bright knots 1 -- 2 kpc west of the nucleus and
2 -- 3 kpc above and below the disk of the galaxy are stars that did not
completely subtract out.  Given an inclination angle of 90$^\circ$,
there appears to be faint (EM = 2 -- 5 pc cm$^{-6}$), widespread
extraplanar gas over the inner 10~kpc of the galaxy, but not nearly as
bright as that which has been detected in NGC~891.  No obvious plumes
or filaments are detected.

The total H$\alpha$ luminosity/D$^2_{25}$ is only 0.8 $\times$
10$^{37}$ erg s$^{-1}$ kpc$^{-2}$, 16\% of which is attributed to
extraplanar gas.  The scale height of the gas is greater on the
southeast side (6.8 kpc) than that on the northwest side
(4.5~kpc). But both of these numbers are uncertain because they are
based on the detection of very faint emission.  The total ionized mass
associated with this extraplanar gas is 2.0 $\times$ 10$^8$
M$_{\odot}$, considerably higher than most galaxies (but then NGC~5529
is one of the larger galaxies in the sample with an optical radius of
78.8 kpc).  Both the large deviations of the extraplanar gas from the
average vertical profile and the strong correlations between disk and
halo emission suggest that the northeast side of the galaxy is
characterized by filamentary eDIG.

\vskip 0.2in
\centerline{NGC~5965}
\vskip 0.2in

At an inclination of 85$^\circ$, NGC~5965 is similar to NGC~5529 in
its measured properties.  A faint spiral arm pattern is apparent in
the H$\alpha$ + [N~II] image (Figure~16).  Very faint emission (EM
$\sim$ 3 pc cm$^{-6}$) is detected extending out about 1 -- 2~kpc from
the bright H~II regions, possibly indicating the presence of eDIG. 

The total H$\alpha$ luminosity/D$^2_{25}$ is $\sim$ 1.0 $\times$
10$^{37}$ erg s$^{-1}$ kpc$^{-2}$, $\sim$ 16\% of which is attributed
to extraplanar gas.  The scale height of the faint extraplanar gas is
$\sim$ 11.6~kpc on the northwest side of the galaxy, and 6.5~kpc on
the southeast side, and the total extraplanar ionized mass is $\sim$
2.3 $\times$ 10$^8$ M$_{\odot}$.  These numbers are uncertain because
they are based on the detection of very faint emission. There is a
strong correlation between the disk and halo emission on both sides of
the galaxy, corresponding to the faint extraplanar emission observed
over the inner regions of the galaxy.

\vskip 0.2in
\centerline{ESO~240-11}
\vskip 0.2in

Obtained at the AAT with the TTF, the H$\alpha$ and [N~II] images of
ESO~240-11 are presented in Figure~17.  An inclination angle of
90$^\circ$ is reported for this galaxy, but looking at the inner
region of the H$\alpha$ image, the disk emission does not run through
the center of the bulge, suggesting that a slightly lower inclination
(around 87$^\circ$) is more accurate.  A number of bright H$\alpha$
knots are visible slightly above the disk (with average EM $\approx$
25 pc cm$^{-6}$), but if the inclination of the galaxy is not exactly
90$^\circ$, given the rather large size of the galaxy, most if not all
of these could very well be H~II regions located in the disk of the
galaxy at moderately large galactocentric radii.  In between these
regions, though, is faint H$\alpha$ emission (EM = 4 - 10 pc
cm$^{-6}$), some of which appears to extend out into the halo of the
galaxy.

The H$\alpha$ luminosity/D$^2_{25}$ ratio in this galaxy is $\sim$ 2.2
$\times$ 10$^{37}$ erg s$^{-1}$ kpc$^{-2}$, 15\% of which is
attributed to extraplanar gas.  A scale height (assuming a published
inclination value of 90$^\circ$) of 15.5~kpc was measured for the
southwest halo, and a scale height of 17.9~kpc was measured for the
northeast halo (the scale heights decrease only slightly if the
inclination is 87$^\circ$, but they are uncertain because they are
based on the detection of very faint emission).  The total ionized
mass in the extraplanar material is 2.4 $\times$ 10$^8$ M$_{\odot}$.
We find that the halo emission is highly correlated with the disk
emission, suggesting a strong link between the extraplanar emission
and the bright H~II regions in the disk. The [N~II]/H$\alpha$
excitation map does not show any obvious vertical gradient, although a
negative radial gradient may be present.

\clearpage

\figcaption{Continuum-subtracted H$\alpha$ and [N~II] $\lambda$6583
images, and [N~II] $\lambda$6583/H$\alpha$ excitation map of NGC~7817
taken with the WHT/TTF.  The bar in the upper-right hand corner of
these images represents a distance of 1~kpc or $\sim$ 10$\farcs$0.
The cross in each panel indicates the position of the continuum
nucleus. All of the images have been rotated counterclockwise through
an angle of 45$^\circ$.  The directions in which each halo extends has
been labeled (in this and all other images).  Contours on the
emission-line images represent flux levels of 5, 10, 20, 50, and 125
$\times$ 10$^{-18}$ erg s$^{-1}$ cm$^{-2}$ arcsec$^{-2}$. The ratio
map is on a linear intensity scale with contours at 0.2, 0.5, 1.0 and
1.4.}

\figcaption{a) Continuum-subtracted H$\alpha$ + [N~II] image of NGC~55
taken at CTIO, b) Continuum-subtracted H$\alpha$ and [N~II]
$\lambda$6583 images, and [N~II] $\lambda$6583/H$\alpha$ excitation
map of the central region of NGC~55 taken with the AAT/TTF.  The bar
in the upper-right hand corner of these images represents a distance
of 1~kpc or $\sim$ 129$\arcsec$.  The cross in each panel indicates
the position of the continuum nucleus. The images have been rotated
clockwise through an angle of 18$^\circ$.  Contours represent flux
levels of 10, 20, 40, 100, and 250 $\times$ 10$^{-18}$ erg s$^{-1}$
cm$^{-2}$ arcsec$^{-2}$ in figure (a) and 5, 10, 20, 50, and 125
$\times$ 10$^{-18}$ erg s$^{-1}$ cm$^{-2}$ arcsec$^{-2}$ in figure
(b). The ratio map is on a linear intensity scale with contours at
0.2, 0.5, 1.0 and 1.4.}

\figcaption{Continuum-subtracted H$\alpha$ and [N~II] $\lambda$6583
images, and [N~II] $\lambda$6583/H$\alpha$ excitation map of NGC~891
taken with the WHT/TTF.  The bar in the upper-right hand corner of
these images represents a distance of 1~kpc or $\sim$ 21$\farcs$7.
The cross in each panel indicates the position of the continuum
nucleus. The images have been rotated counterclockwise through an
angle of 68$^\circ$.  Contours on the emission-line images represent
flux levels of 8, 16, 32, 80, and 200 $\times$ 10$^{-18}$ erg s$^{-1}$
cm$^{-2}$ arcsec$^{-2}$. The ratio map is on a linear intensity scale
with contours at 0.2, 0.5, 1.0 and 1.4.}

\figcaption{Continuum-subtracted H$\alpha$ + [N~II] image of NGC~973
taken at KPNO.  The bar in the upper-right hand corner of this image
represents a distance of 1~kpc or $\sim$ 3$\farcs$3.  The cross
indicates the position of the continuum nucleus. The image has been
rotated counterclockwise through an angle of 132$^\circ$.  Contours
represent flux levels of 10, 20, 40, 100, and 250 $\times$ 10$^{-18}$
erg s$^{-1}$ cm$^{-2}$ arcsec$^{-2}$.}

\figcaption{Continuum-subtracted H$\alpha$ and [N~II] $\lambda$6583
images, and [N~II] $\lambda$6583/H$\alpha$ excitation map of NGC~1507
taken with the WHT/TTF.  The bar in the upper-right hand corner of
these images represents a distance of 1~kpc or $\sim$ 19$\farcs$5.
The cross in each panel indicates the position of the continuum
nucleus. The images have been rotated counterclockwise through an
angle of 79$^\circ$.  Contours on the emission-line images represent
flux levels of 8, 16, 32, 80, and 200 $\times$ 10$^{-18}$ erg s$^{-1}$
cm$^{-2}$ arcsec$^{-2}$. The ratio map is on a linear intensity scale
with contours at 0.2, 0.5, 1.0 and 1.4.}

\figcaption{Continuum subtracted H$\alpha$ and [N~II] $\lambda$6583
images, and [N~II] $\lambda$6583/H$\alpha$ excitation map of
ESO~362-11 taken with the AAT/TTF.  The bar in the upper-right hand
corner of these images represents a distance of 1~kpc or $\sim$
13$\farcs$4.  The cross in each panel indicates the position of the
continuum nucleus. The images have been rotated clockwise through an
angle of 76$^\circ$.  Contours on the emission-line images represent
flux levels of 0.8, 1.6, 3.2, 8, and 20 $\times$ 10$^{-18}$ erg
s$^{-1}$ cm$^{-2}$ arcsec$^{-2}$. The ratio map is on a linear
intensity scale with contours at 0.2, 0.5, 1.0 and 1.4.}

\figcaption{Continuum-subtracted H$\alpha$ and [N~II] $\lambda$6583
images, and [N~II] $\lambda$6583/H$\alpha$ excitation map of NGC~2188
taken with the AAT/TTF.  The bar in the upper-right hand corner of
these images represents a distance of 1~kpc or $\sim$ 13$\farcs$8.  The
cross in each panel indicates the position of the continuum
nucleus. The images have been rotated counterclockwise through an
angle of 85$^\circ$.  Contours on the emission-line images represent
flux levels of 3, 6, 12, 30, and 75 $\times$ 10$^{-18}$ erg s$^{-1}$
cm$^{-2}$ arcsec$^{-2}$. The ratio map is on a linear intensity scale
with contours at 0.2, 0.5, 1.0 and 1.4.}

\figcaption{a) Continuum-subtracted H$\alpha$ + [N~II] image of
NGC~2424 taken at KPNO, b) Continuum-subtracted H$\alpha$ and [N~II]
$\lambda$6583 images, and [N~II] $\lambda$6583/H$\alpha$ excitation
map of NGC~2424 taken with the WHT/TTF.  The bar in the upper-right
hand corner of these images represents a distance of 1~kpc or $\sim$
4$\farcs$7.  The cross in each panel indicates the position of the
continuum nucleus. The images have been rotated counterclockwise
through an angle of 9$^\circ$.  Contours represent flux levels of 8,
16, 32, 80, and 200 $\times$ 10$^{-18}$ erg s$^{-1}$ cm$^{-2}$
arcsec$^{-2}$ in figure (a) and 6, 12, 24, 60, and 150 $\times$
10$^{-18}$ erg s$^{-1}$ cm$^{-2}$ arcsec$^{-2}$ in figure (b). The
ratio map is on a linear intensity scale with contours at 0.2, 0.5,
1.0 and 1.4.}

\figcaption{Continuum-subtracted H$\alpha$ + [N~II] image of ESO~209-9
taken at CTIO.  The bar in the upper-right hand corner of this image
represents a distance of 1~kpc or $\sim$ 16$\farcs$1.  The cross
indicates the position of the continuum nucleus. The image has been
rotated counterclockwise through an angle of 64$^\circ$.  Contours
represent flux levels of 10, 20, 40, 100, and 250 $\times$ 10$^{-18}$
erg s$^{-1}$ cm$^{-2}$ arcsec$^{-2}$. This is the only galaxy in the
sample with no obvious signs of eDIG. }

\figcaption{Continuum-subtracted H$\alpha$ + [N~II] image of NGC~2820
taken at KPNO.  The bar in the upper-right hand corner of this image
represents a distance of 1~kpc or $\sim$ 10$\farcs$3.  The cross
indicates the position of the continuum nucleus. The image has been
rotated counterclockwise through an angle of 31$^\circ$.  Contours
represent flux levels of 15, 30, 60, 150, and 375 $\times$ 10$^{-18}$
erg s$^{-1}$ cm$^{-2}$ arcsec$^{-2}$.}

\figcaption{a) Continuum-subtracted H$\alpha$ + [N~II] image of
NGC~3044 taken at CTIO, b) Continuum-subtracted H$\alpha$ and [N~II]
$\lambda$6583 images, and [N~II] $\lambda$6583/H$\alpha$ excitation
map of NGC~3044 taken with the WHT/TTF.  The bar in the upper-right
hand corner of these images represents a distance of 1~kpc or $\sim$
10$\farcs$0.  The cross in each panel indicates the position of the
continuum nucleus. The images have been rotated clockwise through an
angle of 13$^\circ$.  Contours represent flux levels of 10, 20, 40,
100, and 250 $\times$ 10$^{-18}$ erg s$^{-1}$ cm$^{-2}$ arcsec$^{-2}$
in figure (a) and 5, 10, 20, 50, and 125 $\times$ 10$^{-18}$ erg
s$^{-1}$ cm$^{-2}$ arcsec$^{-2}$ in figure (b). The ratio map is on a
linear intensity scale with contours at 0.2, 0.5, 1.0 and 1.4. }

\figcaption{Continuum-subtracted H$\alpha$ + [N~II] image of NGC~3432
taken at KPNO.  The bar in the upper-right hand corner of this image
represents a distance of 1~kpc or $\sim$ 26$\farcs$5.  The cross
indicates the position of the continuum nucleus. The image has been
rotated counterclockwise through an angle of 52$^\circ$.  Contours
represent flux levels of 15, 30, 60, 150, and 375 $\times$ 10$^{-18}$
erg s$^{-1}$ cm$^{-2}$ arcsec$^{-2}$.}

\figcaption{Continuum-subtracted H$\alpha$ and [N~II] $\lambda$6583
images, and [N~II] $\lambda$6583/H$\alpha$ excitation map of NGC~4013
taken with the WHT/TTF.  The bar in the upper-right hand corner of
these images represents a distance of 1~kpc or $\sim$ 12$\farcs$1.
The cross in each panel indicates the position of the continuum
nucleus. The images have been rotated counterclockwise through an
angle of 24$^\circ$.  Contours on the emission-line images represent
flux levels of 10, 20, 40, 100, and 250 $\times$ 10$^{-18}$ erg
s$^{-1}$ cm$^{-2}$ arcsec$^{-2}$. The ratio map is on a linear
intensity scale with contours at 0.2, 0.5, 1.0 and 1.4. }

\figcaption{Continuum-subtracted H$\alpha$ + [N~II] image of NGC~4197
taken at KPNO.  The bar in the upper-right hand corner of this image
represents a distance of 1~kpc or $\sim$ 8$\farcs$9.  The cross
indicates the position of the continuum nucleus. The image has been
rotated counterclockwise through an angle of 54$^\circ$.  Contours
represent flux levels of 15, 30, 60, 150, and 375 $\times$ 10$^{-18}$
erg s$^{-1}$ cm$^{-2}$ arcsec$^{-2}$.}

\figcaption{Continuum-subtracted H$\alpha$ + [N~II] image of NGC~5529
taken at KPNO.  The bar in the upper-right hand corner of this image
represents a distance of 1~kpc or $\sim$ 4$\farcs$7.  The cross
indicates the position of the continuum nucleus. The image has been
rotated clockwise through an angle of 25$^\circ$.  Contours represent
flux levels of 10, 20, 40, 100, and 250 $\times$ 10$^{-18}$ erg
s$^{-1}$ cm$^{-2}$ arcsec$^{-2}$.}

\figcaption{Continuum-subtracted H$\alpha$ + [N~II] image of NGC~5965
taken at KPNO.  The bar in the upper-right hand corner of this image
represents a distance of 1~kpc or $\sim$ 4$\farcs$5.  The cross
indicates the position of the continuum nucleus. The image has been
rotated clockwise through an angle of 37$^\circ$.  Contours represent
flux levels of 10, 20, 40, 100, and 250 $\times$ 10$^{-18}$ erg
s$^{-1}$ cm$^{-2}$ arcsec$^{-2}$.}

\figcaption{Continuum-subtracted H$\alpha$ and [N~II] $\lambda$6583
images, and [N~II] $\lambda$6583/H$\alpha$ excitation map of
ESO~240-11 taken with the AAT/TTF.  The bar in the upper-right hand
corner of these images represents a distance of 1~kpc or $\sim$
5$\farcs$9.  The cross in each panel indicates the position of the
continuum nucleus. The images have been rotated counterclockwise
through an angle of 141$^\circ$.  Contours on the emission-line images
represent flux levels of 10 and 50 $\times$ 10$^{-18}$ erg s$^{-1}$
cm$^{-2}$ arcsec$^{-2}$. The ratio map ranges from 0.3 (light grey
areas) to about 0.9 -- 1.0 (darker areas). }

\figcaption{Distributions of the quantities derived from the narrow-band images. 
($a$) total H$\alpha$ luminosity;
($b$) ratio of the total H$\alpha$ luminosity to the square of the
galaxy diameter at $\mu_B$ = 25 B mag arcsec$^{-2}$; 
($c$) diffuse gas fraction or the fraction of the H$\alpha$-emitting material which is in the diffuse phase (see text for more detail); 
($d$) H$\alpha$ luminosity from the eDIG; 
($e$) fraction of the total H$\alpha$ luminosity which is produced by the eDIG; 
($f$) scale height of the eDIG based on a two-exponential fit to the emission; 
($g$) emission measure of the eDIG in the midplane derived from a two-exponential fit; 
($h$) electron density of the eDIG in the midplane derived assuming a
constant filling factor $f$ = 0.2 and a line-of-sight length $l$ = 5 kpc
for all the galaxies; 
($i$) one-sided ionized mass of the eDIG derived from the scale height and
the midplane electron density;
($j$) peak value of the correlation at zero radial shift between the disk and the eDIG emission (see \S 6.1 for more detail); 
($k$) 1-$\sigma$ deviation of the eDIG from the average vertical profile
(see \S 6.2 for more detail). }

\figcaption{Exponential fits to the vertical profile of H$\alpha$
images.  The dashed line represents a one-exponential fit while the
dot-dash line represents a two-exponential fit.  Both linear (top) and
log (bottom) flux scales are shown for each galaxy.}

\figcaption{(Top) The radial distribution of the extraplanar gas is
compared to the radial distribution of the disk emission.  The dark
solid line represents the summed emission within the disk while the
light solid line and the dashed line represent emission from both
sides of the halo (as labeled in each graph).  The halo emission has
been scaled such that the maximum intensity has been set to unity and
the relative intensities of both sides have been maintained.  The disk
emission has been scaled separately such that the maximum intensity of
the disk has been set to unity.  (Bottom) The panel on the left
displays the correlation coefficient of the disk emission with one
side of the halo emission while the panel on the right represents the
correlation of the disk emission with the other side.  A correlation
value of 1 represents a perfect correlation while a value of --1
represents a perfect anti-correlation.}

\clearpage

\epsscale{1.0}
\begin{figure}[htbp]
\figurenum{18}
\plotone{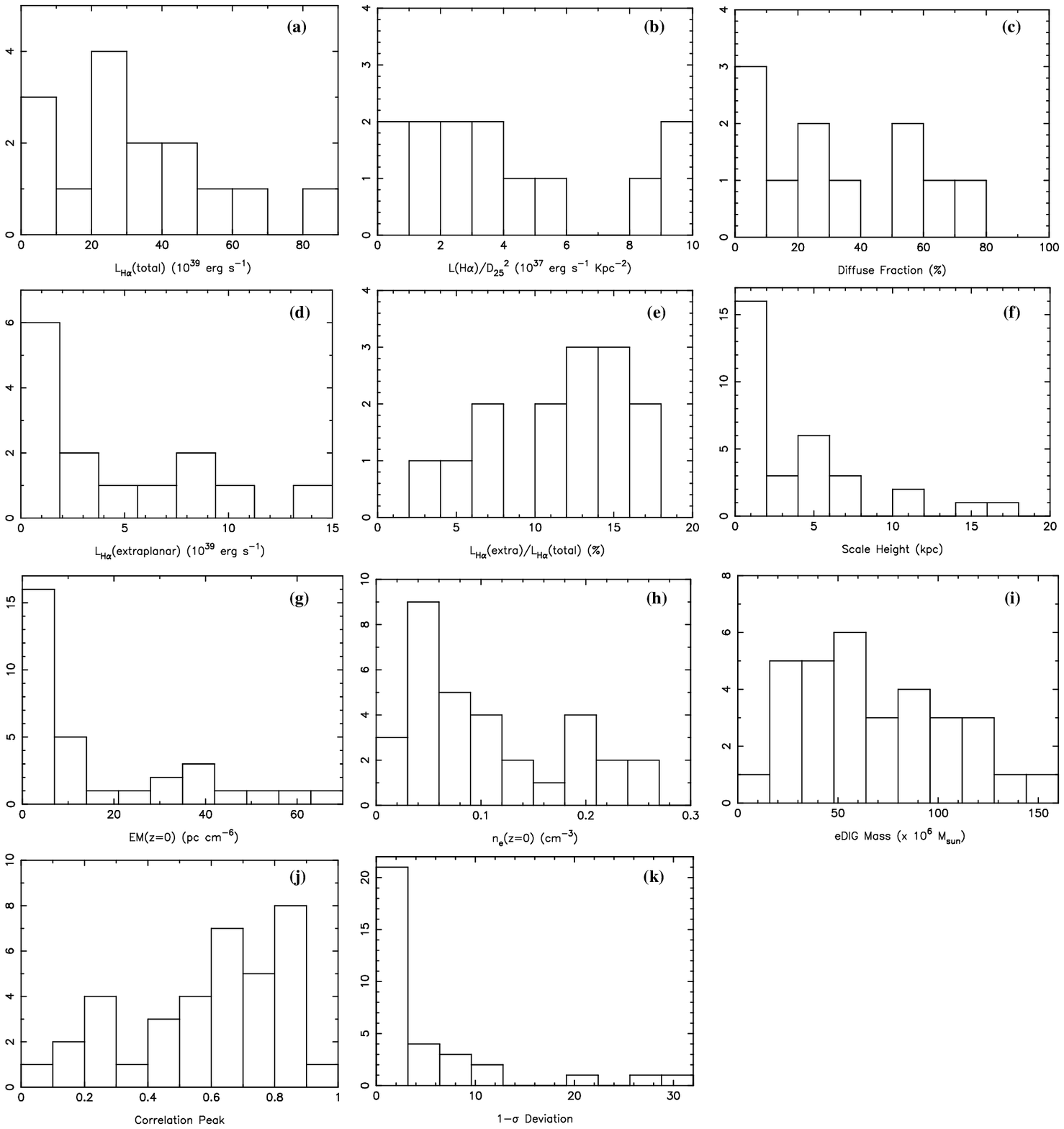}
\caption{}
\end{figure}

\clearpage

\epsscale{0.65}
\begin{figure}[htbp]
\figurenum{19ab}
\plottwo{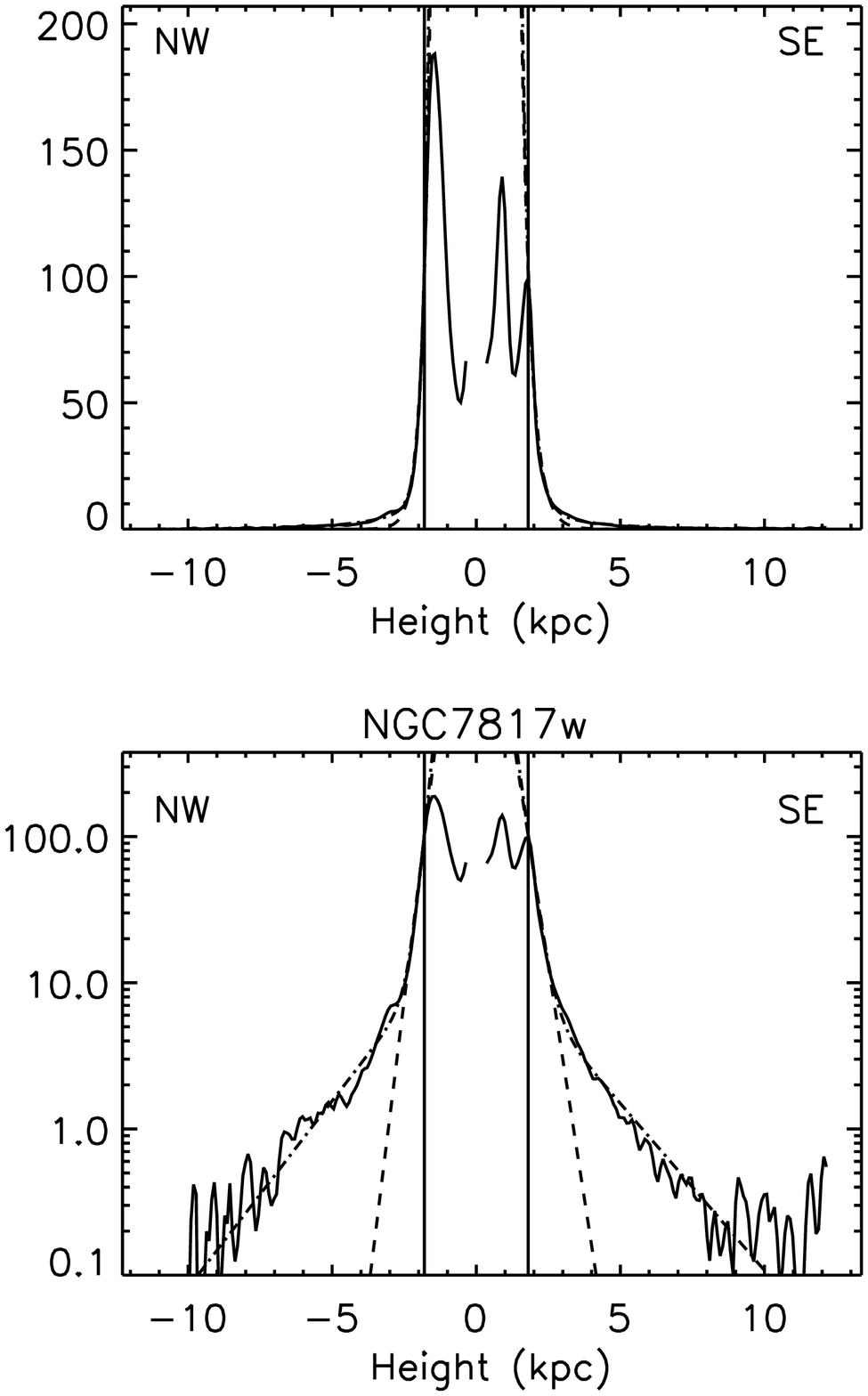}{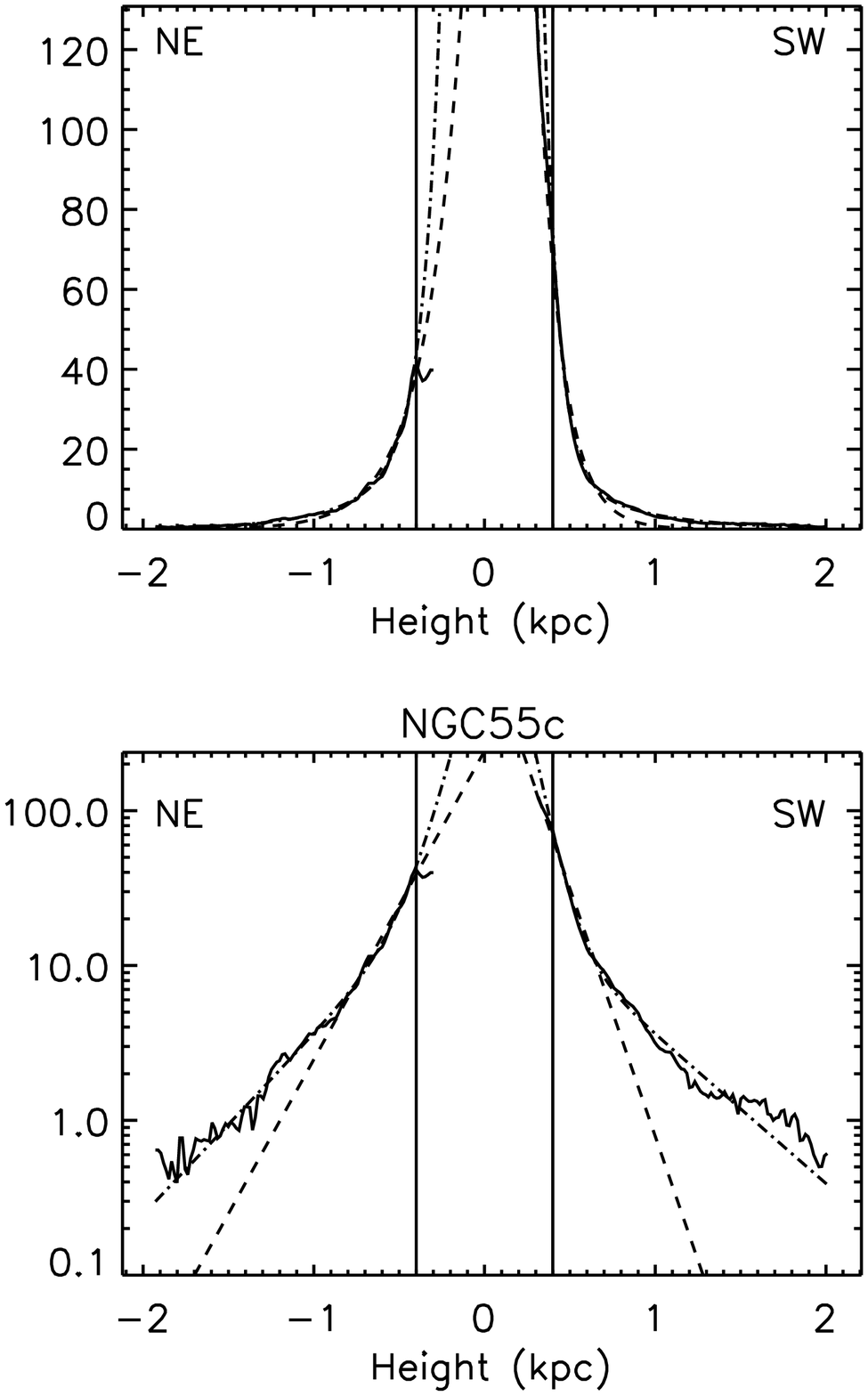}
\caption{}
\end{figure}

\epsscale{0.60}
\begin{figure}[htbp]
\figurenum{19cd}
\plottwo{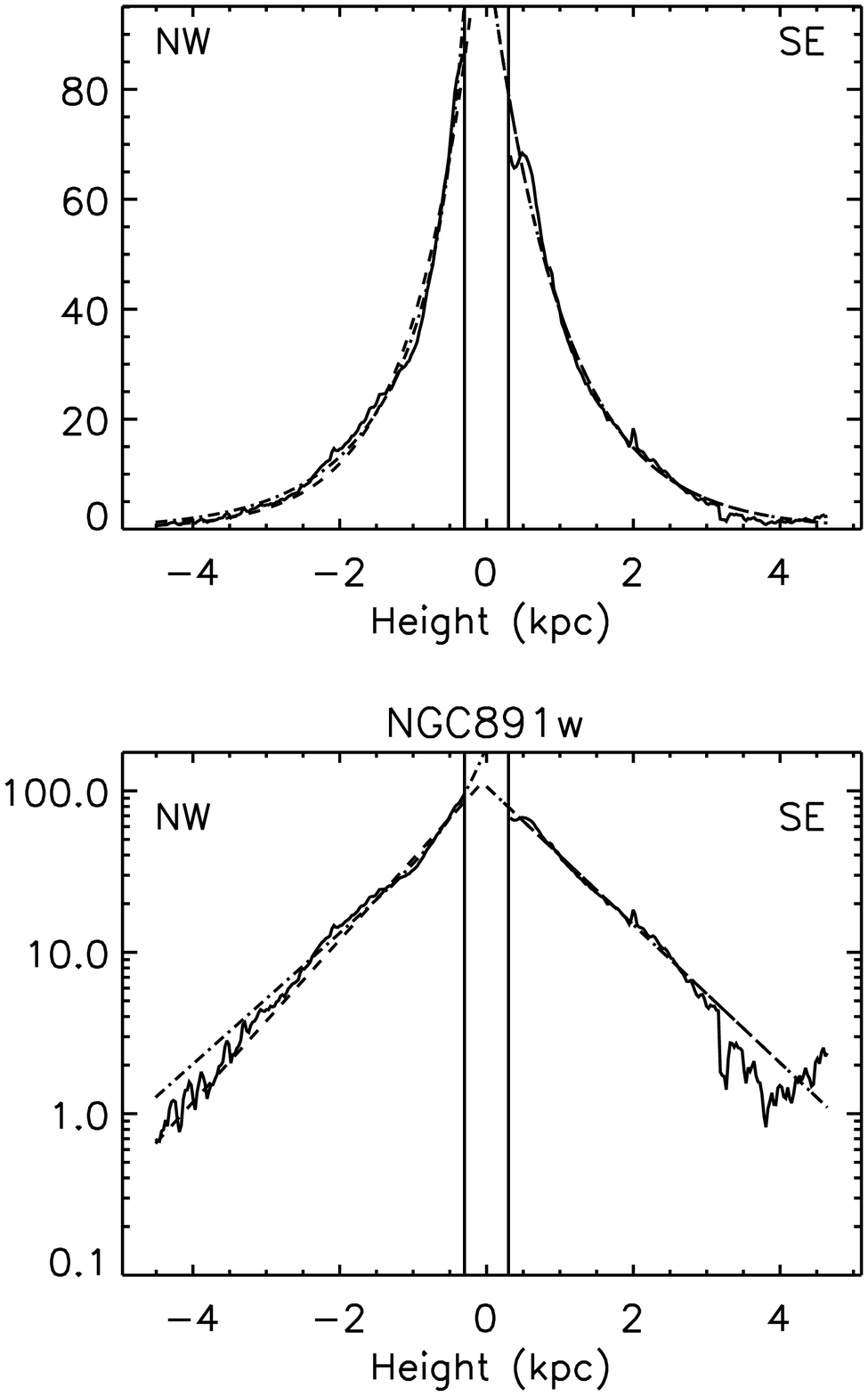}{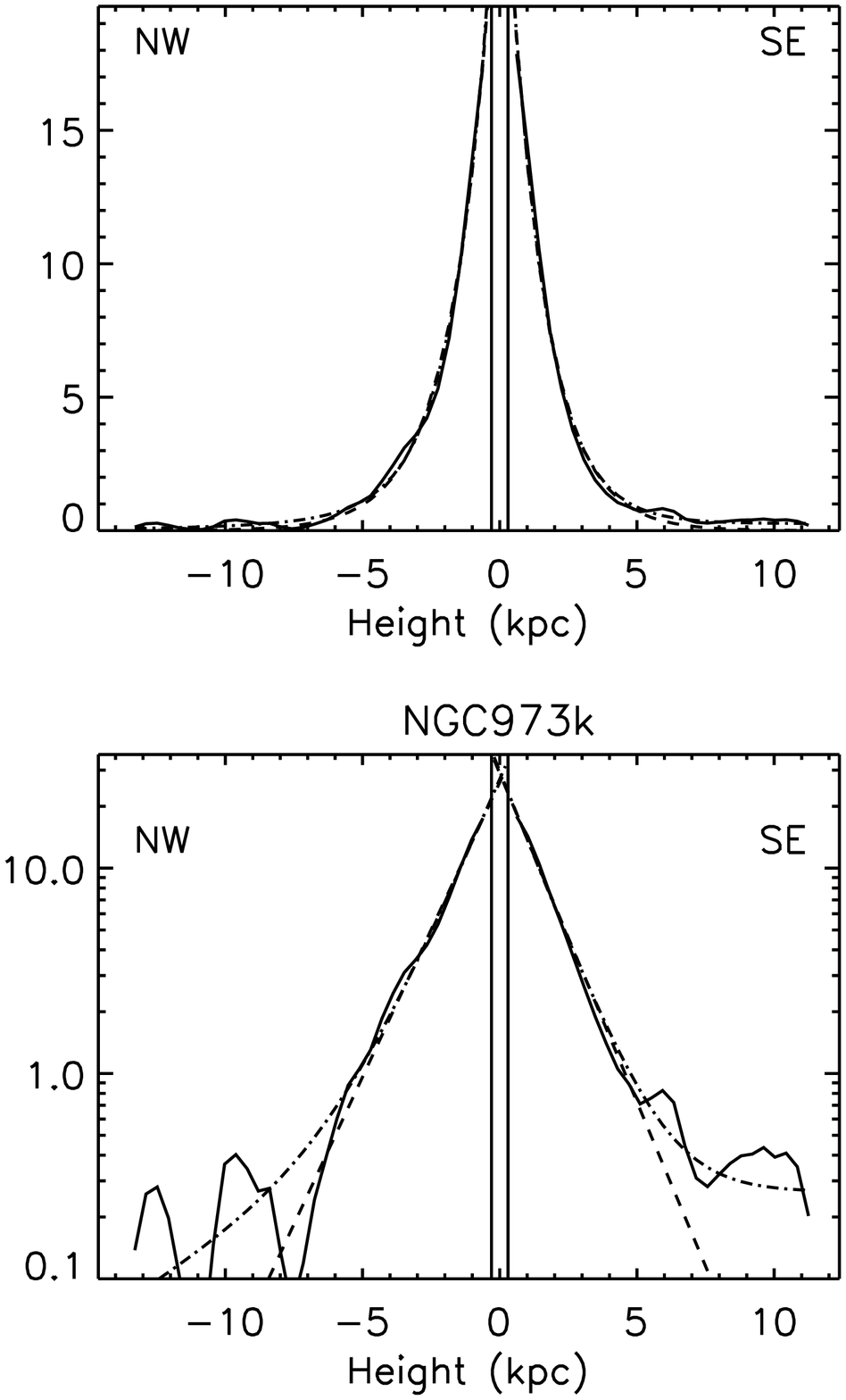}
\caption{(Cont'd)}
\end{figure}

\epsscale{0.60}
\begin{figure}[htbp]
\figurenum{19ef}
\plottwo{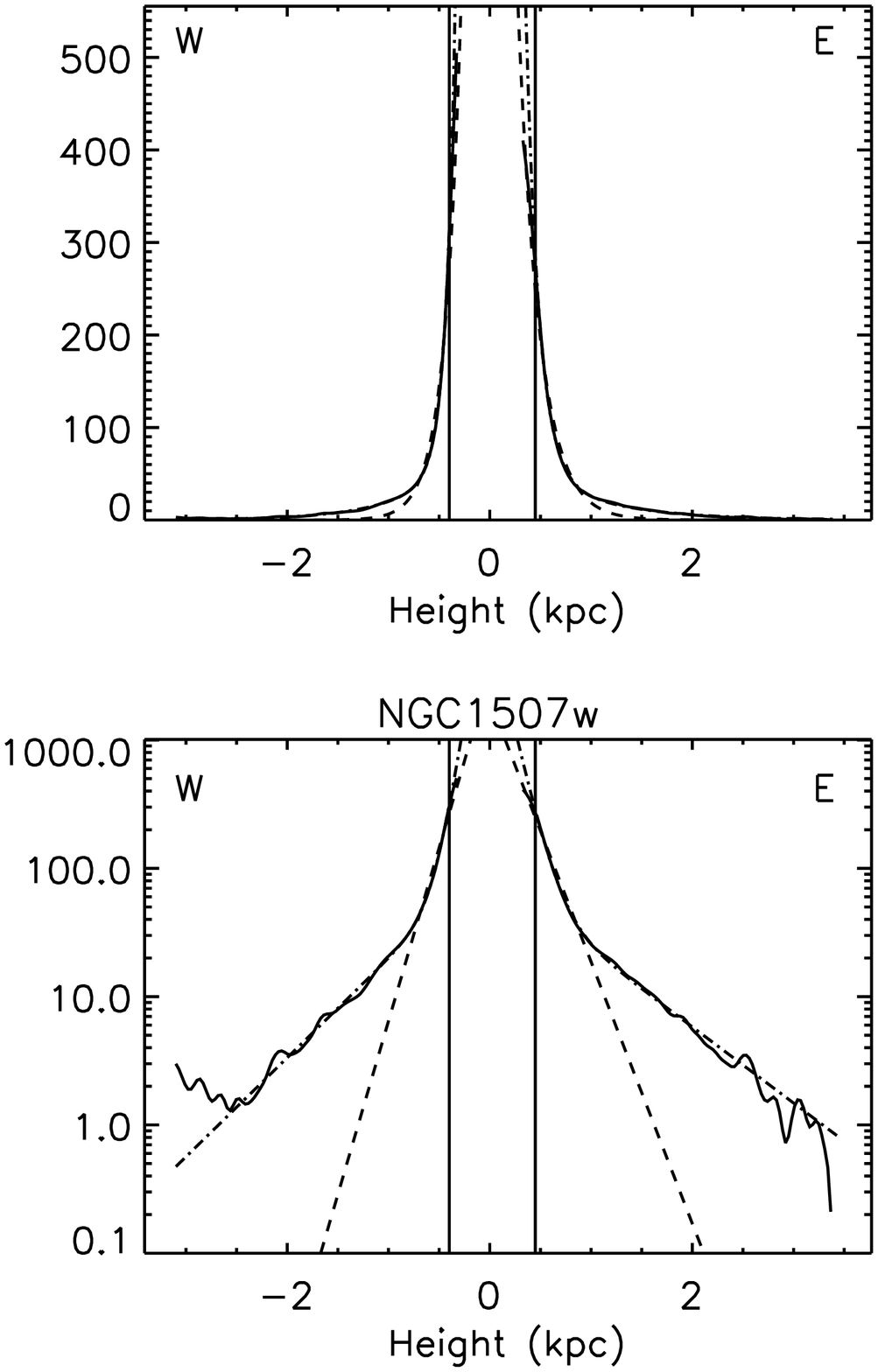}{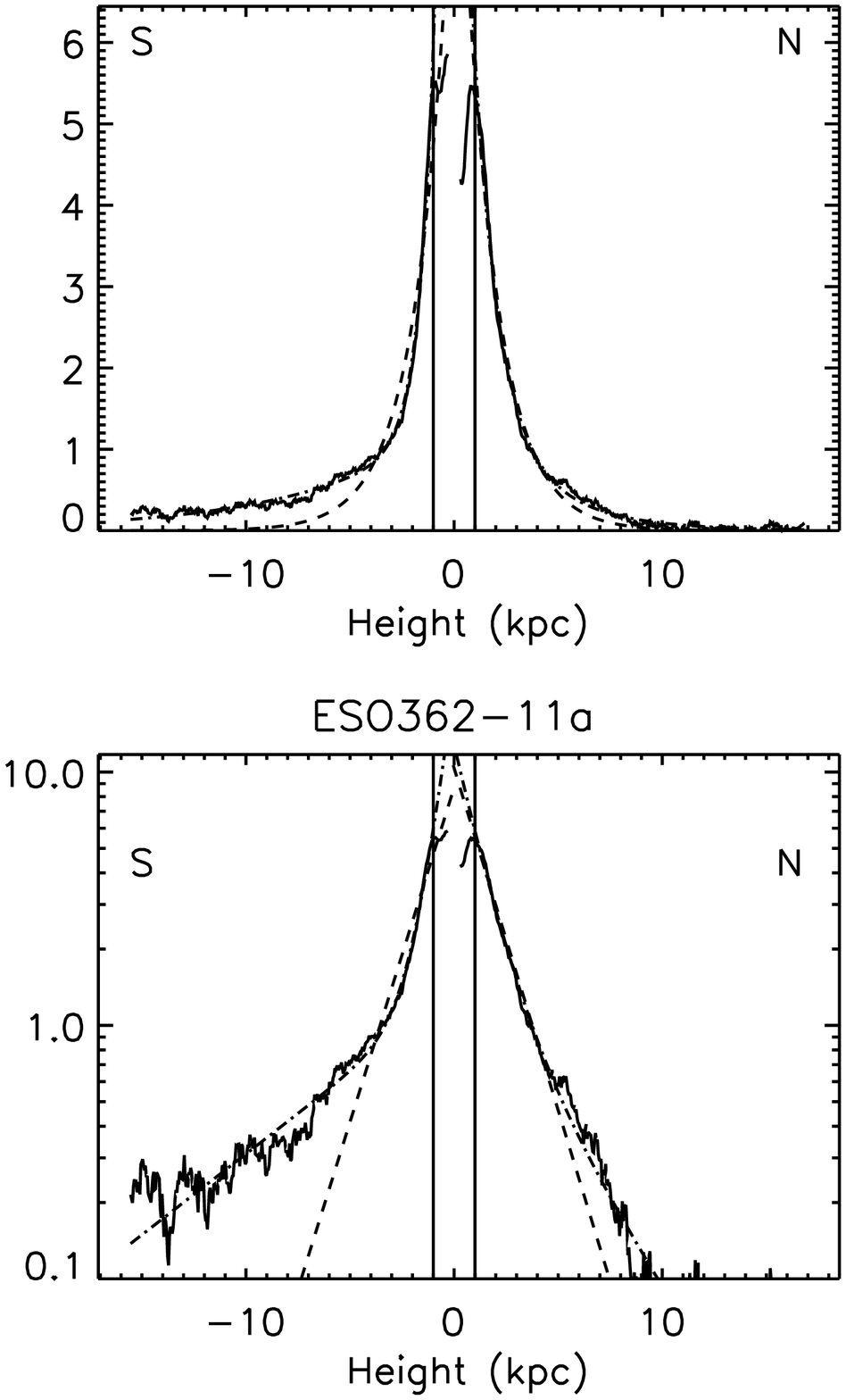}
\caption{(Cont'd)}
\end{figure}

\epsscale{0.60}
\begin{figure}[htbp]
\figurenum{19gh}
\plottwo{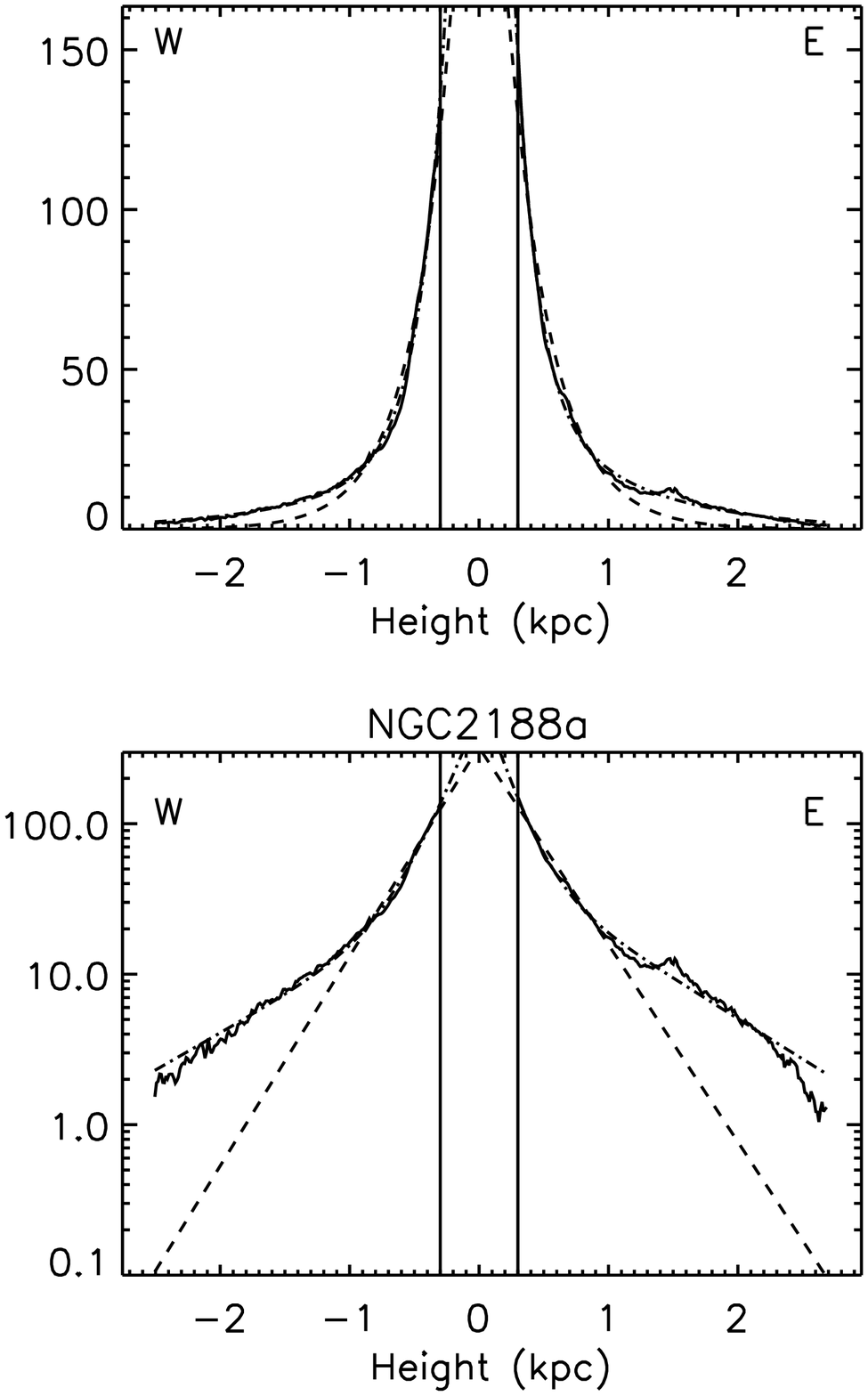}{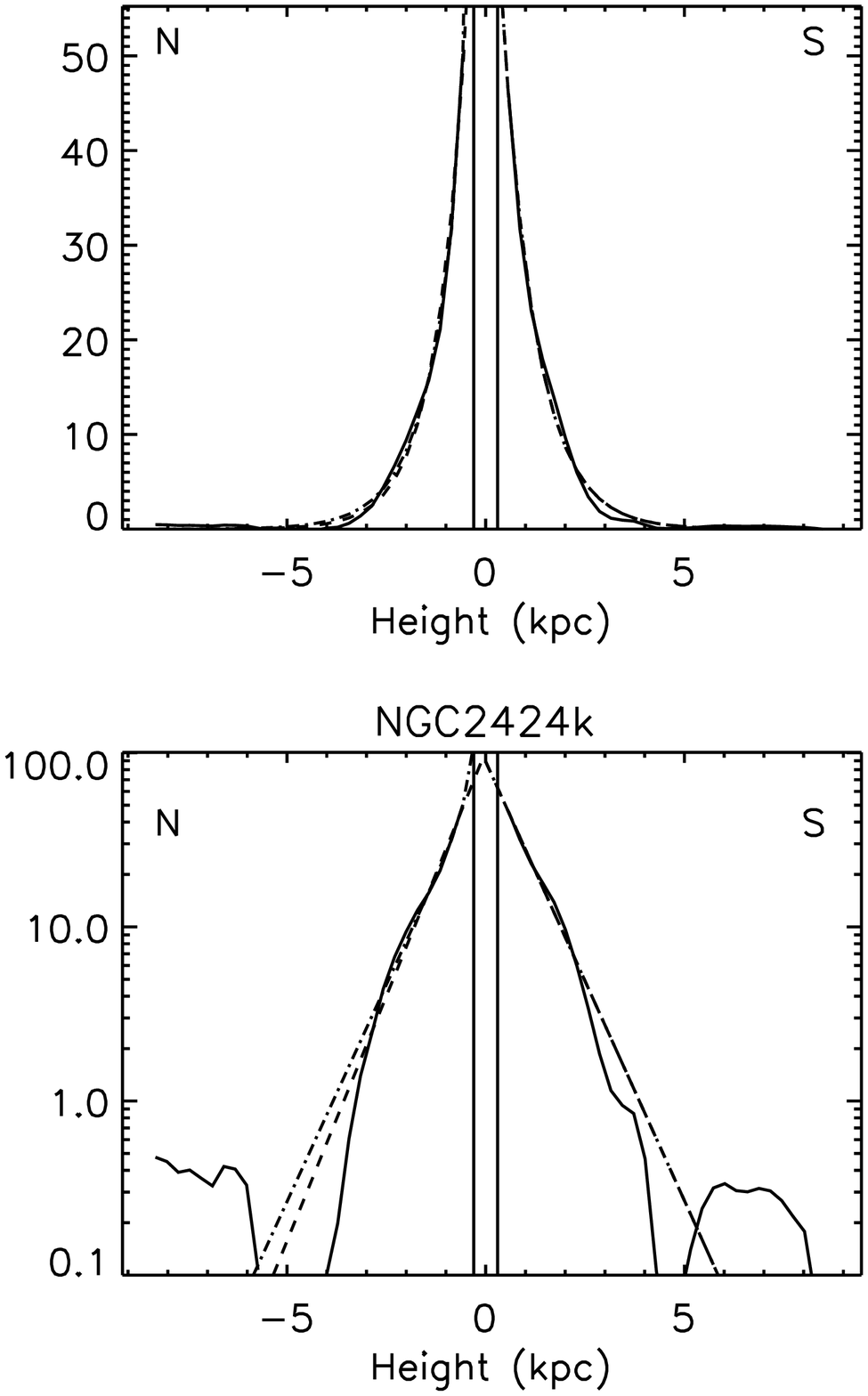}
\caption{(Cont'd)}
\end{figure}

\epsscale{0.60}
\begin{figure}[htbp]
\figurenum{19ij}
\plottwo{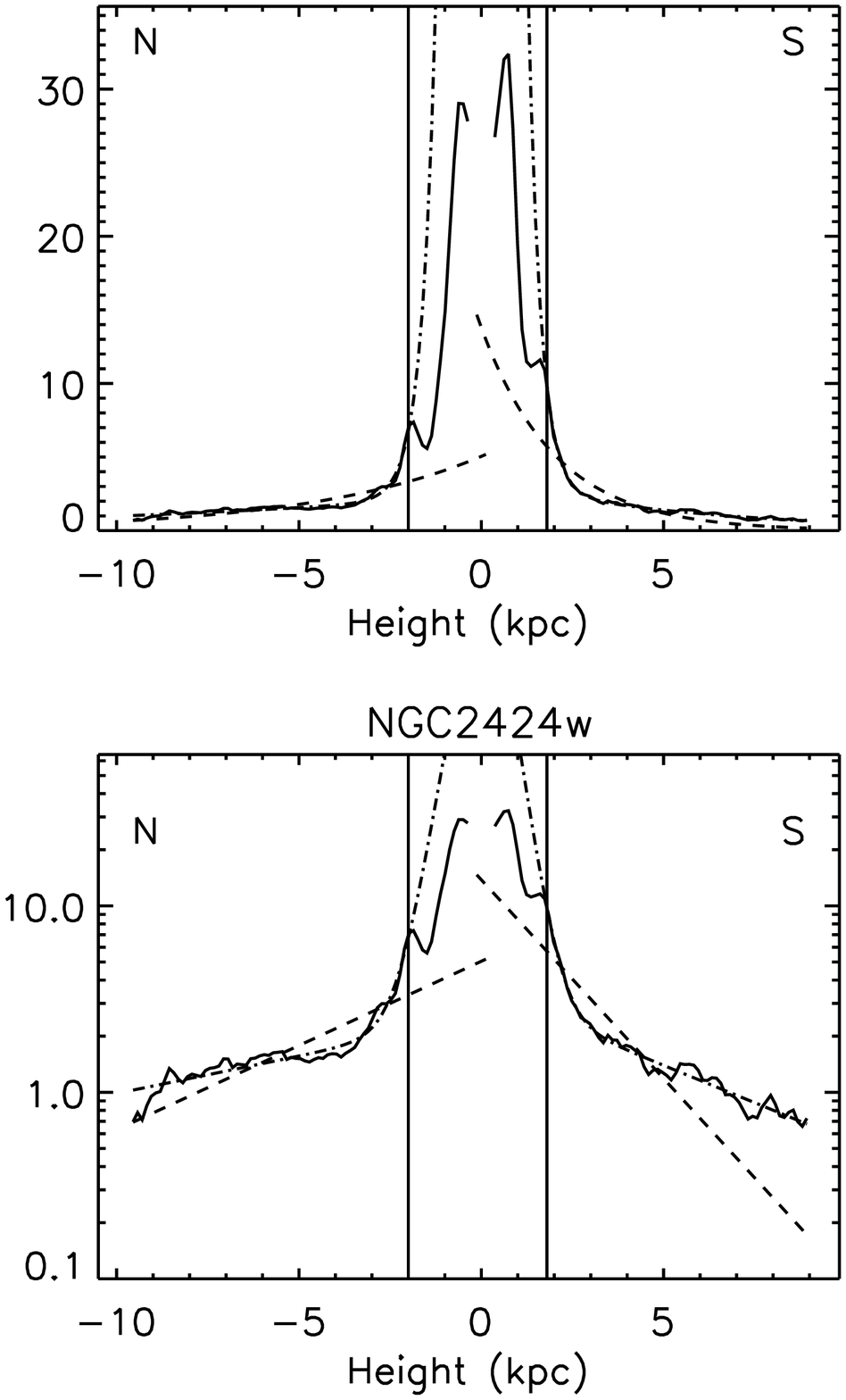}{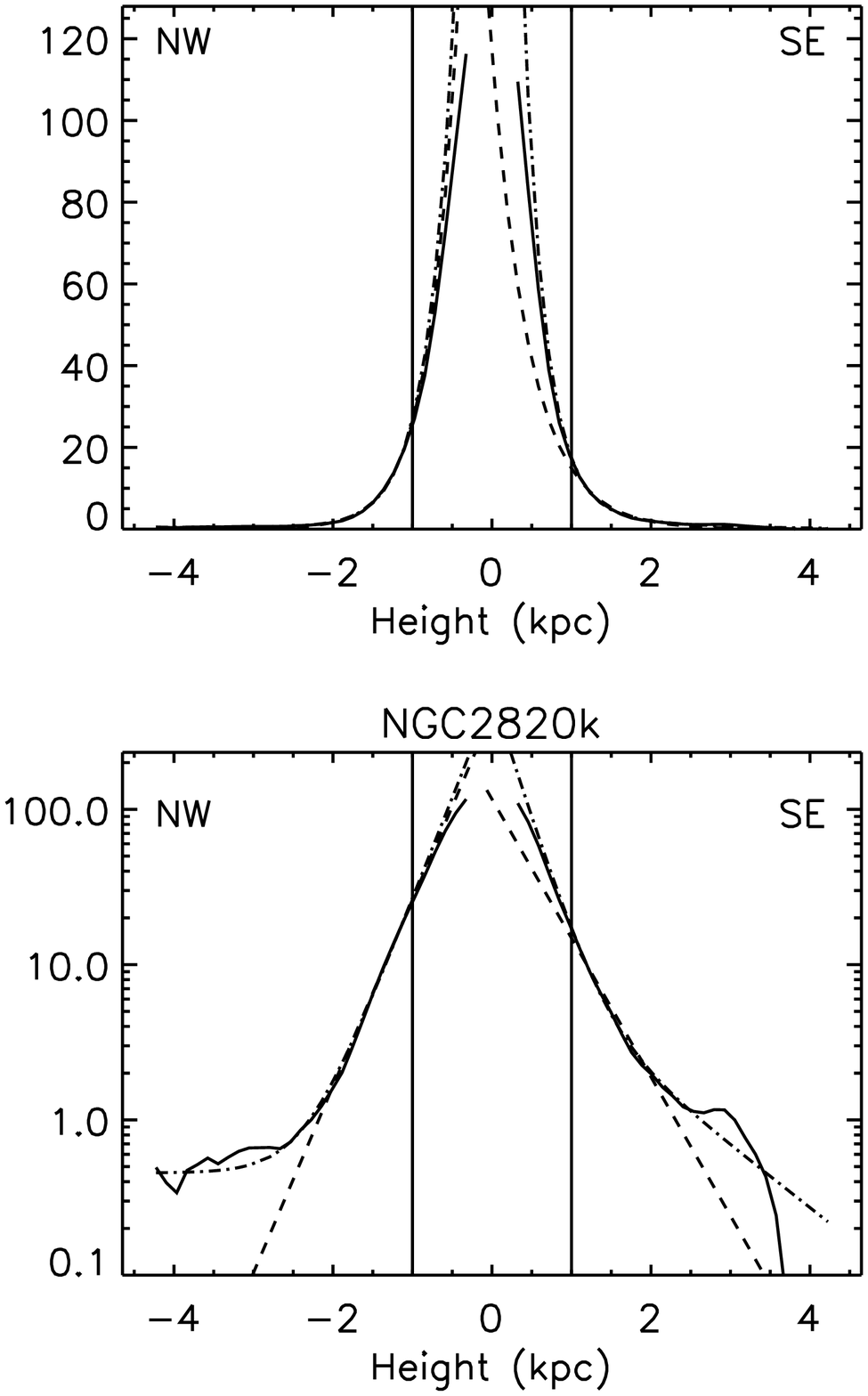}
\caption{(Cont'd)}
\end{figure}

\epsscale{0.60}
\begin{figure}[htbp]
\figurenum{19kl}
\plottwo{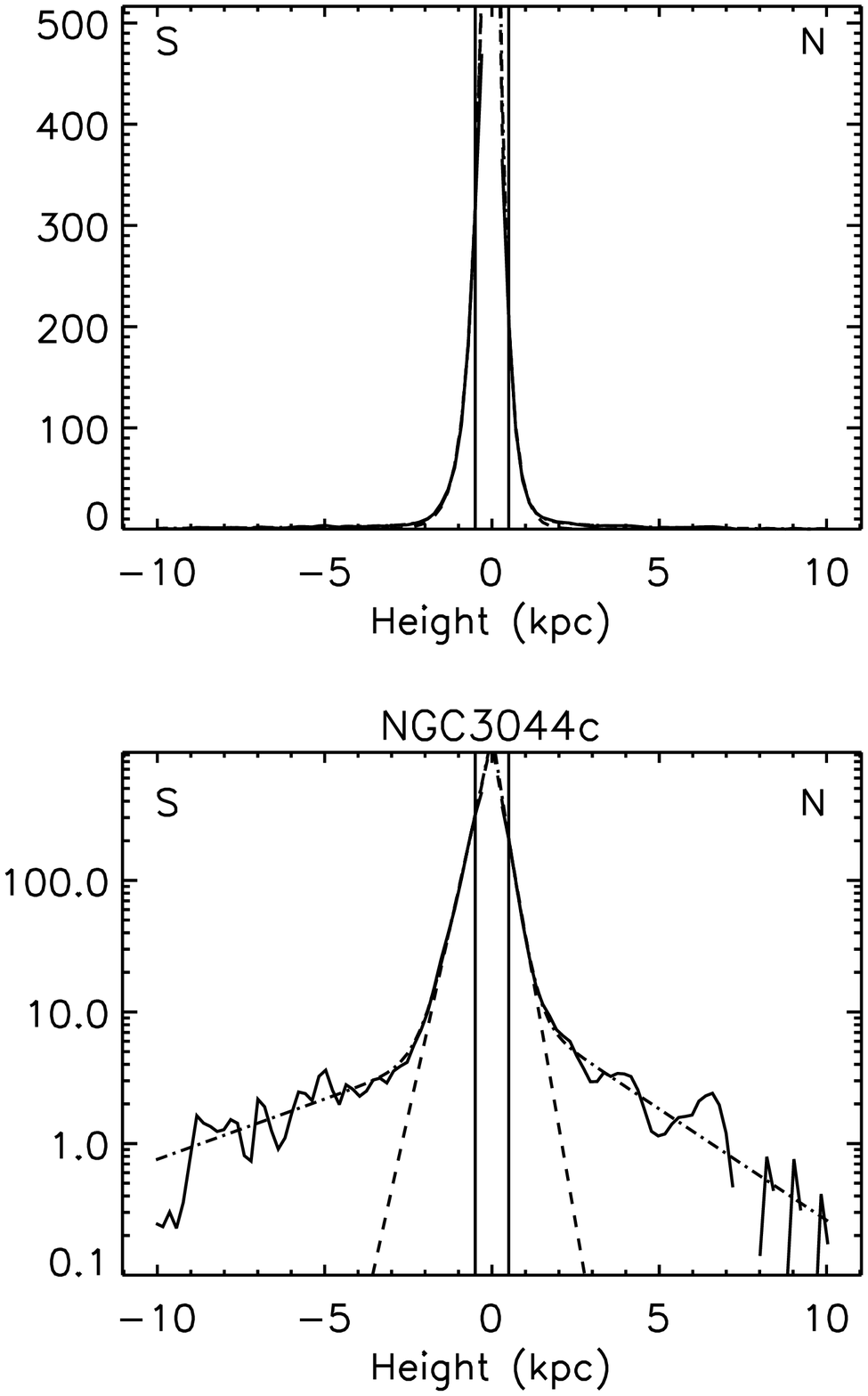}{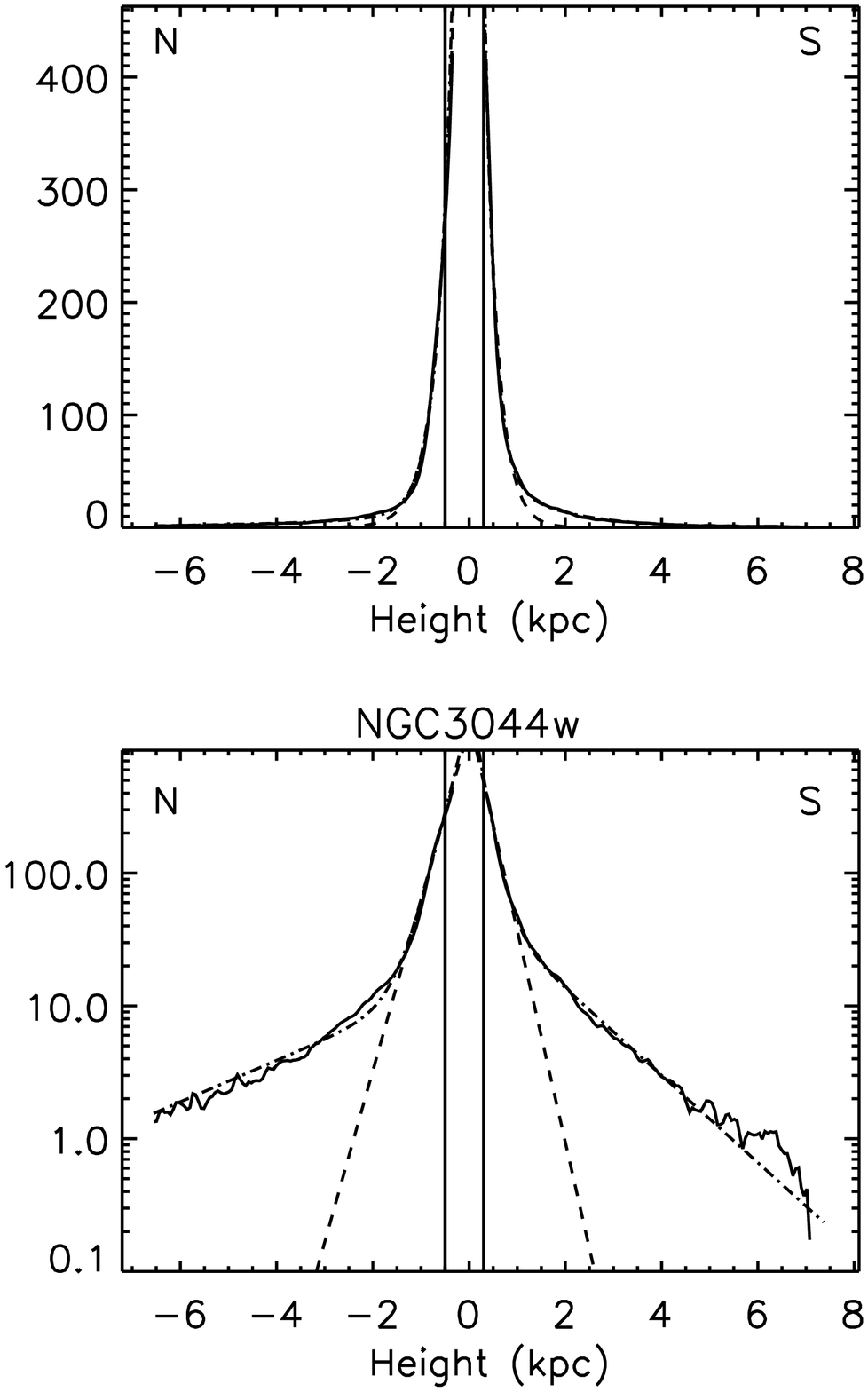}
\caption{(Cont'd)}
\end{figure}

\epsscale{0.60}
\begin{figure}[htbp]
\figurenum{19mn}
\plottwo{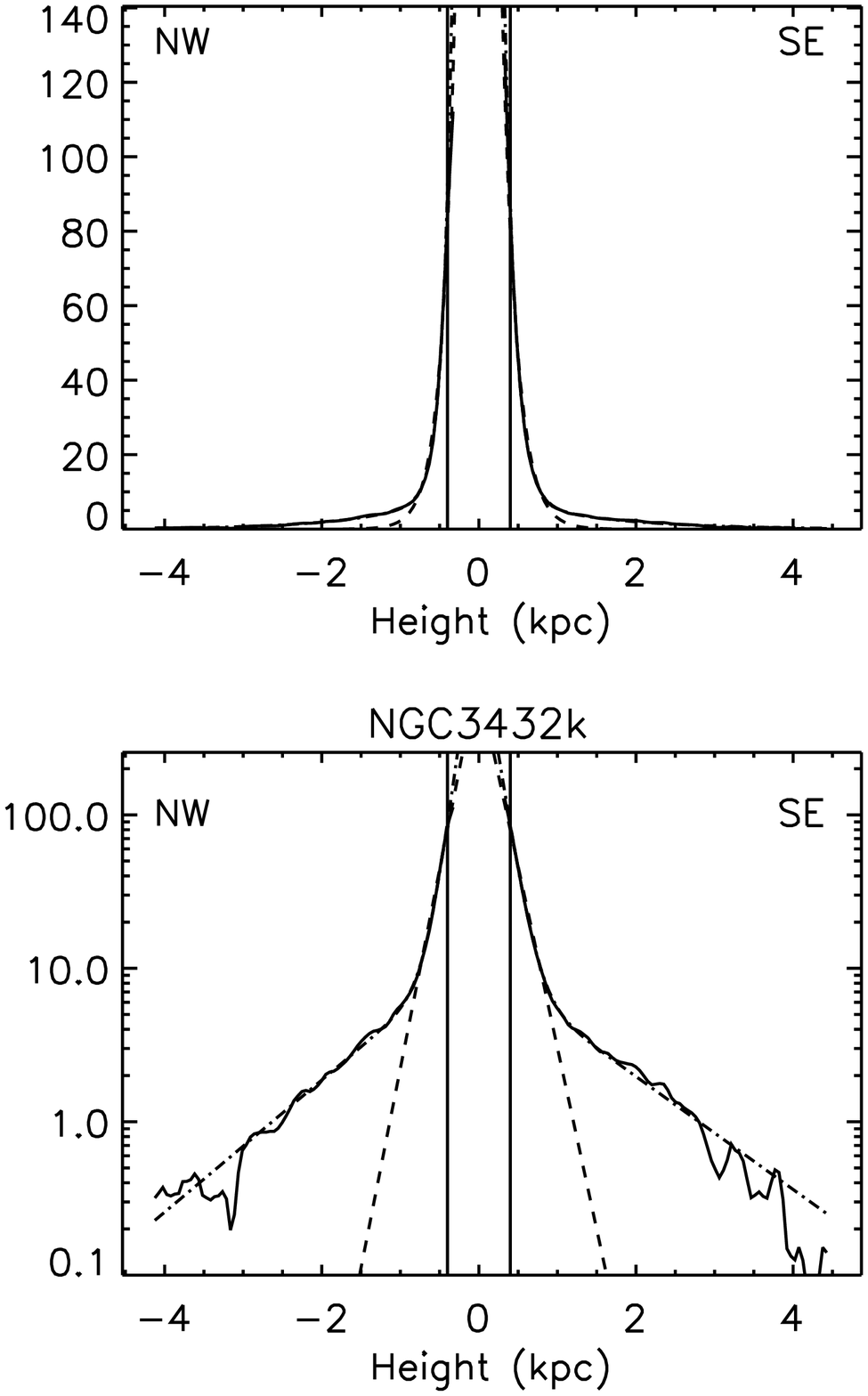}{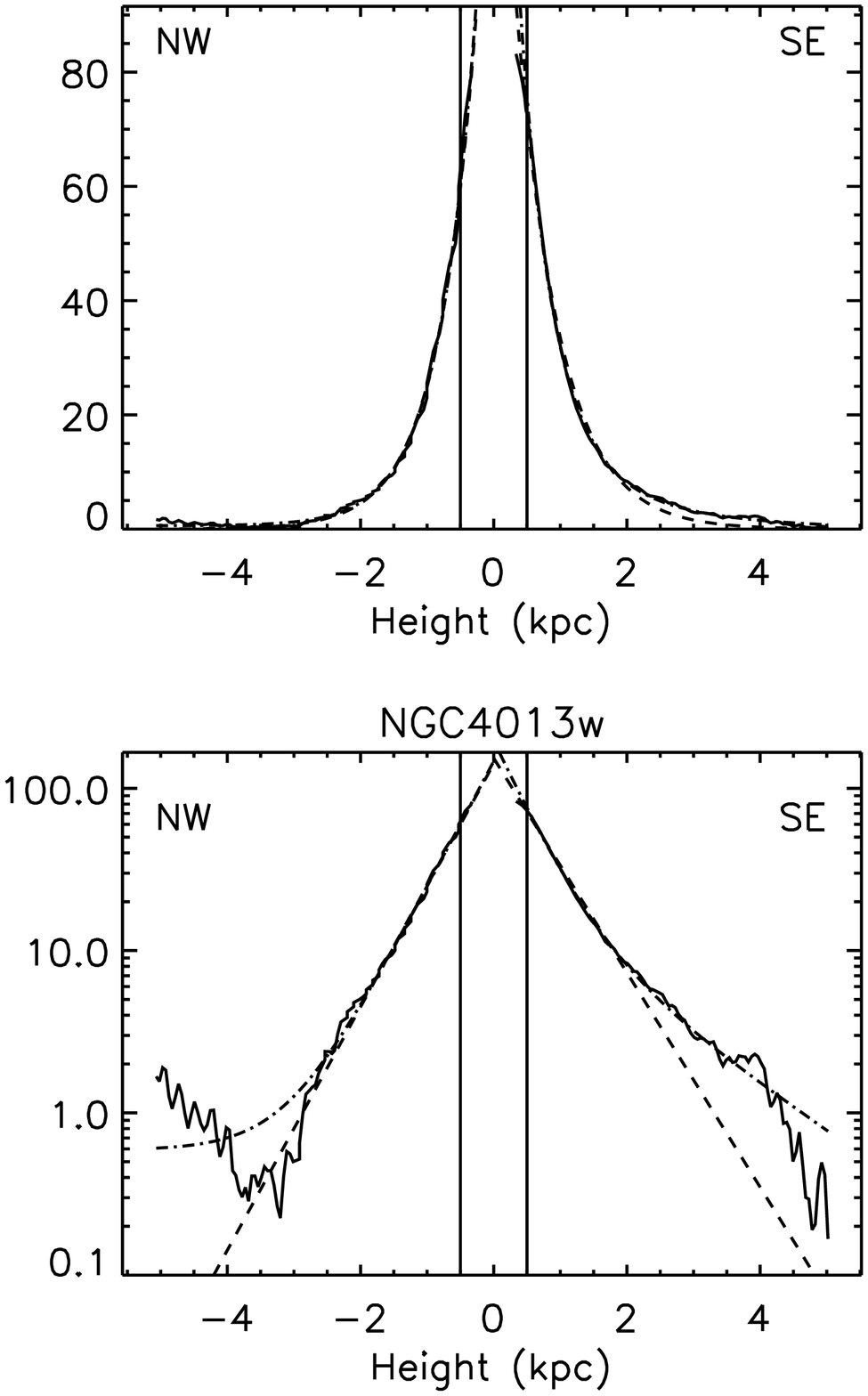}
\caption{(Cont'd)}
\end{figure}

\epsscale{0.60}
\begin{figure}[htbp]
\figurenum{19op}
\plottwo{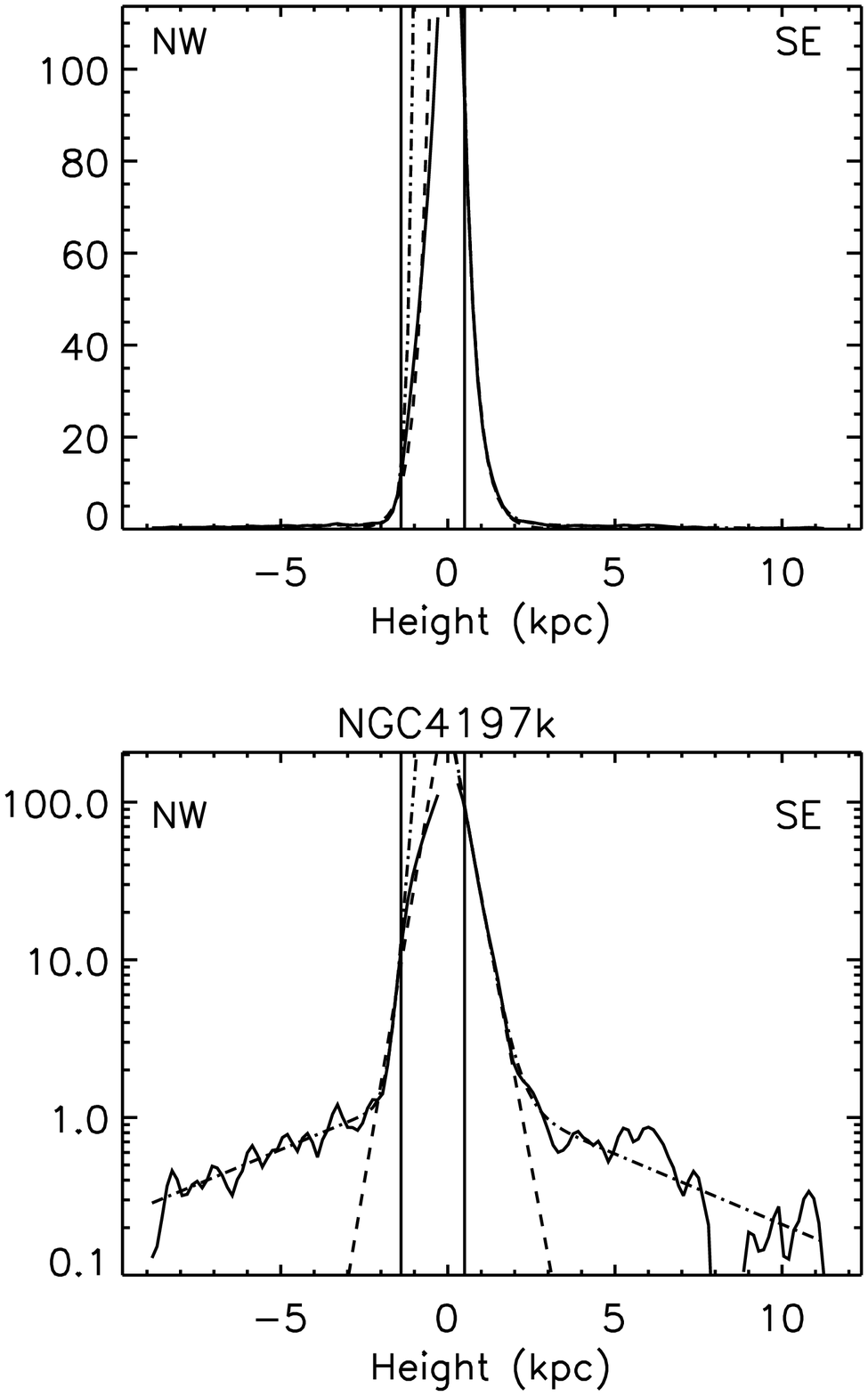}{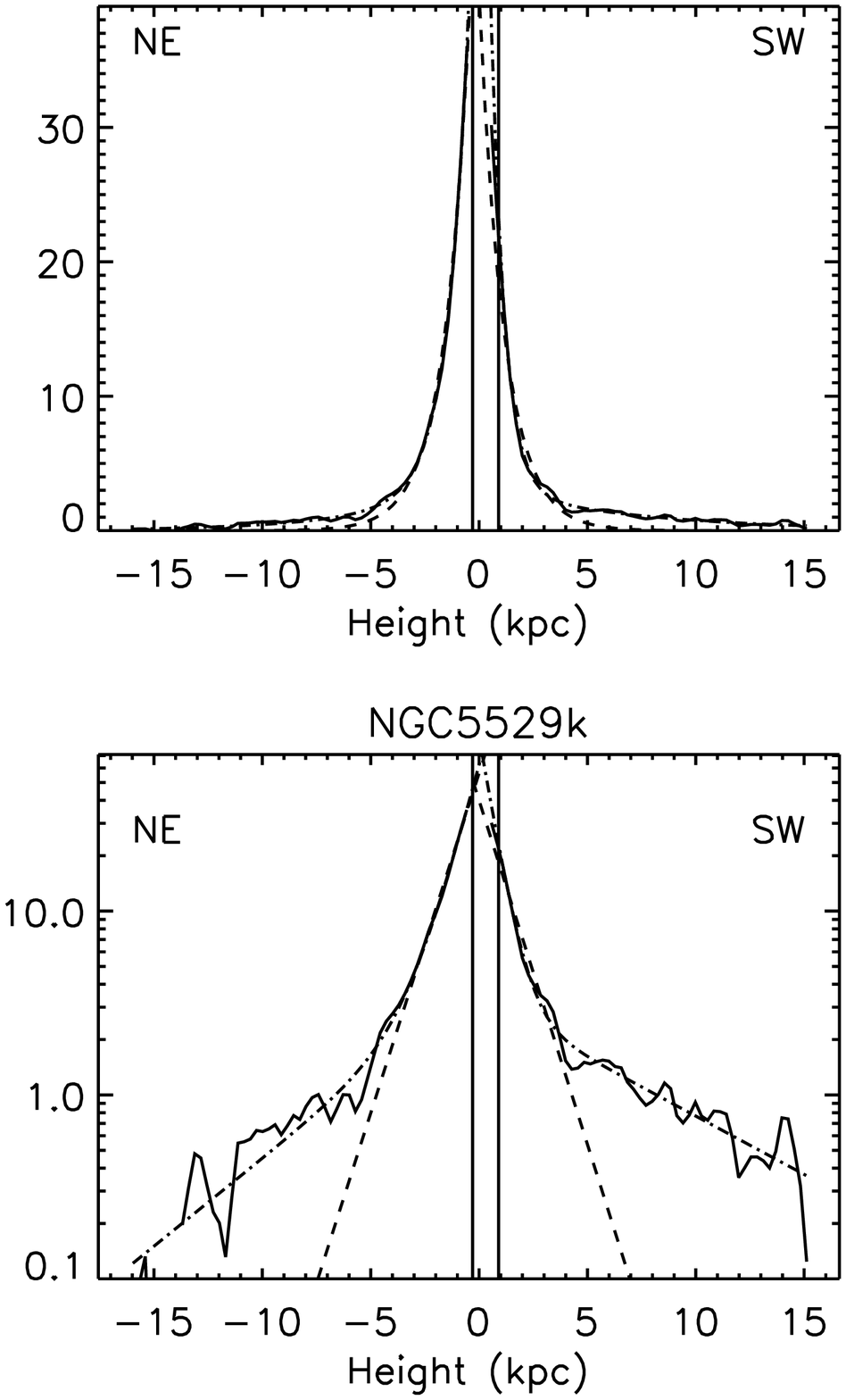}
\caption{(Cont'd)}
\end{figure}

\epsscale{0.60}
\begin{figure}[htbp]
\figurenum{19qr}
\plottwo{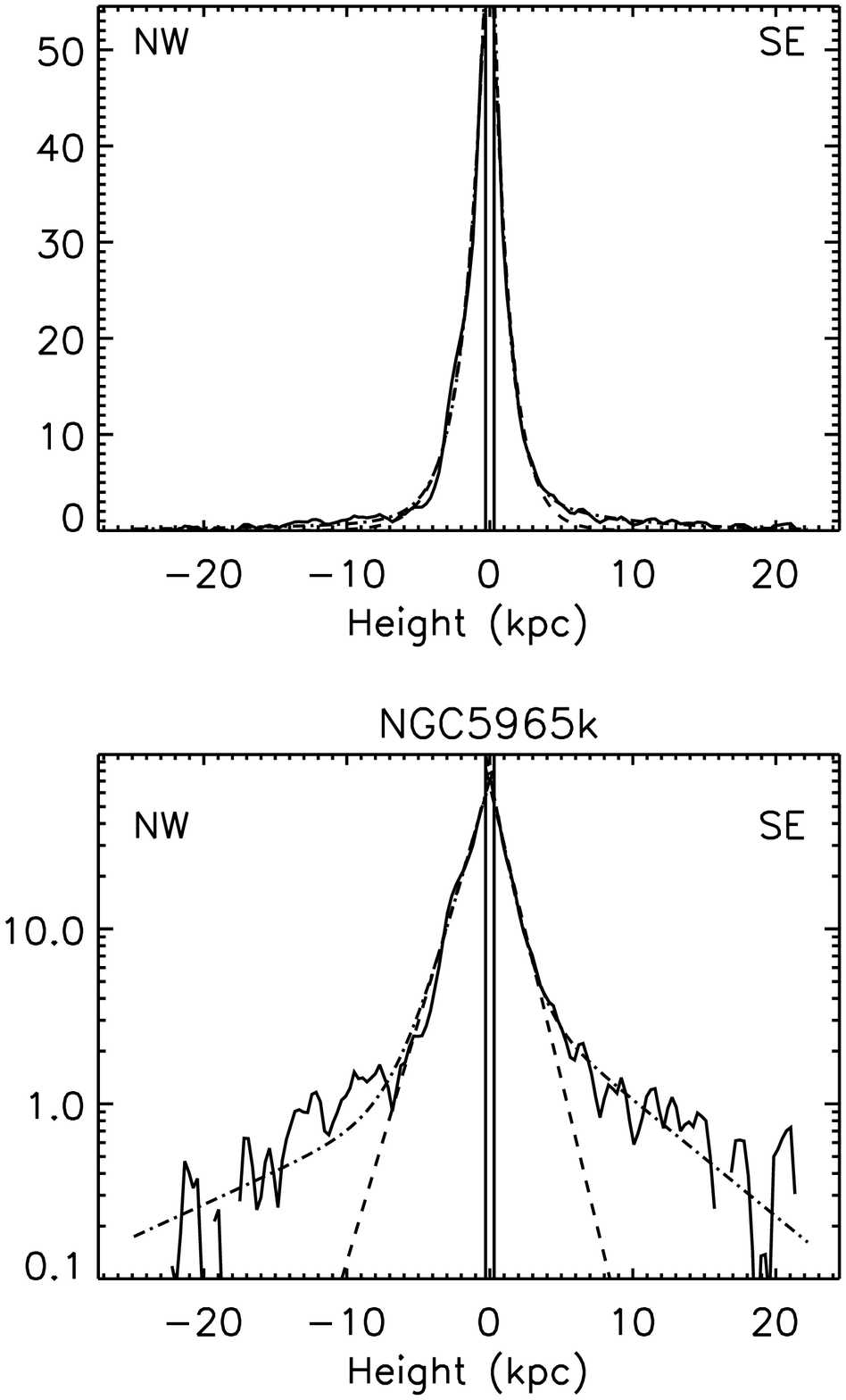}{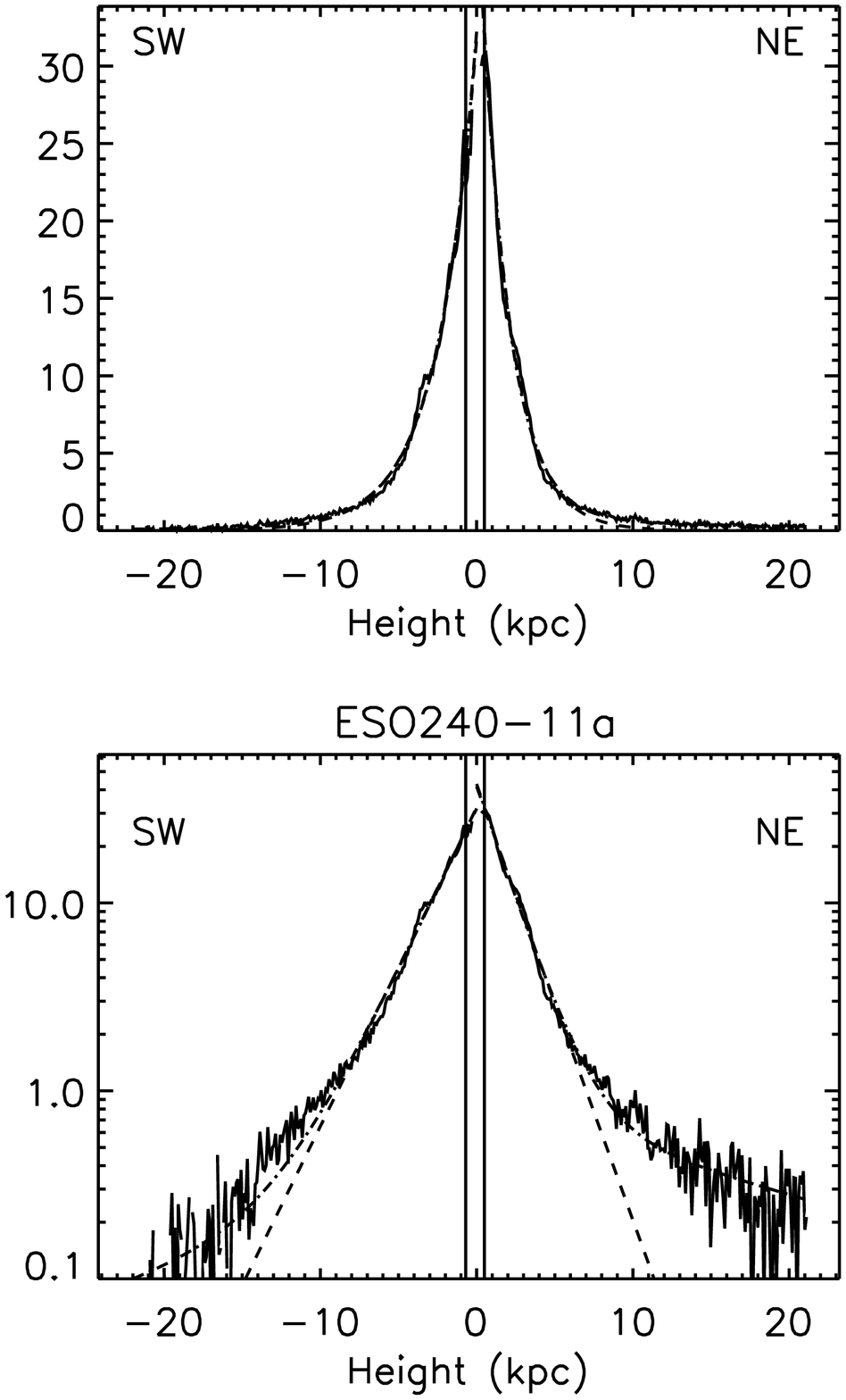}
\caption{(Cont'd)}
\end{figure}

\clearpage

\epsscale{1.0}
\begin{figure}[htbp]
\figurenum{20ab}
\plottwo{f20a.eps}{f20b.eps}
\caption{}
\end{figure}

\begin{figure}[htbp]
\figurenum{20cd}
\plottwo{f20c.eps}{f20d.eps}
\caption{(Cont'd)}
\end{figure}

\begin{figure}[htbp]
\figurenum{20ef}
\plottwo{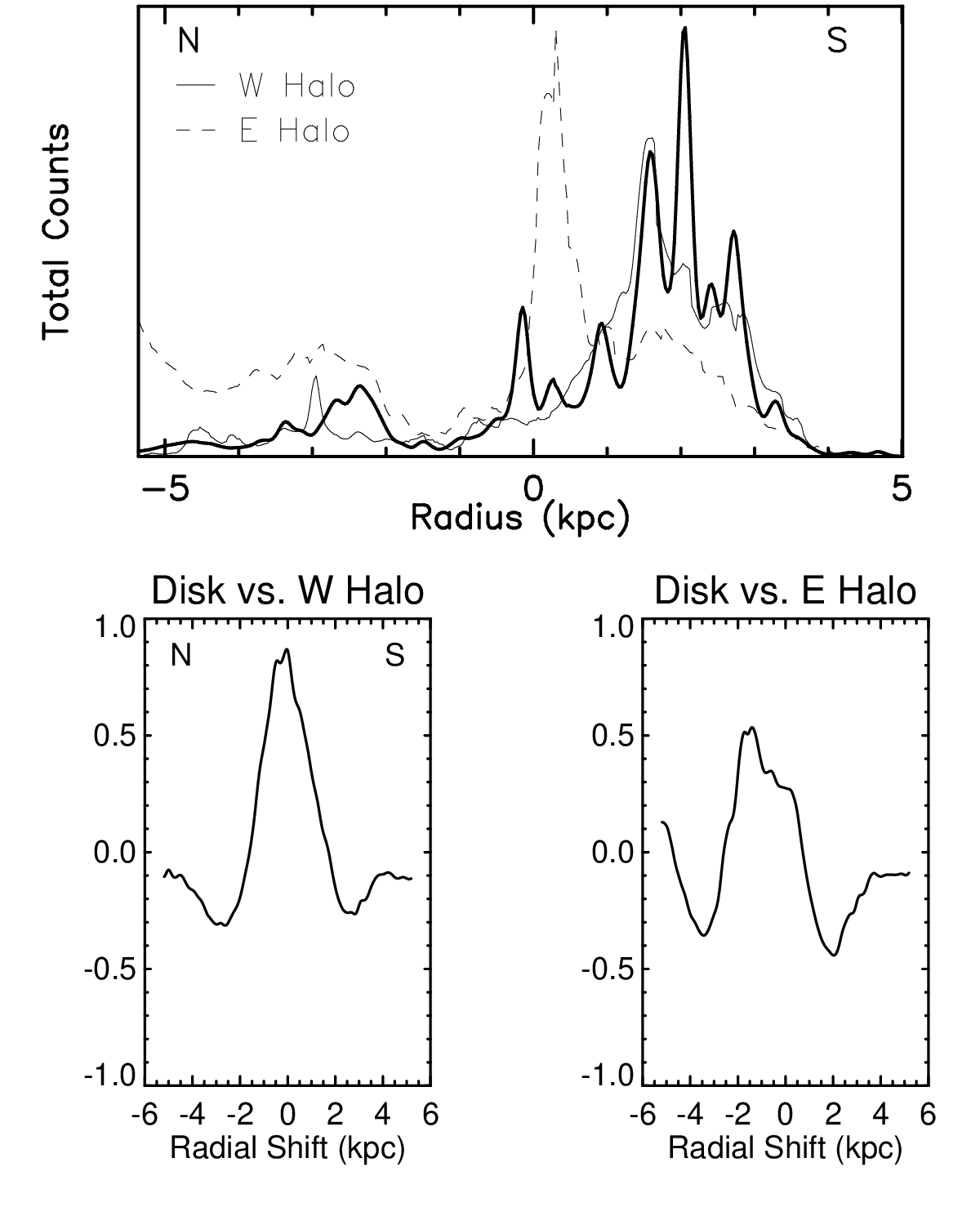}{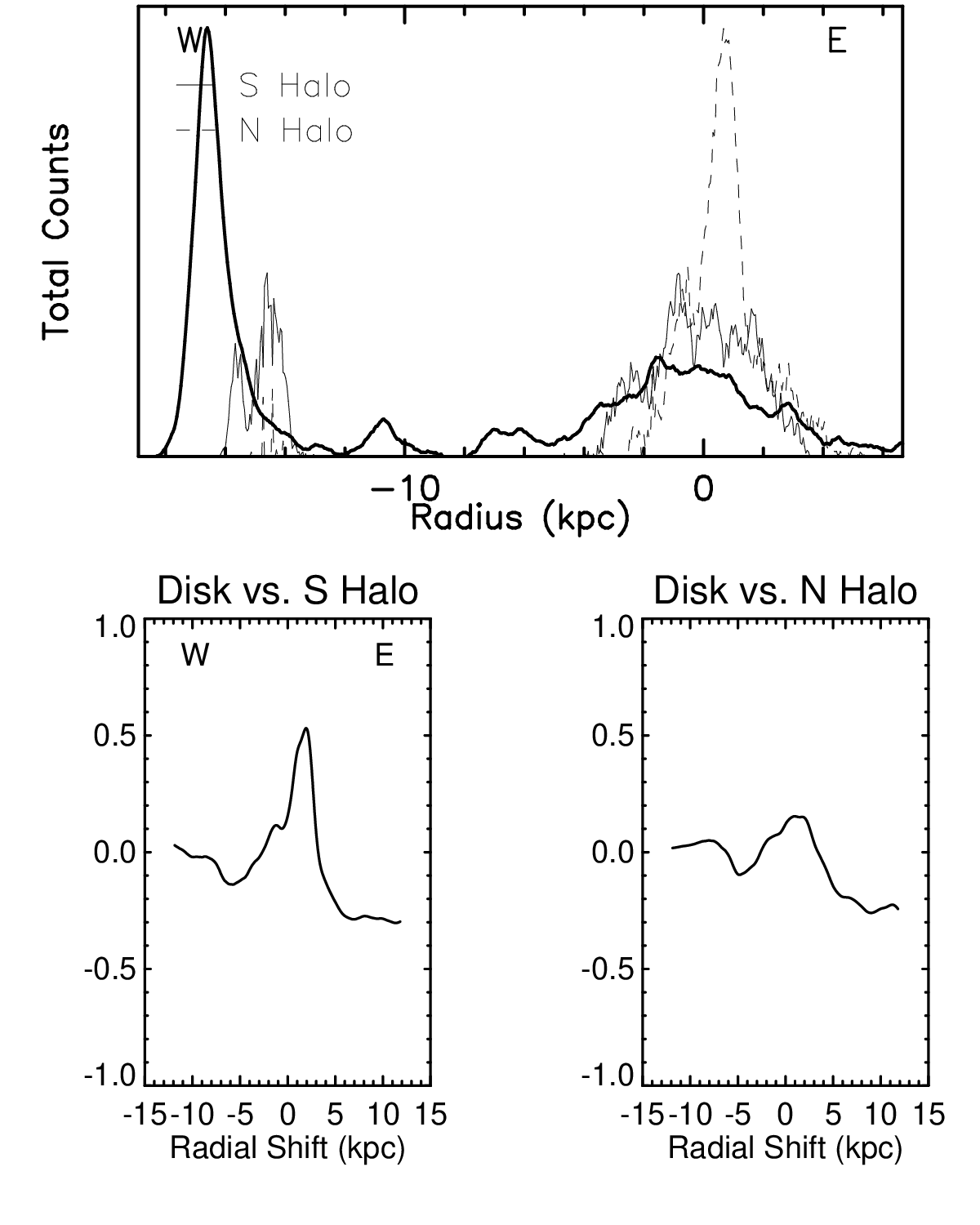}
\caption{(Cont'd)}
\end{figure}

\begin{figure}[htbp]
\figurenum{20gh}
\plottwo{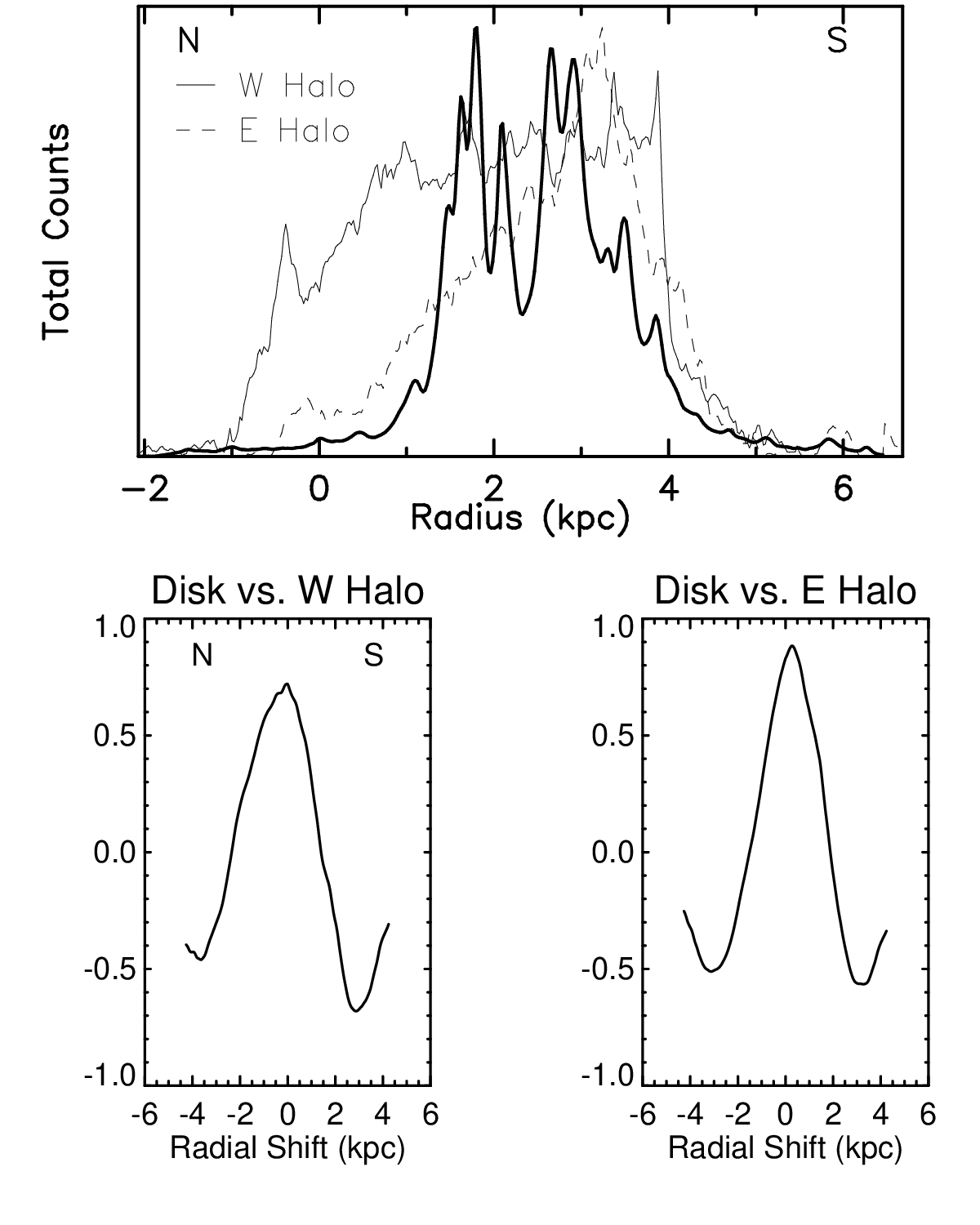}{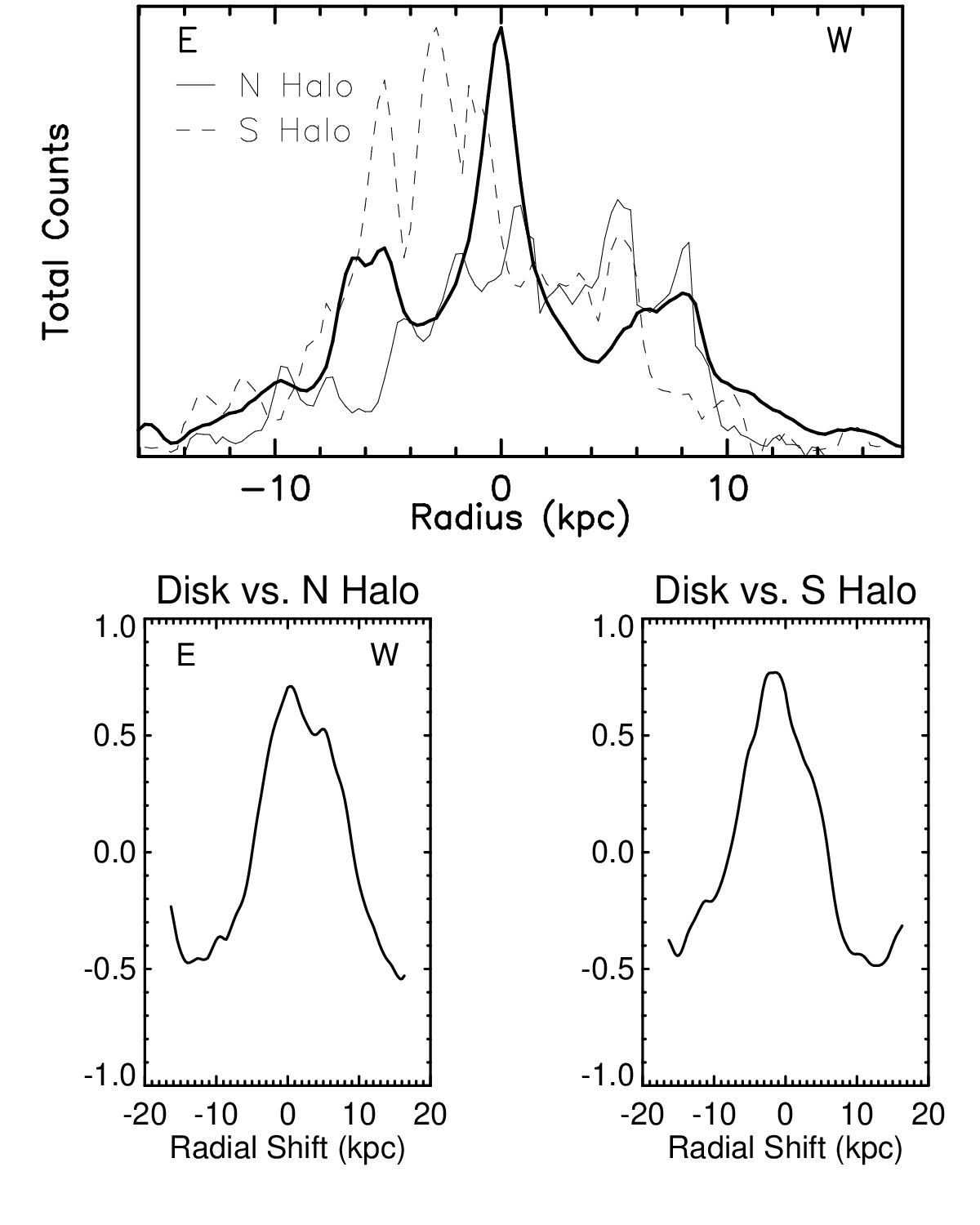}
\caption{(Cont'd)}
\end{figure}

\begin{figure}[htbp]
\figurenum{20ij}
\plottwo{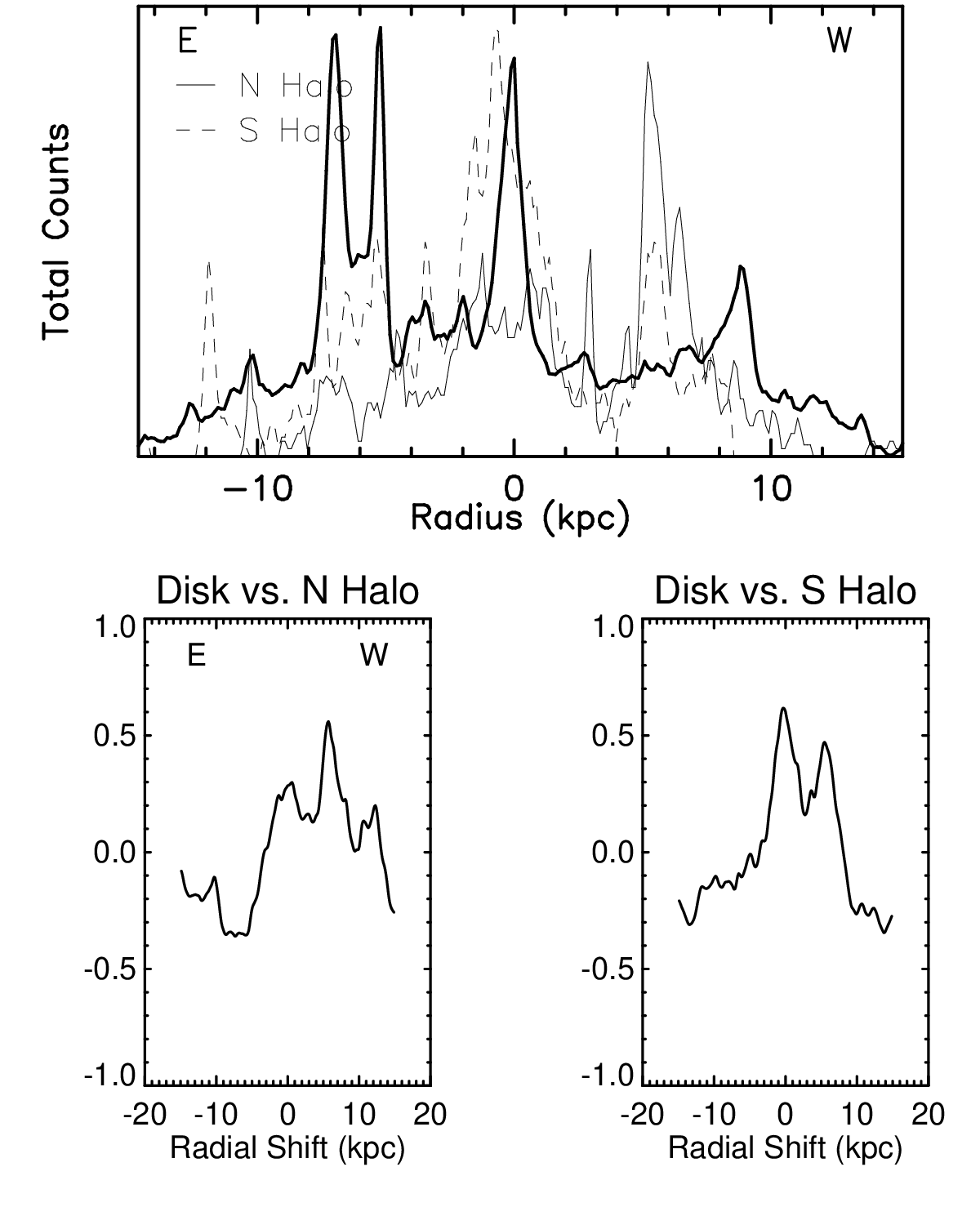}{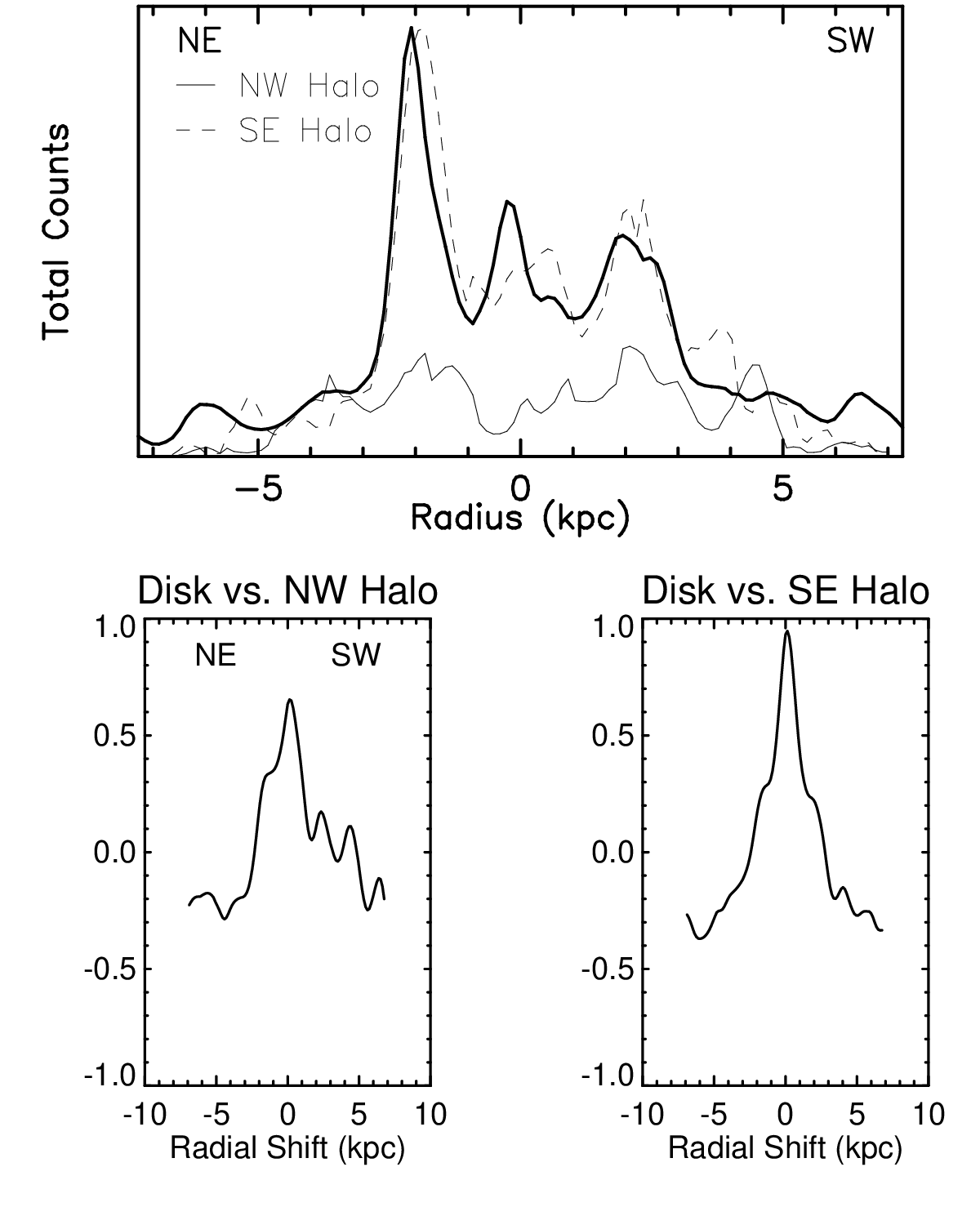}
\caption{(Cont'd)}
\end{figure}

\begin{figure}[htbp]
\figurenum{20kl}
\plottwo{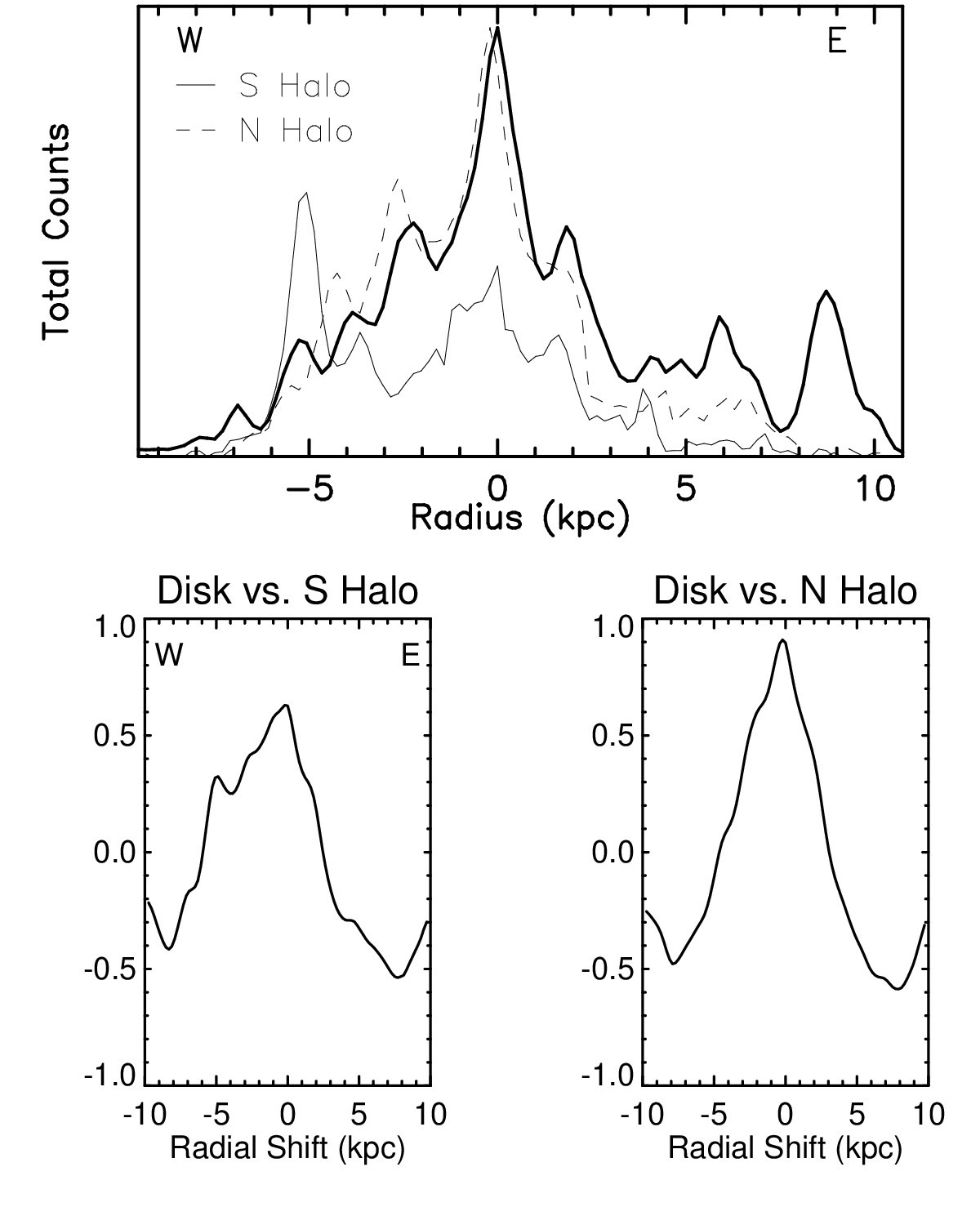}{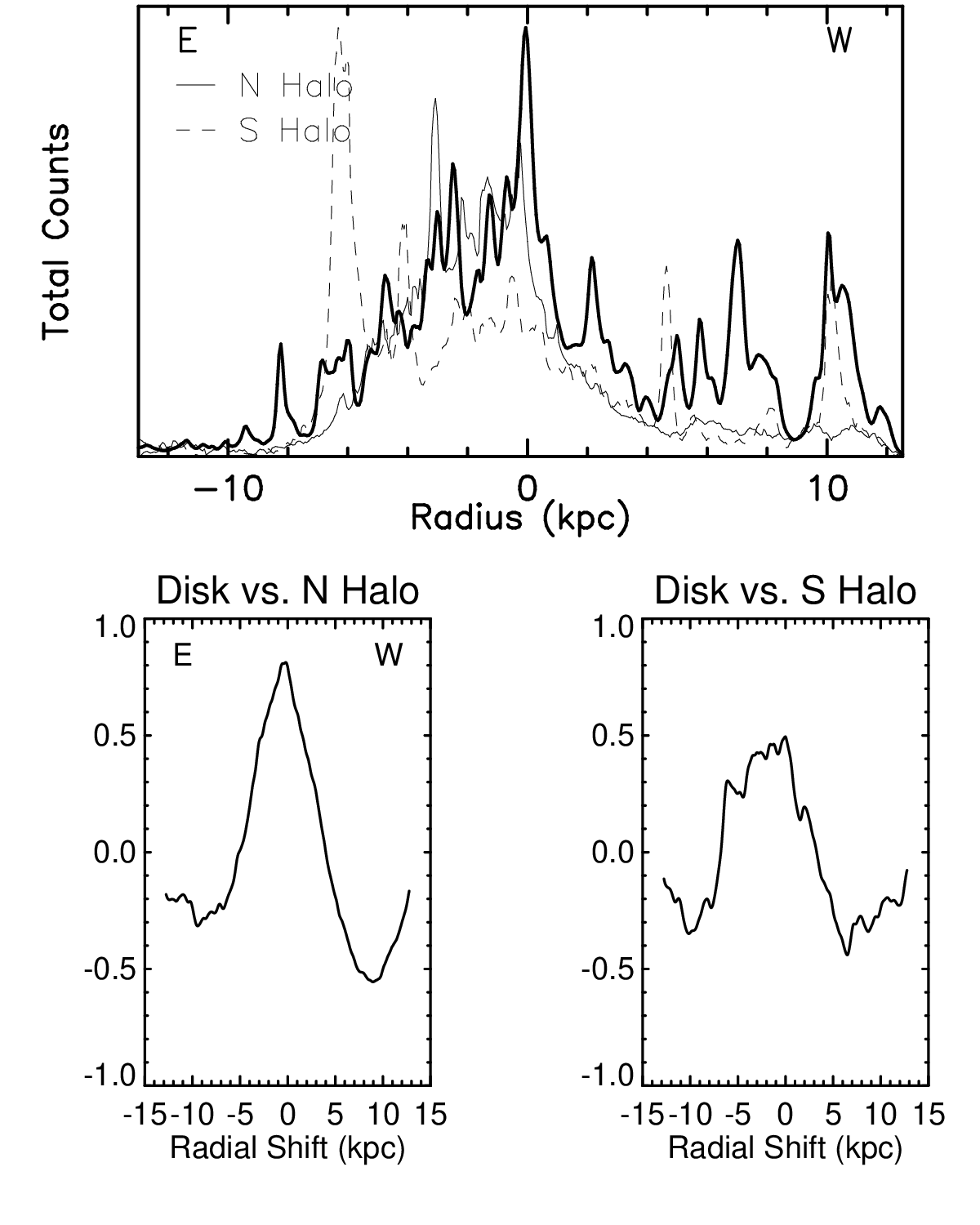}
\caption{(Cont'd)}
\end{figure}

\begin{figure}[htbp]
\figurenum{20mn}
\plottwo{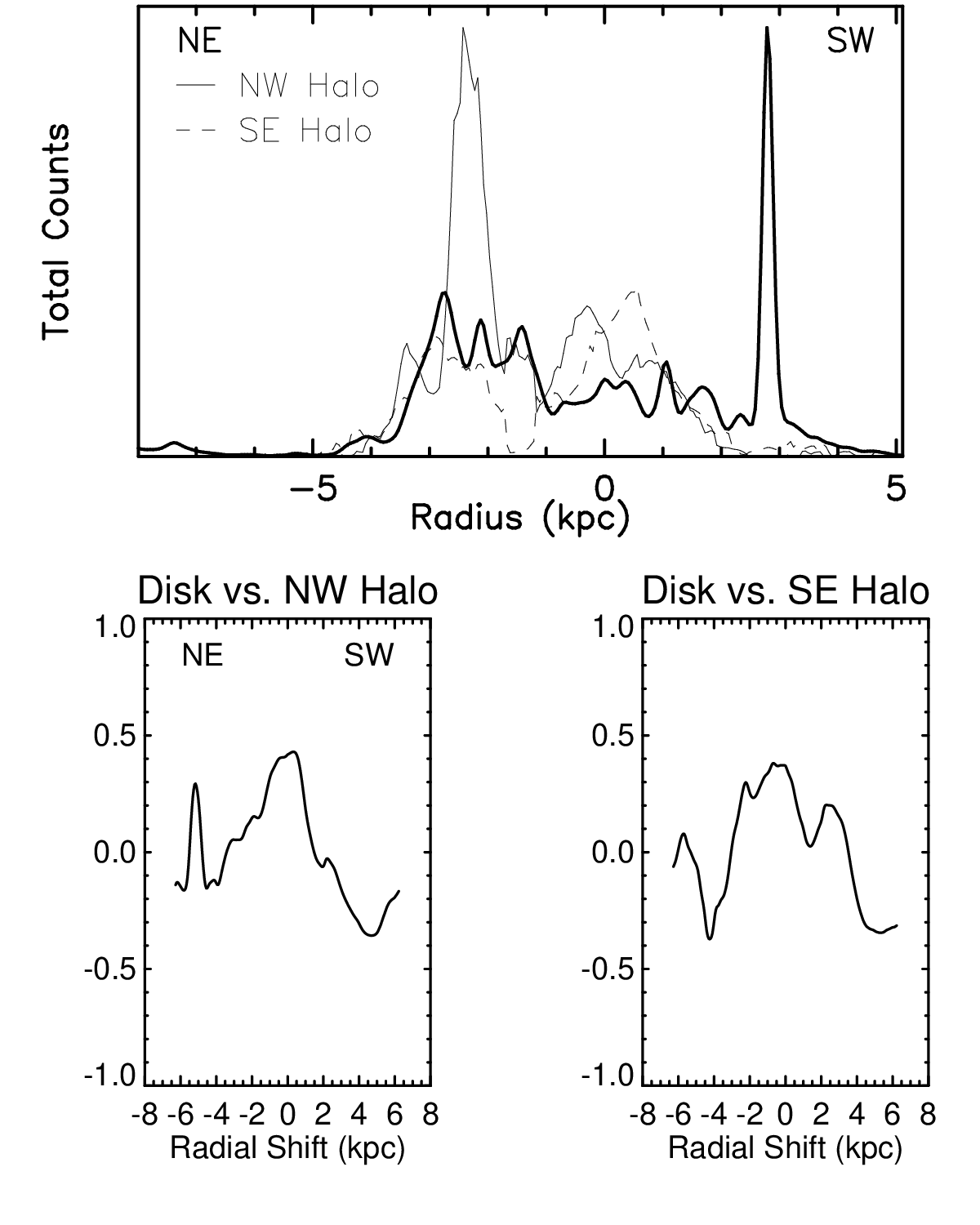}{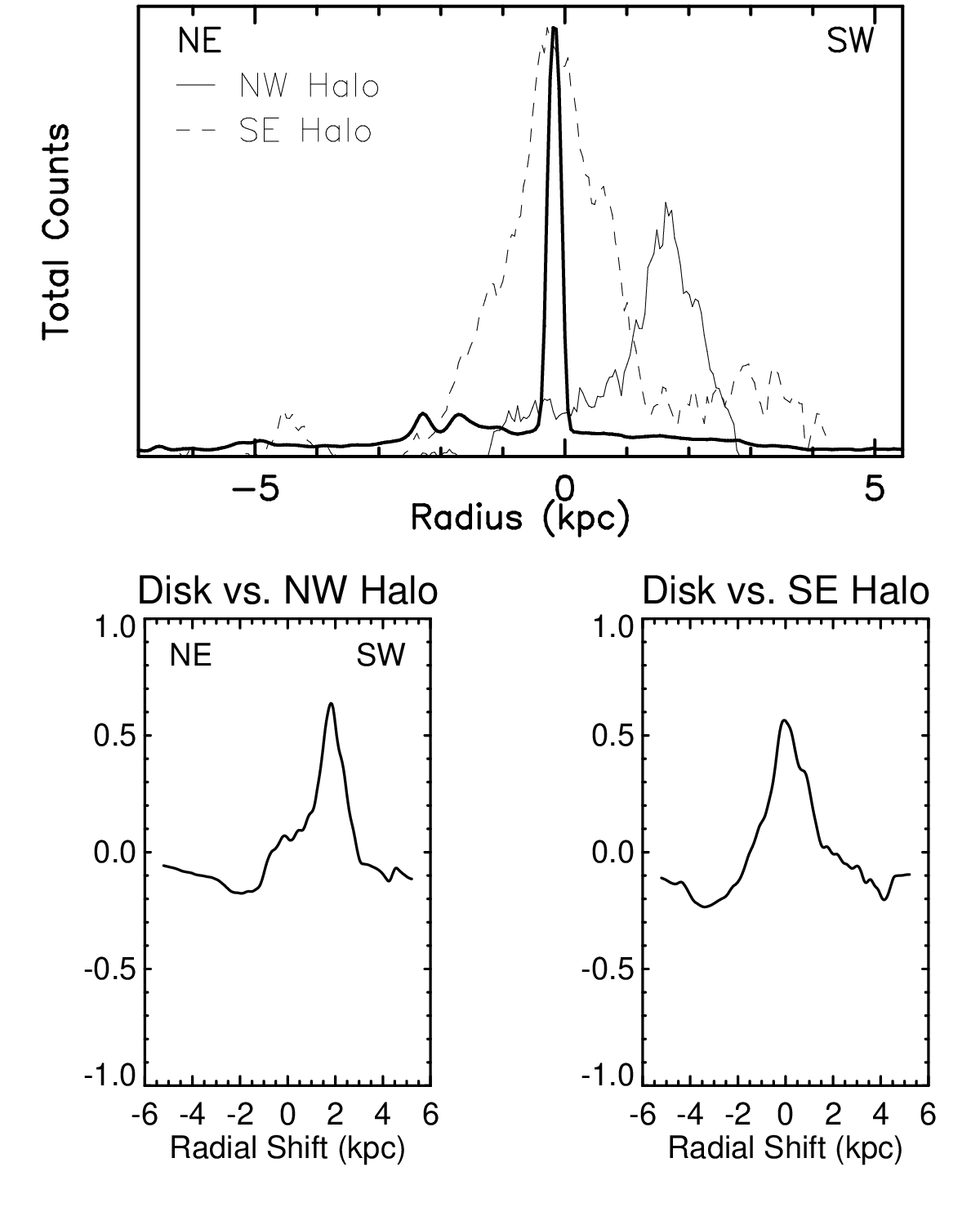}
\caption{(Cont'd)}
\end{figure}

\begin{figure}[htbp]
\figurenum{20op}
\plottwo{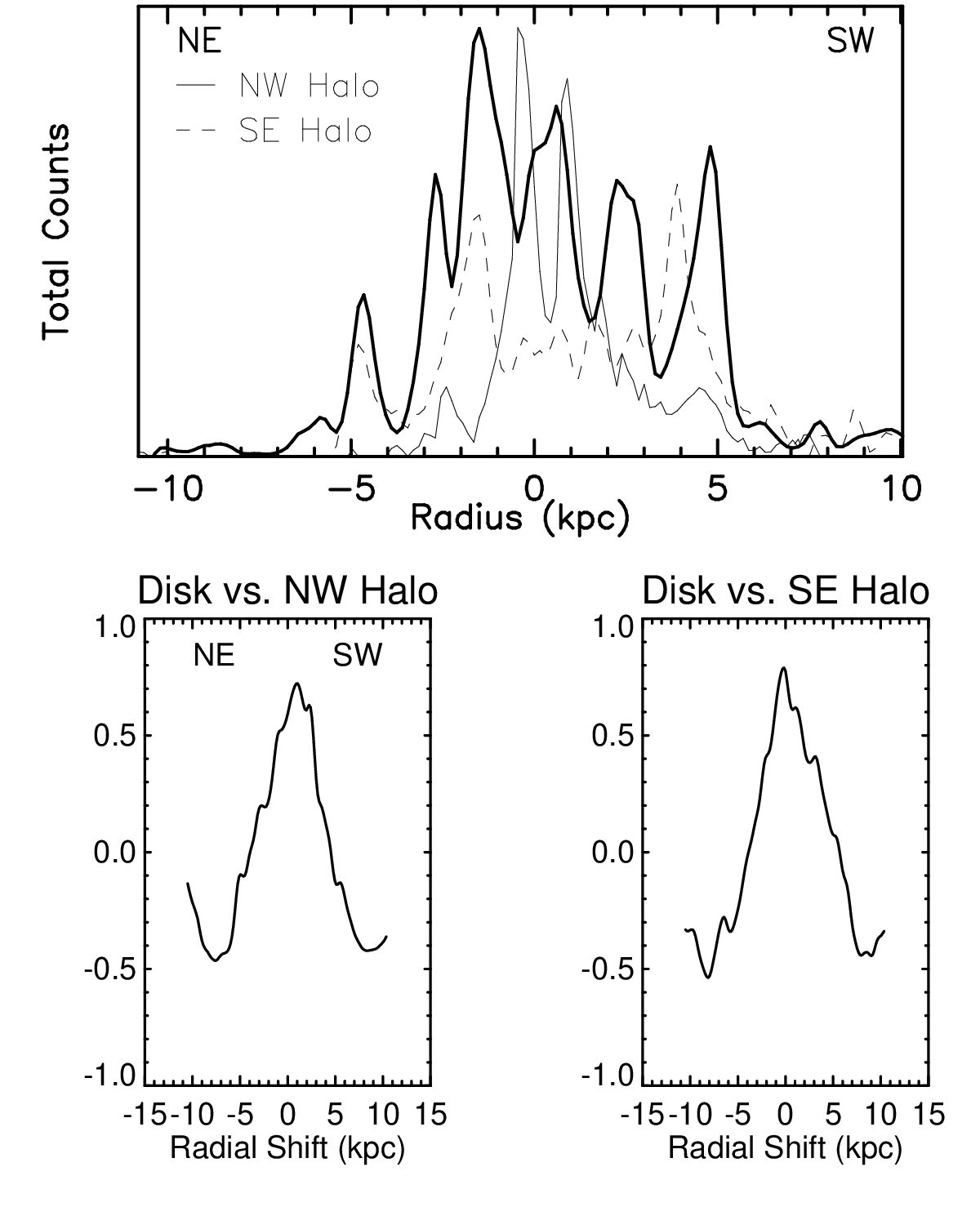}{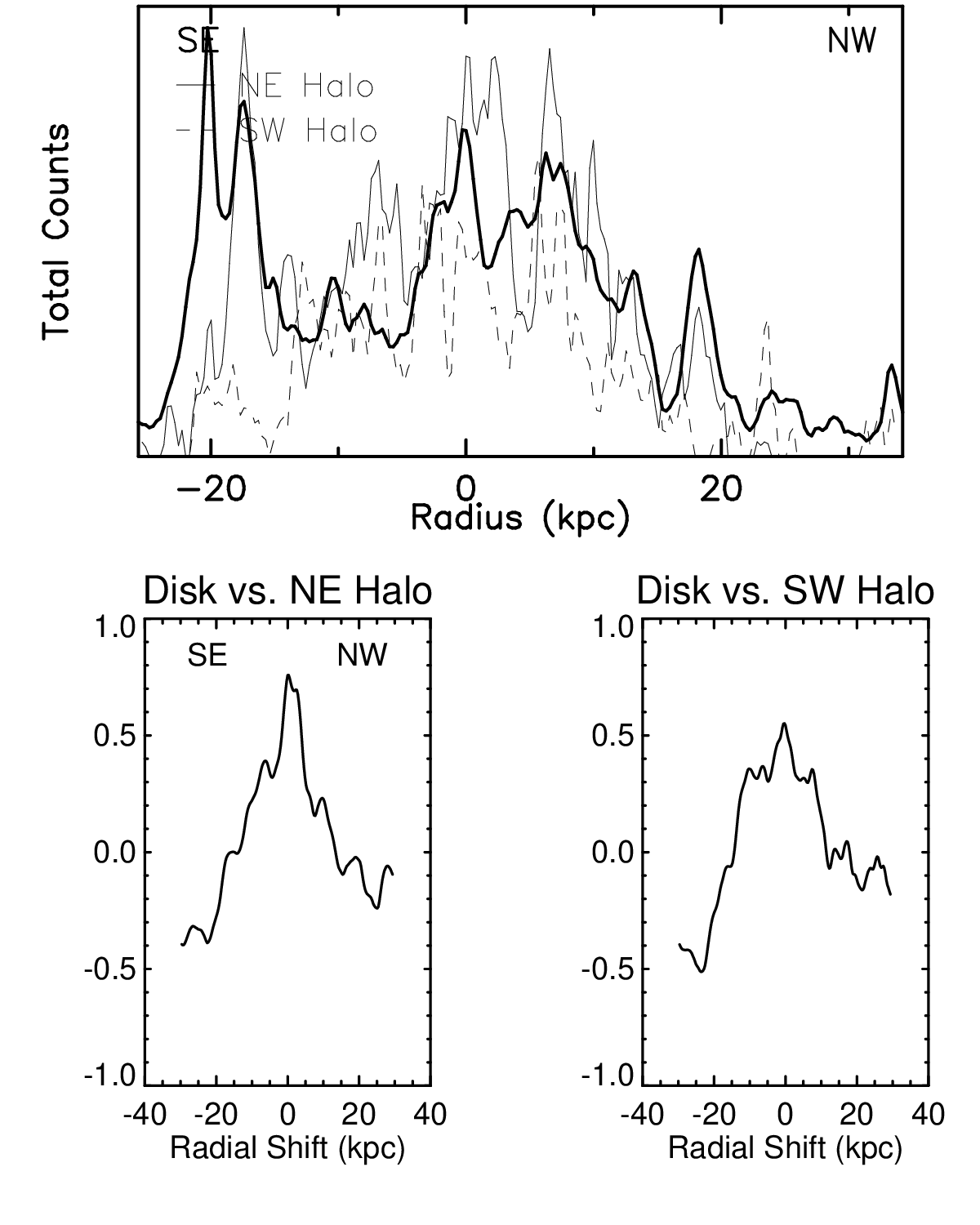}
\caption{(Cont'd)}
\end{figure}

\begin{figure}[htbp]
\figurenum{20qr}
\plottwo{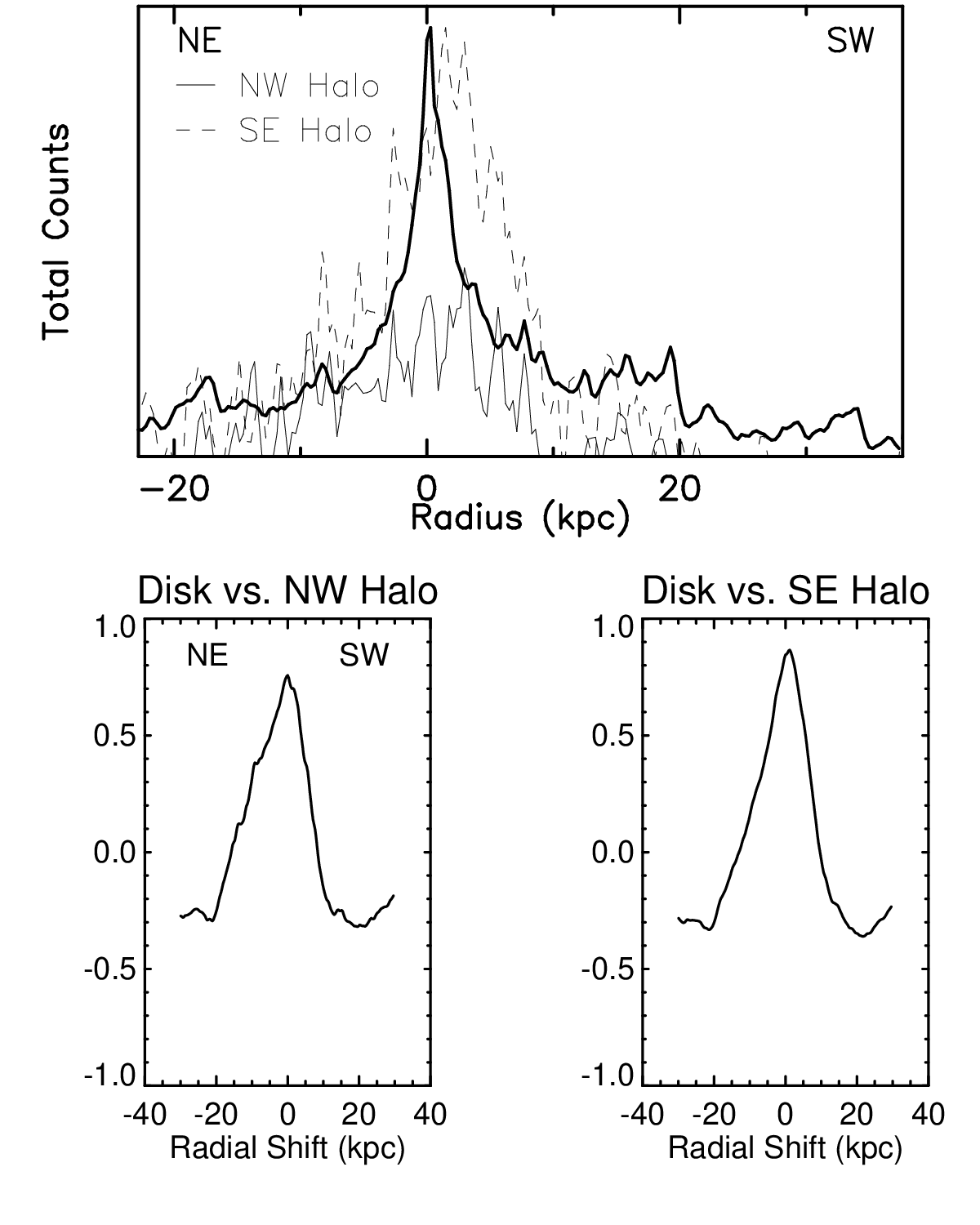}{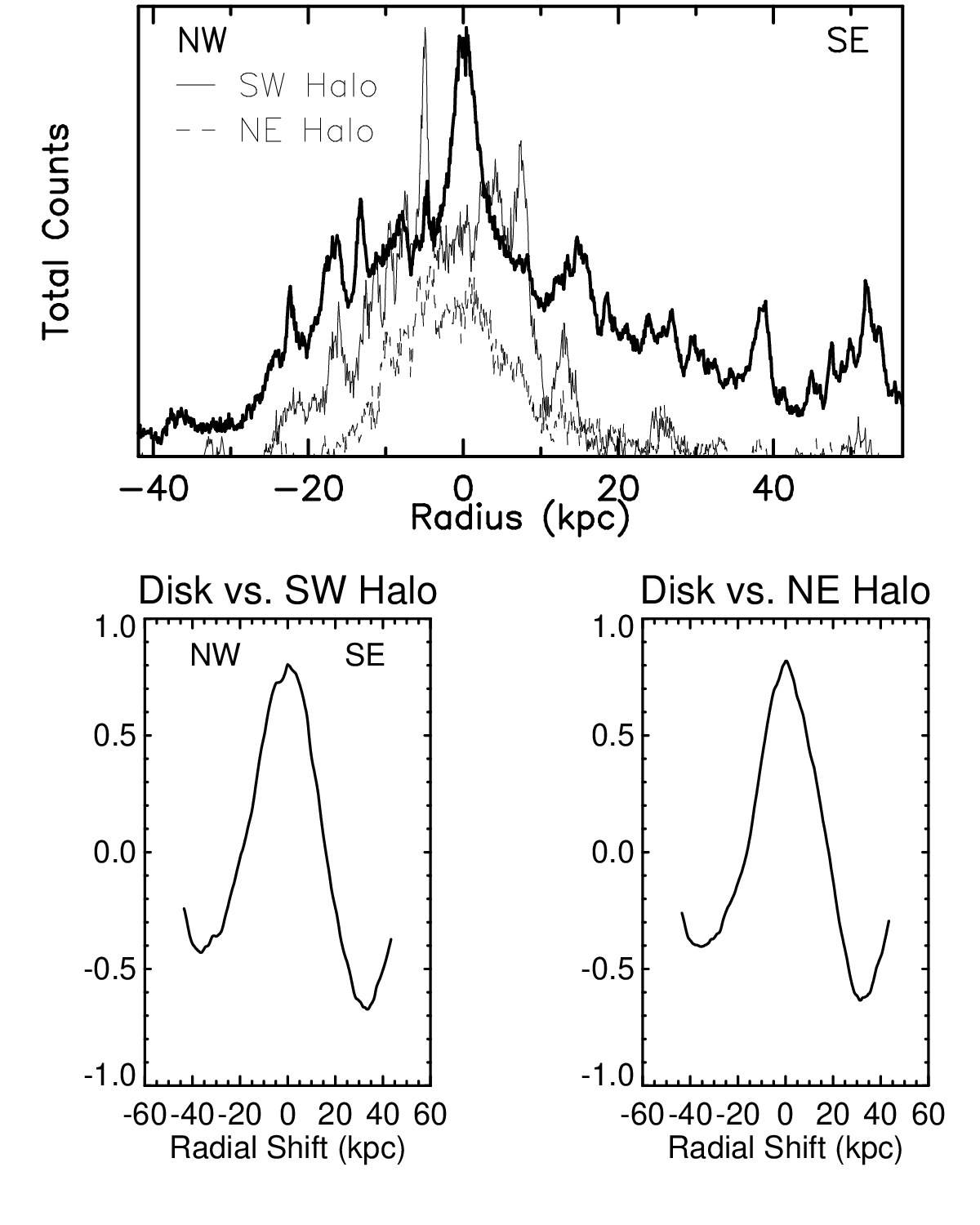}
\caption{(Cont'd)}
\end{figure}

\clearpage

\begin{table*}[hp]
\tiny
\tablenum{1}
\caption{Sample.}

\begin{tabular}{lrrrrrrrrrrrrrrr}
\tableline
\tableline
Galaxy &
R.A. (J2000)\tablenotemark{a} &
Dec (J2000)\tablenotemark{a} &
D$_{25}$\tablenotemark{a} &
Position\tablenotemark{a} &
Galaxy\tablenotemark{a} & 			
Distance\tablenotemark{b} &                  
Incl.\tablenotemark{a} &
L$_{\rm FIR}$/D$_{25}^2$\tablenotemark{c} &
L$_{\rm IR}$/D$_{25}^2$\tablenotemark{c} &
$\frac{f_{25}}{f_{60}}$\tablenotemark{c} &
$\frac{f_{60}}{f_{100}}$\tablenotemark{c} \\
 &
hh~mm~ss &
dd~mm~ss &
(arcmin) &
Angle &                               
Type &                               
(Mpc) &                            
(deg.) &                               
\multicolumn{2}{c}{(10$^{40}$ erg s$^{-1}$ kpc$^{-2}$)} &
 & \\
\tableline
 NGC~7817    & 00~03~59.0 &  +20~45~00 &  3.55 &  45 & SAbc    & 20.6\tablenotemark{d} & 79                  & 8.40 &15.88 & 0.11 & 0.33 \\
 NGC~55      & 00~15~08.5 & --39~13~13 & 32.36 & 108 &  SBm    &  1.6                  & 90                  & 0.04 & 0.09 & 0.14 & 0.22 \\
 NGC~891     & 02~22~33.1 &  +42~20~48 & 13.49 &  22 &  Sb     &  9.6                  & 89\tablenotemark{e} & 4.76 & 7.27 & 0.02 & 0.23 \\
 NGC~973     & 02~34~20.2 &  +32~30~20 &  3.72 &  48 &  Sbc    & 63\tablenotemark{d}   & 84\tablenotemark{e} & 2.13 & 4.92 & 0.18 & 0.45 \\
 NGC~1507    & 04~04~27.3 & --02~11~17 &  3.63 &  11 & SB(s)m  & 10.6                  & 80\tablenotemark{e} & 2.12 & 4.88 & 0.17 & 0.39 \\
 ESO~362-11  & 05~16~39.2 & --37~06~00 &  4.47 &  76 &  Sbc    & 15.4                  & 90                  & 2.92 & 5.61 & 0.12 & 0.33 \\
 NGC~2188    & 06~10~09.6 & --34~06~22 &  4.37 & 175 &  SB(s)m &  7.9                  & 86                  & 1.88 & 3.97 & 0.12 & 0.43 \\
 NGC~2424    & 07~40~39.8 &  +39~13~59 &  3.80 &  81 &  SB(r)b & 44.1\tablenotemark{d} & 82\tablenotemark{e} & 1.11 & 3.30 & 0.36 & 0.29 \\
 ESO~209-9   & 07~58~14.9 & --49~51~06 &  6.17 & 152 & SB(s)cd & 12.8                  & 90                  & 4.60 & 7.50 & 0.05 & 0.30 \\
 NGC~2820    & 09~21~47.1 &  +64~15~29 &  4.07 &  59 & SB(s)c  & 20.0\tablenotemark{d} & 90                  & 4.09 & 7.30 & 0.07 & 0.34 \\
 NGC~3044    & 09~53~39.8 &  +01~34~46 &  4.90 &  13 & SB(s)c  & 20.6                  & 90                  & 6.75 &12.32 & 0.12 & 0.46 \\
 NGC~3432    & 10~52~31.3 &  +36~37~08 &  6.76 &  38 &  SB(s)m &  7.8                  & 84\tablenotemark{e} & 1.92 & 3.27 & 0.06 & 0.33 \\
 NGC~4013    & 11~58~31.7 &  +43~56~48 &  5.25 &  66 &  Sb     & 17.0                  & 90\tablenotemark{e} & 4.60 & 7.46 & 0.05 & 0.24 \\
 NGC~4197    & 12~14~38.3 &  +05~48~21 &  3.39 &  36 &  Sc     & 23.1\tablenotemark{d} & 79\tablenotemark{e} & 3.04 & 6.66 & 0.18 & 0.37 \\
 NGC~5529    & 14~15~34.1 &  +36~13~36 &  6.17 & 115 &  Sc     & 43.9                  & 90                  & 1.17 & 2.74 & 0.15 & 0.24 \\
 NGC~5965    & 15~34~02.1 &  +56~41~10 &  5.25 &  53 &  Sb     & 45.6\tablenotemark{d} & 85\tablenotemark{e} & 0.44 & 1.52 & 0.46 & 0.28 \\
 ESO~240-11  & 23~37~49.4 & --47~43~35 &  5.25 & 129 &  Sb     & 34.9                  & 90                  & 1.00 & 2.32 & 0.27 & 0.18 \\
\tableline
\end{tabular}

\tablenotetext{a}{Values taken from de Vaucouleurs et al. (1991).}

\tablenotetext{b}{Values taken from Tully (1988) unless otherwise noted.}

\tablenotetext{c}{L$_{\rm IR}$ and L$_{\rm FIR}$ are calculated using
the prescriptions listed in Sanders \& Mirabel (1996).  A value of C =
1 was used in the calculation of L$_{\rm FIR}$. The $IRAS$ fluxes used
for the calculations are from Moshir et al. (1995). }

\tablenotetext{d}{References for distances: NGC~7817 (Tully, Shaya, \&
Pierce 1992), NGC~973 (Gourgoulhon, Chamaraux, \& Fouqu\'e 1992),
NGC~2424 (Vallee~1991), NGC~2820 (Hummel \& van der Hulst 1989),
NGC~4197 (Guthrie~1992), NGC~5965 (Vallee~1991). The distances for
NGC~973, NGC~2424, and NGC~5965 assume $H_0$ = 75 km s$^{-1}$
Mpc$^{-1}$, while the others are based on the Tully-Fisher relation.}

\tablenotetext{e}{References for inclinations: NGC~891 (Swaters~1994),
NGC~973 (Guthrie~1992), NGC~1507 (Aaronson et al.~1982),
NGC~2424 (Guthrie~1992), NGC~3432 (Aaronson et al.~1982),
NGC~4013 (Bottema~1995), NGC~4197 (Yasuda et al.~1997), NGC~5965 
(Guthrie~1992).}

\normalsize
\end{table*}

\vskip 0.25in

\begin{table*}[h]
\tiny
\tablenum{2}
\caption{Observing Logs for Conventional Narrowband Imaging.}

\begin{tabular}{lrrrrrrcr}
\tableline
\tableline
Galaxy   & Date of & Telescope\tablenotemark{a} & Filter & Exposure & \# of & Resolution\tablenotemark{b} & FOV & Flux Limit \\
  & Observation &   & CWL/FWHM(\AA) & Time  & Exp. & ($\arcsec$ pixel$^{-1}$) & & (erg s$^{-1}$ cm$^{-2}$ arcsec$^{-2}$) \\
\tableline
NGC~55   & Dec. 1997 & CTIO & 6567/68 & 4.0 hours & 20 & 2.03 & 69$\farcm$0 $\times$ 69$\farcm$0 & 11 $\times$ 10$^{-18}$ \\
NGC~973  & Dec. 1997 & KPNO & 6653/68 & 4.2 hours & 26 & 1.34 & 22$\farcm$9 $\times$ 22$\farcm$9 & 9 $\times$ 10$^{-18}$ \\
NGC~2424 & Dec. 1997 & KPNO & 6618/74 & 3.8 hours & 23 & 1.34 & 22$\farcm$9 $\times$ 22$\farcm$9 & 6 $\times$ 10$^{-18}$ \\
ESO~209-9& Dec. 1997 & CTIO & 6567/68 & 4.1 hours & 17 & 2.03 & 69$\farcm$0 $\times$ 69$\farcm$0 & 18 $\times$ 10$^{-18}$ \\
NGC~2820 & Dec. 1997 & KPNO & 6573/67 & 4.4 hours & 27 & 1.34 & 22$\farcm$9 $\times$ 22$\farcm$9 & 5 $\times$ 10$^{-18}$ \\
NGC~3044 & Dec. 1997 & CTIO & 6600/67 & 2.8 hours & 15 & 2.03 & 69$\farcm$0 $\times$ 69$\farcm$0 & 13 $\times$ 10$^{-18}$ \\
NGC~3432 & Dec. 1997 & KPNO & 6573/67 & 3.7 hours & 22 & 1.34 & 22$\farcm$9 $\times$ 22$\farcm$9 & 8 $\times$ 10$^{-18}$ \\
NGC~4197 & Apr. 1998 & KPNO & 6618/74 & 4.2 hours & 24 & 1.34 & 22$\farcm$9 $\times$ 22$\farcm$9 & 7 $\times$ 10$^{-18}$ \\
NGC~5529 & Apr. 1998 & KPNO & 6618/74 & 3.8 hours & 23 & 1.34 & 22$\farcm$9 $\times$ 22$\farcm$9 & 8 $\times$ 10$^{-18}$ \\
NGC~5965 & Apr. 1998 & KPNO & 6618/74 & 3.8 hours & 23 & 1.34 & 22$\farcm$9 $\times$ 22$\farcm$9 & 10 $\times$ 10$^{-18}$ \\
\tableline

\end{tabular}

\tablenotetext{a}{The KPNO telescope used for these observations was
the 0.9-meter, while the 0.9-meter Curtis Schmidt telescope was used
at Cerro Telolo.}

\tablenotetext{b}{After binning. }

\normalsize
\end{table*}

\vskip 0.25in

\begin{table*}[h]
\tiny
\tablenum{3}
\caption{Observing Logs for TTF Imaging.}

\begin{tabular}{lrrrrrrrcr}

\tableline
\tableline
Galaxy   & Date of & Telescope & Filter & Bandpass & Exposure & \# of & Resolution & FOV & Flux Limit \\
  & Observation &   & CWL/FWHM(\AA) & FWHM(\AA) & Time  & Exp. & ($\arcsec$ pixel$^{-1}$) &  & (erg s$^{-1}$ cm$^{-2}$ arcsec$^{-2}$) \\
\tableline
NGC~7817  & Jan. 1998 & WHT & 6680/280 & 14.9 & 1.0 hours & 2 & 0.58  & 7$\farcm$7 $\times$ 7$\farcm$7 & 5 $\times$ 10$^{-18}$ \\
NGC~55    & Dec. 1997 & AAT & 6680/210 & 13.9 & 1.5 hours & 3 & 0.315 & 8$\farcm$8 $\times$ 7$\farcm$2 & 2 $\times$ 10$^{-18}$ \\
NGC~891   & Jan. 1998 & WHT & 6680/280 & 15.7 & 0.8 hours & 3 & 0.58  & 7$\farcm$7 $\times$ 7$\farcm$7 & 5 $\times$ 10$^{-18}$ \\
NGC~1507  & Jan. 1998 & WHT & 6680/280 & 16.1 & 1.5 hours & 3 & 0.58  & 7$\farcm$7 $\times$ 7$\farcm$7 & 5 $\times$ 10$^{-18}$ \\
ESO~362-11& Dec. 1997 & AAT & 6680/210 & 13.5 & 1.5 hours & 3 & 0.594 & 16$\farcm$5 $\times$ 13$\farcm$5 & 4 $\times$ 10$^{-18}$ \\
NGC~2188  & Dec. 1997 & AAT & 6680/210 & 13.9 & 1.3 hours & 3 & 0.315 & 8$\farcm$8 $\times$ 7$\farcm$2 & 3 $\times$ 10$^{-18}$ \\
NGC~2424  & Jan. 1998 & WHT & 6680/280 & 14.9 & 1.0 hours & 2 & 0.58  & 7$\farcm$7 $\times$ 7$\farcm$7 & 5 $\times$ 10$^{-18}$ \\
NGC~3044  & Jan. 1998 & WHT & 6680/280 & 15.7 & 1.0 hours & 3 & 0.58  & 7$\farcm$7 $\times$ 7$\farcm$7 & 5 $\times$ 10$^{-18}$ \\
NGC~4013  & Jan. 1998 & WHT & 6680/280 & 16.1 & 1.0 hours & 2 & 0.58  & 7$\farcm$7 $\times$ 7$\farcm$7 & 7 $\times$ 10$^{-18}$ \\
ESO~240-11& Dec. 1997 & AAT & 6680/210 & 13.5 & 1.5 hours & 3 & 0.594 & 16$\farcm$5 $\times$ 13$\farcm$5 & 5 $\times$ 10$^{-18}$ \\
\tableline

\end{tabular}

\normalsize
\end{table*}

\vskip 0.25in

\begin{table*}[h]
\tiny
\tablenum{4}
\caption{Calculated Properties of the Galaxies}
\begin{tabular}{lrrrrr}
\tableline
\tableline
Galaxy &
Telescope &
L$_{H\alpha}$(total)\tablenotemark{a} &              
L$_{H\alpha}$(total)/D$_{25}^2$ &      
L$_{H\alpha}$(eDIG)\tablenotemark{a}  &         
L$_{H\alpha}$(eDIG)/L$_{H\alpha}$(total) \\
 &                               
 &
(10$^{39}$ erg s$^{-1}$) &                      
(10$^{37}$ erg s$^{-1}$ kpc$^{-2}$) &                
(10$^{39}$ erg s$^{-1}$) &                     
(\%) \\
\tableline
NGC~7817  &  WHT &22.4 &  5.0 &  2.4  & 11 \\
NGC~55    & CTIO &   8.9 &  3.9 &  0.17 &  5 \\
NGC~55\tablenotemark{b} & AAT & -- & -- & -- & -- \\
NGC~891\tablenotemark{c}& WHT  &  15.1 &  1.1 & -- & -- \\
NGC~973\tablenotemark{d}   & KPNO &  39.6  &  0.8 &  6.3  & 16 \\
NGC~1507  &  WHT &49.8  & 39.8 &  5.4  & 11 \\
ESO~362-11&  AAT &   3.4 &  0.8 &  0.12 &  3 \\
NGC~2188\tablenotemark{b} &  AAT & 9.9 &  9.8 &  1.3  & 13 \\
NGC~2424  & KPNO &  30.3  &  1.3 &  4.1  & 13 \\
NGC~2424  &  WHT &33.2  &  1.4 &  4.5  & 14 \\
NGC~2820  & KPNO & 11.3 &  2.0 &  1.8  & 16 \\
NGC~3044  & CTIO & 89.4  & 10.4 & 14.3  & 16 \\
NGC~3044  &  WHT &83.2 &  9.6 & 13.3   & 16 \\
NGC~3432  & KPNO & 20.3 &  8.6 &  1.3  &  6 \\
NGC~4013  &  WHT &21.9 &  3.2 &  1.4  &  6 \\
NGC~4197  & KPNO & 24.7 &  4.8 &  3.2  & 13 \\
NGC~5529  & KPNO & 48.3  &  0.8 &  7.6  & 16 \\
NGC~5965  & KPNO & 47.3 &  1.0 &  7.7  & 16 \\
ESO~240-11&  AAT &62.6 &  2.2 &  9.5  & 15 \\
\tableline
\end{tabular}
\tablenotetext{a}{Typical error bars on the total luminosities are $\sim$
10\% but can reach 25\% due to uncertainties in the continuum
subtraction. The corresponding uncertainties on the eDIG luminosities
are $\sim$ 15\% and $\sim$ 30\%, respectively}

\tablenotetext{b}{The AAT observations of NGC~55 and NGC~2188 do not
fully encompass the galaxy, so it is not possible to determine
the global quantities from these observations.}

\tablenotetext{c}{The presence of reflective ghosts in the image of
NGC~891 prevents us from calculating some of the eDIG properties for
this object.}

\tablenotetext{d}{The presence of a bright star close to the NE side of
NGC~973 makes sky subtraction and detection of faint emission
difficult in this area.  Therefore, some of the calculated galaxy
properties for NGC~973 are lower limits, as determined by the
unaffected portion of the galaxy.}

\normalsize
\end{table*}

\clearpage

\begin{table*}[h]
\tiny
\tablenum{5}
\caption{Calculated Properties of Galaxies based on Exponential Fits}

\begin{tabular}{lrrrrr|rrrr|r|r}
\tableline
\tableline
 &
 &
\multicolumn{4}{c}{One-exponential Fit\tablenotemark{a}} &
\multicolumn{4}{c}{Two-exponential Fit} &
 \\
 &
 &
EM\tablenotemark{b} &              
Scale &      
$n_e$ &         
``Disk'' &     
EM\tablenotemark{b}  &              
Scale &      
$n_e$ &         
eDIG &
Total eDIG & 
Recombination\\  
Galaxy &
Side &
@ z = 0 &              
Height &      
@ z = 0  &         
Mass\tablenotemark{c} &     
@ z = 0 &              
Height &      
@ z = 0 &         
Mass\tablenotemark{c} &     
Mass\tablenotemark{c} &
Rate\tablenotemark{d}\\  
 &
 &
(pc cm$^{-6}$) &                      
(kpc) &                
(cm$^{-3}$) &                     
(10$^6$ M$_\odot$) &                               
(pc cm$^{-6}$) &                      
(kpc) &                
(cm$^{-3}$) &                     
(10$^6$ M$_\odot$) &                              
(10$^6$ M$_\odot$) &
(10$^6$ s$^{-1}$ cm$^{-2}$) \\
\tableline
             NGC~7817 & NW &  520.4  & 0.272 & 0.721 &  57.49&       &       &       &       &         &      \\
                      & NW &  806.4  & 0.208 & 0.897 &  54.73& 13.32 & 1.698 & 0.115 &  57.43&         & 3.57 \\
                      & SE &  378.5  & 0.341 & 0.615 &  61.51&       &       &       &       &         &      \\  
                      & SE &  470.3  & 0.282 & 0.685 &  56.67& 10.15 & 1.902 & 0.101 &  56.17& 113.6   & 3.08 \\
             NGC~55   & NE &  243.4  & 0.218 & 0.493 &  31.54&       &       &       &       &         &      \\  
                      & NE & 1455.   & 0.099 & 1.205 &  35.00& 52.43 & 0.373 & 0.229 &  25.01&         & 3.11 \\
                      & SW & 1252.   & 0.136 & 1.118 &  44.61&       &       &       &       &         &      \\  
                      & SW & 9146.   & 0.080 & 3.022 &  71.00& 33.03 & 0.451 & 0.182 &  24.00& 49.01   & 2.37 \\
             NGC~891  & NW &   93.71 & 0.864 & 0.306 &  77.28&       &       &       &       &         &      \\  
                      & NW &   43.88 & 0.258 & 0.209 &  15.83& 69.01 & 1.072 & 0.263 &  82.31&         &11.8  \\
                      & SE &   85.71 & 1.015 & 0.293 &  87.34&       &       &       &       &         &      \\  
                      & SE &   44.48 & 1.015 & 0.211 &  62.77& 41.23 & 1.015 & 0.203 &  60.45& 142.8   & 6.65 \\
             NGC~973  & NW &    5.45 & 1.513 & 0.074 &  32.74&       &       &       &       &         &      \\  
                      & NW &    4.62 & 1.374 & 0.068 &  27.39&  0.71 & 5.047 & 0.027 &  39.40&         & 0.585\\
                      & SE &    5.02 & 1.357 & 0.071 &  28.21&       &       &       &       &         &      \\  
       &SE\tablenotemark{e}&    4.67 & 1.292 & 0.068 &  25.89&  --   &  --   &  --   &  --   & 39.40   & --   \\
             NGC~1507 & W  &   57.80 & 0.161 & 0.240 &  11.37&       &       &       &       &         &      \\  
                      & W  &   16.08 & 0.090 & 0.127 &   3.35& 36.97 & 0.563 & 0.192 &  31.73&         & 3.30 \\
                      & E  &   97.96 & 0.213 & 0.313 &  19.54&       &       &       &       &         &      \\  
                      & E  &   50.11 & 0.127 & 0.224 &   8.36& 36.79 & 0.732 & 0.192 &  41.16& 72.89   & 4.29 \\
           ESO~362-11 & S  &    8.69 & 1.645 & 0.093 &  44.99&       &       &       &       &         &      \\  
                      & S  &   20.31 & 0.693 & 0.142 &  28.97&  1.35 & 6.802 & 0.037 &  73.52&         & 1.48 \\
                      & N  &   10.35 & 1.608 & 0.102 &  47.94&       &       &       &       &         &      \\  
                      & N  &   10.85 & 1.115 & 0.104 &  34.05&  1.96 & 3.267 & 0.044 &  42.41& 115.9   & 1.01 \\
NGC~2188\tablenotemark{f}&W&  154.3  & 0.312 & 0.392 &  35.94&       &       &       &       &         &      \\  
                      & W  &  152.7  & 0.194 & 0.390 &  22.24& 29.97 & 0.884 & 0.173 &  44.86&         & 4.20 \\
                      & E  &  158.5  & 0.331 & 0.398 &  38.64&       &       &       &       &         &      \\  
                      & E  &  161.4  & 0.163 & 0.401 &  19.23& 44.70 & 0.811 & 0.211 &  50.27& 95.13   & 5.74 \\
NGC~2424\tablenotemark{g}&N&    3.15 & 4.806 & 0.056 &  79.17&       &       &       &       &         &      \\ 
                      & N  &    2.34 & 0.380 & 0.048 &   5.39&  2.02 & 10.834& 0.045 & 142.6 &         & 3.49 \\
                      & S  &    4.55 & 2.037 & 0.067 &  40.28&       &       &       &       &         &      \\  
                      & S  &    1.86 & 0.338 & 0.043 &   4.27&  2.31 & 5.431 & 0.048 &  76.66& 219.3   & 1.99 \\
             NGC~2820 & NW &  415.3  & 0.362 & 0.644 &  68.49&       &       &       &       &         &      \\  
       &NW\tablenotemark{e}&  529.1  & 0.334 & 0.727 &  71.00&  --   &  --   &  --   &  --   &         & --   \\
                      & SE &  116.3  & 0.485 & 0.341 &  48.51&       &       &       &       &         &      \\  
                      & SE &  537.6  & 0.268 & 0.733 &  57.62& 11.09 & 1.079 & 0.105 &  33.30& 33.30   & 1.89 \\      
NGC~3044\tablenotemark{h}&S& 1176.   & 0.381 & 1.084 & 121.3 &       &       &       &       &         &      \\  
                      & S  & 1242.   & 0.362 & 1.114 & 118.1 &  6.26 & 4.742 & 0.079 & 110.0 &         & 4.70 \\
                      & N  & 1101.   & 0.298 & 1.049 &  91.73&       &       &       &       &         &      \\  
                      & N  & 1347.   & 0.260 & 1.160 &  88.59& 12.82 & 2.569 & 0.113 &  85.4 & 195.4   & 5.21 \\
NGC~3044\tablenotemark{i}&N& 1235.   & 0.337 & 1.111 & 110.0 &       &       &       &       &         &      \\  
                      & N  & 1522.   & 0.292 & 1.233 & 105.6 & 16.37 & 2.781 & 0.128 & 104.3 &         & 7.24 \\
                      & S  & 1461.   & 0.272 & 1.208 &  96.13&       &       &       &       &         &      \\  
                      & S  & 2073.   & 0.204 & 1.439 &  86.08& 61.51 & 1.323 & 0.248 &  96.14& 200.4   &12.9  \\
             NGC~3432 & NW &   35.81 & 0.167 & 0.189 &   9.24&       &       &       &       &         &      \\  
                      & NW &   24.89 & 0.116 & 0.158 &   5.37&  7.99 & 1.009 & 0.089 &  26.45&         & 1.27 \\
                      & SE &   38.28 & 0.184 & 0.195 &  10.56&       &       &       &       &         &      \\  
                      & SE &   29.19 & 0.135 & 0.171 &   6.79&  6.82 & 1.180 & 0.083 &  28.59& 55.04   & 1.29 \\
\tableline

\end{tabular}
\end{table*}

\clearpage

\begin{table*}[h]
\tiny
\tablenum{5}
\caption{Calculated Properties of Galaxies based on Exponential Fits (cont'd)}

\begin{tabular}{lrrrrr|rrrr|r|r}
\tableline
\tableline
 &
 &
\multicolumn{4}{c}{One-exponential Fit\tablenotemark{a}} &
\multicolumn{4}{c}{Two-exponential Fit} &
 \\
 &
 &
EM\tablenotemark{b} &              
Scale &      
$n_e$ &         
``Disk'' &     
EM\tablenotemark{b}  &              
Scale &      
$n_e$ &         
eDIG &
Total eDIG & 
Recombination\\  
Galaxy &
Side &
@ z = 0 &              
Height &      
@ z = 0  &         
Mass\tablenotemark{c} &     
@ z = 0 &              
Height &      
@ z = 0 &         
Mass\tablenotemark{c} &     
Mass\tablenotemark{c} &
Rate\tablenotemark{d}\\  
 &
 &
(pc cm$^{-6}$) &                      
(kpc) &                
(cm$^{-3}$) &                     
(10$^6$ M$_\odot$) &                               
(pc cm$^{-6}$) &                      
(kpc) &                
(cm$^{-3}$) &                     
(10$^6$ M$_\odot$) &                              
(10$^6$ M$_\odot$) &
(10$^6$ s$^{-1}$ cm$^{-2}$) \\
\tableline
             NGC~4013 & NW &  141.7  & 0.579 & 0.376 &  64.09&       &       &       &       &         &      \\  
       &NW\tablenotemark{e}&  146.3  & 0.558 & 0.382 &  62.58&  --   &  --   &  --   &  --   &         & --   \\
                      & SE &  153.4  & 0.657 & 0.391 &  75.40&       &       &       &       &         &      \\  
                      & SE &  175.2  & 0.470 & 0.418 &  57.68& 21.15 & 1.513 & 0.145 &  64.72& 64.72   & 5.06 \\
             NGC~4197 & NW &    8.12 & 0.346 & 0.090 &   9.11&       &       &       &       &         &      \\  
                      & NW &    8.78 & 0.164 & 0.094 &   4.51&  1.28 & 4.947 & 0.036 &  51.90&         & 1.02 \\
                      & SE &    7.76 & 0.381 & 0.088 &   9.86&       &       &       &       &         &      \\  
                      & SE &    6.71 & 0.359 & 0.082 &   8.61&  1.22 & 4.860 & 0.035 &  49.77& 101.7   & 0.946\\
             NGC~5529 & NE &   56.16 & 1.175 & 0.237 &  81.68&       &       &       &       &         &      \\  
                      & NE &   58.32 & 0.948 & 0.241 &  67.23&  4.06 & 4.550 & 0.064 &  84.83&         & 2.96 \\
                      & SW &   39.94 & 1.161 & 0.200 &  67.86&       &       &       &       &         &      \\  
                      & SW &   85.55 & 0.624 & 0.292 &  53.53&  3.31 & 6.845 & 0.057 & 115.6 & 200.4   & 3.53 \\
             NGC~5965 & NW &   19.30 & 1.583 & 0.139 &  64.72&       &       &       &       &         &      \\  
                      & NW &   17.65 & 1.467 & 0.133 &  57.11&  1.26 & 11.577& 0.035 & 120.0 &         & 2.25 \\
                      & SE &   13.73 & 1.294 & 0.117 &  44.42&       &       &       &       &         &      \\  
                      & SE &    9.08 & 0.986 & 0.095 &  27.52&  3.58 & 6.523 & 0.060 & 114.4 & 234.4   & 3.73 \\
            ESO~240-11& SW &   31.77 & 2.569 & 0.178 & 134.5 &       &       &       &       &         &      \\  
                      & SW &   31.85 & 2.473 & 0.178 & 129.4 &  0.39 & 15.536& 0.020 &  90.48&         & 0.988\\
                      & NE &   41.53 & 1.883 & 0.204 & 112.5 &       &       &       &       &         &      \\  
                      & NE &   42.04 & 1.741 & 0.205 & 104.9 &  0.86 & 17.863& 0.029 & 153.3 & 243.8   & 2.39 \\
\tableline

\end{tabular}
\tablenotetext{a}{The second line for each object in the
``one-exponential fit'' column refers to the ``two-exponential fit''
thin disk. }

\tablenotetext{b}{The emission measure in the midplane of the galaxy
is based on the $H\alpha$ intensity at $\vert z \vert$ = 0, which is
derived from an extrapolation of the one- or two-exponential fit to
the vertical emission profiles.  Equation (1) is used for the
calculations. This quantity is an emission measure determined in the
midplane of the disk from the observed, edge-on orientation and is an
average over all radii since it is derived from the average vertical
profile of each galaxy.}

\tablenotetext{c}{The ionized mass is derived using equation (4). }

\tablenotetext{d}{The recombination rate in a column perpendicular to
the galactic plane is calculated using equation (5). }

\tablenotetext{e}{A two-exponential fit is not possible in this case.}

\tablenotetext{f}{The AAT observation of NGC~2188 does not fully
encompass the galaxy.  The values listed here only take a portion of
the galaxy into account, so the total ionized mass listed here should
be considered a lower limit.}

\tablenotetext{g}{The quantities for NGC 2424 were derived from the
WHT data. The quantities derived from the KPNO data are not reliable.}

\tablenotetext{h}{These quantities were derived from the CTIO data.}

\tablenotetext{i}{These quantities were derived from the WHT data.}

\normalsize
\end{table*}

\clearpage

\begin{table*}[h]
\tablenum{6}
\caption{Peak Correlation Coefficient between Disk and Extraplanar Line Emission.}

\begin{tabular}{lcccccccc}

\tableline
\tableline
        & & \multicolumn{3}{c}{Side 1} & & \multicolumn{3}{c}{Side 2} \\
Galaxy  & Telescope & & Peak & FWHM/2 (kpc) & & & Peak & FWHM/2 (kpc) \\
\tableline
NGC~7817                 & WHT  & NW & 0.70  & 3.1  & &  SE & 0.70  & 3.0 \\
NGC~55                   & CTIO & NE & 0.41  & 1.2  & &  SW & 0.56  & 1.1 \\
NGC~55                   & AAT  & NE & 0.24  & 0.3  & &  SW & 0.28  & 0.2 \\
NGC~891\tablenotemark{a} & WHT  & -- &  --   & --   & &  -- &  --   & --  \\
NGC~973                  & KPNO & NW & 0.61  & 8.9  & &  SE & 0.66  & 7.4 \\
NGC~1507                 & WHT  & W  & 0.86  & 1.1  & &  E  & 0.27  & 1.6 \\
ESO~362-11               & AAT  & S  & 0.15  & 1.6  & &  N  & 0.12  & 3.0 \\
NGC~2188                 & AAT  & W  & 0.72  & 1.7  & &  E  & 0.83  & 1.8 \\
NGC~2424                 & KPNO & N  & 0.70  & 6.1  & &  S  & 0.68  & 6.3 \\
NGC~2424                 & WHT  & N  & 0.28  & 4.4  & &  S  & 0.60  & 3.5 \\
NGC~2820                 & KPNO & NW & 0.63  & 1.6  & &  SE & 0.93  & 2.0 \\
NGC~3044                 & CTIO & S  & 0.63  & 4.1  & &  N  & 0.90  & 3.1 \\
NGC~3044                 & WHT  & S  & 0.49  & 5.0  & &  N  & 0.79  & 2.9 \\
NGC~3432                 & KPNO & NW & 0.42  & 1.7  & &  SE & 0.37  & 2.2 \\
NGC~4013                 & WHT  & NW & 0.06  & 0.8  & &  SE & 0.56  & 1.1 \\
NGC~4197                 & KPNO & NW & 0.59  & 3.9  & &  SE & 0.84  & 3.5 \\
NGC~5529                 & KPNO & NE & 0.76  & 10.  & &  SW & 0.54  & 10. \\
NGC~5965                 & KPNO & NW & 0.88  & 8.2  & &  SE & 0.90  & 8.0 \\
ESO~240-11               & AAT  & SW & 0.84  & 13.  & &  NE & 0.80  & 15. \\
\tableline

\end{tabular}

\tablenotetext{a}{The presence of reflective ghosts in the image of
NGC~891 prevents us from determining the radial profile of the
extraplanar material in this object.}

\normalsize
\end{table*}

\vskip 0.25in

\begin{table*}[h]
\tablenum{7}
\caption{Deviations from Average Vertical Profile}

\begin{tabular}{lccccc}

\tableline
\tableline
Galaxy     & Telescope & Side & $\sigma$ & Side & $\sigma$ \\
\tableline
NGC~7817    & WHT  & NW &  41.4 & SE &   2.7  \\
NGC~55      & CTIO & NE &   2.6 & SW &   3.9  \\
NGC~55\tablenotemark{a}   & -- & -- & -- &   --    \\
NGC~891     & WHT  & NW &   0.94 & SE &   1.4  \\
NGC~973     & KPNO & NW &   3.5 & SE &   1.9  \\
NGC~1507    & WHT  & W  &   9.9 & E  &   4.2  \\
ESO~362-11  & AAT  & S  &   0.82 & N  &  33.9  \\
NGC~2188    & AAT  & W  &   0.82 & E  &   0.63  \\
NGC~2424    & KPNO & N  &   3.7 & S  &   4.3  \\
NGC~2424    & WHT  & N  &   0.41 & S  &   0.56  \\
NGC~2820    & KPNO & NW &   2.1 & SE &  30.9  \\
NGC~3044    & CTIO & S  &   4.6 & N  &  13.2  \\
NGC~3044    & WHT  & S  &   2.6 & N  &   0.42 \\
NGC~3432    & KPNO & NW &   1.0 & SE &   4.9  \\
NGC~4013    & WHT  & NW &  22.4 & SE &   4.4  \\
NGC~4197    & KPNO & NW &   2.0 & SE &  13.3  \\
NGC~5529    & KPNO & NE &  11.3 & SW &   1.1  \\
NGC~5965    & KPNO & NW &   8.3 & SE &   6.4  \\
ESO~240-11  & AAT  & SW & 116.6 & NE &   4.4  \\
\tableline    

\end{tabular}

\tablenotetext{a}{The AAT data of NGC~55 focus only on a small
portion of the galaxy, and so it is not possible to derive 
these quantities from this set of observations.}

\end{table*}

\vskip 0.25in

\begin{table*}[h]
\tiny
\tablenum{8}
\caption{Pearson's Correlation Coefficients and P[null] Values between eDIG and Galaxy Properties.}

\begin{tabular}{lllllllllll}

\tableline 
\tableline 
 &
 Quantity &
 Galaxy Type &
 L$_{FIR}$/D$_{25}^2$ &
 L$_{IR}$/D$_{25}^2$ &
 $\frac{f_{25}}{f_{60}}$ &
 $\frac{f_{60}}{f_{100}}$ &
 Distance &
 Inclination &
 L$_{H\alpha}$(total)/D$_{25}^2$ \\
 eDIG Properties & & & & & & & & & \\
\tableline 
\multicolumn{10}{c}{Sample: all 17 objects } \\
\tableline 
L$_{H\alpha}$(total) &  r & 0.20   &     0.36  &      0.38   &    0.14   &     0.36   &     0.31   &     0.097  &      0.313 \\
              &  P[null]  & 0.43   &     0.16  &      0.13   &    0.58   &     0.15   &     0.23   &     0.71   &      0.22  \\
L$_{H\alpha}$(total)/D$_{25}^2$ & r &0.56&0.49 &      0.47   &    0.50   &     0.76   &     0.62   &     0.335  &      ----  \\
              &  P[null]  & 0.028  &     0.062 &      0.079  &    0.057  &     0.0010 &     0.015  &     0.22   &      ----  \\
L$_{H\alpha}$(eDIG)  &  r & 0.25   &     0.34  &      0.36   &    0.19   &     0.36   &     0.36   &     0.174  &      0.184 \\
              &  P[null]  & 0.34   &     0.19  &      0.16   &    0.46   &     0.16   &     0.15   &     0.51   &      0.48  \\
L$_{H\alpha}$(eDIG)/$_{H\alpha}$L(tot)& r &0.12&0.006&0.032  &    0.36   &     0.12   &     0.44   &     0.042  &      0.097 \\
              &  P[null]  & 0.66   &     0.98  &      0.90   &    0.15   &     0.65   &     0.080  &     0.87   &      0.71  \\
EM (z=0)             &  r & 0.14   &     0.54  &      0.53   &    0.28   &     0.37   &     0.30   &     0.195  &      0.501 \\
              &  P[null]  & 0.60   &     0.024 &      0.030  &    0.28   &     0.14   &     0.24   &     0.45   &      0.024 \\
Scale Height         &  r & 0.69   &     0.37  &      0.34   &    0.59   &     0.24   &     0.72   &     0.126  &      0.365 \\
              &  P[null]  & 0.0021 &     0.14  &      0.19   &    0.012  &     0.36   &     0.00098&     0.63   &      0.11  \\
$n_e$ (z=0)          &  r & 0.18   &     0.60  &      0.57   &    0.40   &     0.31   &     0.40   &     0.173  &      0.541 \\
              &  P[null]  & 0.48   &     0.011 &      0.017  &    0.011  &     0.22   &     0.11   &     0.51   &      0.014 \\
eDIG Ionized Mass    &  r & 0.25   &     0.88  &      0.85   &    0.40   &     0.32   &     0.22   &     0.287  &      0.259 \\
              &  P[null]  & 0.33   &  0.0000041&    0.0000014&    0.011  &     0.20   &     0.45   &     0.26   &      0.27  \\
Sigma\tablenotemark{a} & r& 0.10   &     0.005 &      0.009  &    0.079  &     0.27   &     0.10   &     0.135  &      0.111 \\
              &  P[null]  & 0.63   &     0.98  &      0.97   &    0.71   &     0.20   &     0.64   &     0.57   &      0.62  \\
\tableline 
\multicolumn{10}{c}{Sample: 10 objects (excluding NGC~55 and galaxies with $i < 85^\circ$)} \\
\tableline 
L$_{H\alpha}$(total) &  r & 0.313  &     0.474 &      0.518  &    0.243  &     0.317  &     0.443  &     0.123  &      0.516 \\
A              &  P[null]  & 0.38   &     0.17  &      0.13   &    0.50   &     0.37   &     0.20   &     0.73   &      0.13  \\
L$_{H\alpha}$(total)/D$_{25}^2$ & r &0.580&0.634&     0.668  &    0.328  &     0.836  &     0.591  &     0.065  &      ----  \\
              &  P[null]  & 0.10   &     0.067 &      0.049  &    0.39   &     0.0050 &     0.093  &     0.87   &      ----  \\
L$_{H\alpha}$(eDIG)  &  r & 0.255  &     0.432 &      0.483  &    0.281  &     0.345  &     0.448  &     0.091  &      0.503 \\
              &  P[null]  & 0.48   &     0.21  &      0.16   &    0.43   &     0.33   &     0.19   &     0.80   &      0.14  \\
L$_{H\alpha}$(eDIG)/$_{H\alpha}$L(tot)& r &0.393&0.164&0.133 &    0.325  &     0.058  &     0.238  &     0.140  &      0.104 \\
              &  P[null]  & 0.26   &     0.65  &      0.71   &    0.36   &     0.87   &     0.51   &     0.70   &      0.78  \\
EM (z=0)             &  r & 0.147  &     0.434 &      0.370  &    0.507  &     0.232  &     0.625  &     0.092  &      0.664 \\
              &  P[null]  & 0.56   &     0.072 &      0.13   &    0.032  &     0.35   &     0.0056 &     0.72   &      0.0050\\
Scale Height         &  r & 0.172  &     0.568 &      0.537  &    0.672  &     0.550  &     0.632  &     0.009  &      0.488 \\
              &  P[null]  & 0.50   &     0.014 &      0.022  &    0.0023 &     0.018  &     0.0049 &     0.97   &      0.055 \\
$n_e$ (z=0)          &  r & 0.208  &     0.515 &      0.455  &    0.580  &     0.341  &     0.674  &     0.094  &      0.729 \\
              &  P[null]  & 0.41   &     0.029 &      0.058  &    0.012  &     0.17   &     0.0022 &     0.71   &      0.0013\\
eDIG Ionized Mass    &  r & 0.292  &     0.592 &      0.618  &    0.579  &     0.234  &     0.639  &     0.001  &      0.115 \\
              &  P[null]  & 0.24   &     0.0096&      0.0062 &    0.012  &     0.35   &     0.0043 &     1.00   &      0.67  \\
Sigma\tablenotemark{a} & r& 0.042  &     0.233 &      0.229  &    0.190  &     0.337  &     0.200  &     0.203  &      0.248 \\
              &  P[null]  & 0.86   &     0.32  &      0.33   &    0.42   &     0.15   &     0.40   &     0.39   &      0.32  \\
\tableline
\end{tabular}
\normalsize \tablenotetext{a}{Sigma represents the deviation from the
average profile listed in Table 7.}
\end{table*}

\end{document}